%% file: main.tex
\documentclass[preprint,12pt,nopreprintline]{elsarticle}

\usepackage{amsmath,amssymb,amsfonts,mathtools,bm}
\usepackage{amsthm}
\usepackage{booktabs}
\usepackage{graphicx}
\usepackage{microtype}
\usepackage{etoolbox}
\usepackage{threeparttable}
\usepackage{tabularx}
\usepackage{array}
\usepackage{enumitem}
\usepackage[hidelinks]{hyperref}
\usepackage{xcolor}
\usepackage{xspace}
\usepackage{float}
\usepackage{pdflscape}
\usepackage{longtable}
\usepackage{capt-of}

\makeatletter
\appto\ps@pprintTitle{%
  \apptocmd\@oddfoot{\unskip}{}{}%
  \let\@evenfoot\@oddfoot}
\makeatother

\newcommand{\Tpre}{\mathcal{T}_0}
\newcommand{\Tpost}{\mathcal{T}_1}
\newcommand{\Oi}{\mathcal{O}_i}

\newcommand{\GraphSCM}{Graph-SCM\xspace}
\newcommand{\GeoEco}{GeoEco-SCM\xspace}
\newcommand{\R}{\mathbb{R}}
\newcommand{\1}{\bm{1}}

\newcommand{\tr}{\operatorname{tr}}
\newcommand{\argmin}{\operatorname*{arg\,min}}

\newtheorem{assumption}{Assumption}
\newtheorem{condition}{Condition}
\newtheorem{proposition}{Proposition}

\begin{document}
\begin{frontmatter}

\title{Graph-regularized synthetic control for spatial panels with post-only treated locations}

\author[1]{Hajime Seya}
\author[1]{Ryosei Moriwaki}
\author[2]{Daisuke Murakami}
\address[1]{Kobe University, Kobe, Japan}
\address[2]{The Institute of Statistical Mathematics, Tokyo, Japan}

\begin{abstract}
Spatial policy panels can contain treated locations that are observed only after an intervention and therefore have no pre-treatment outcome history. Without a pre-treatment trajectory to match, choosing donor weights for such \emph{post-only} locations requires an additional source of information. We propose \GraphSCM, which estimates donor weights jointly across treated locations and encourages geographically neighboring locations to use similar donor combinations. \GeoEco keeps the same geographic graph but adjusts the strength of its existing edges using pre-treatment economic similarity. Both methods handle complete, partial, and absent histories within one convex weight-estimation problem. Monte Carlo experiments compare them with covariate-balanced synthetic controls, equally weighted donor averages, ordinary synthetic control where estimable, and nonseparable spatio-temporal imputation and direct prediction. No method dominates. Graph-based borrowing improves on covariate balancing in several settings, but a selective-coverage design favors covariate balancing. Economic adjustment helps when geography alone misses relevant donor relationships. The results support graph regularization as a way to extend synthetic-control coverage when connected locations share similar donor representations, with its usefulness depending on the structure of the available information.
\end{abstract}


\end{frontmatter}

\section{Introduction}
\label{sec:intro}

Synthetic control methods (SCM) estimate an untreated counterfactual by choosing a convex combination of untreated donor units whose pre-treatment outcomes resemble those of a treated unit \citep{abadie2010synthetic,abadie2021using}. The pre-treatment trajectory is therefore not merely a descriptive input: it is the main source of information used to determine the donor representation, meaning the target's vector of donor weights. This creates a difficult but common problem in spatial panels. A treated location may enter an outcome database only after an intervention. Examples include newly sampled land-price points, newly observed monitoring locations, or spatial records whose historical geocoding is incomplete. Such a location can be observed in the post-treatment period while having no own pre-treatment outcome trajectory. We refer to this observation pattern as \emph{post-only}. We use PRE and POST to denote the pre-treatment and post-treatment periods, respectively.

The post-only case is qualitatively different from ordinary intermittent missingness. If target $i$ has no observed outcome in any pre-treatment period, the usual criterion for matching pre-treatment outcomes contains no target-specific fitting information. Consequently, the donor weights are not uniquely determined: every feasible weight vector satisfies the same target-specific criterion. Here we focus on weights learned from pre-treatment outcomes; observed covariates can provide a separate balancing criterion. Dropping such locations avoids the need to construct their counterfactuals, but changes the population covered by the analysis. This becomes particularly important when historical data availability is associated with spatial or economic characteristics. In such settings, excluding post-only locations changes the target population and, when treatment effects are heterogeneous, may also change the average treatment effect being estimated.

The key idea is to borrow information about donor weights across locations. Because the graph links targets, local PRE information can guide combinations of geographically distant, unaffected donors. Nearby treated locations with observed pre-treatment histories can provide information about which donor combinations are plausible if donor relationships vary smoothly over space.

We exploit this idea through graph-regularized synthetic control (Graph-SCM). Treated locations are connected by a geographic graph, and the donor weights for all locations are estimated jointly. The usual pre-treatment matching criterion determines the weights at locations with observed histories, while a graph penalty discourages large differences in donor weights between geographically connected locations. Consequently, a post-only location can borrow information from connected locations with observed pre-treatment outcomes rather than being dropped from the analysis or having its missing outcomes imputed first.

Geographic proximity alone, however, need not imply that two locations have similar donor relationships. We therefore also propose GeoEco-SCM, an economically reweighted extension of Graph-SCM. GeoEco-SCM retains the same geographic graph but adjusts the strength of its existing connections using economic characteristics measured before treatment. Connections between geographically neighboring locations that are also economically similar receive greater relative weight, whereas connections between economically dissimilar neighbors are weakened. Thus, economic information changes how strongly donor weights are shared across geographically connected locations without changing which locations are connected.

Complete, partial, and absent PRE histories are handled within the same convex estimation problem: each location contributes the outcome-matching terms available for it, and the graph links the donor representations across locations. Economic covariates then adjust this sharing locally along existing edges. This provides a common constrained formulation for heterogeneous historical coverage.

This construction has three useful features for spatial statistics. First, local similarity is represented by a graph Laplacian, a standard way to impose smoothness on irregular spatial supports \citep{belkin2006manifold}. Second, if a location has no pre-treatment outcomes, the graph penalty makes its donor weights a weighted combination of donor weights propagated from connected locations with observed histories. In graph terminology this weighted-average interpolation is a discrete harmonic extension \citep{zhu2003harmonic}, but its substantive meaning is simply that observed-history locations provide the information that is passed through the graph. For graph-based borrowing to reach a post-only location, that location must be connected, directly or indirectly, to a location with observed pre-treatment outcomes. Third, the graph penalty is imposed on the \emph{donor weights} across locations. It therefore makes a different modeling assumption from spatial kriging or spatial matrix completion. Section~\ref{sec:graphscm} gives the precise interpolation conditions and the corresponding equations.

The paper makes four contributions. First, we formulate the post-only SCM problem and propose a joint graph-regularized estimator for all treated locations. Second, we explain how information reaches a post-only location through the graph and state the connectivity condition needed for this interpolation to be well defined. Third, we extend the method so that pre-treatment economic similarity can weaken or strengthen existing geographic connections without changing the graph topology. Fourth, we use a structured Monte Carlo study to identify when graph-based borrowing helps and when it does not. Five primary data-generating processes (DGPs) examine geographic smoothness, economic information, selective observation, complex donor dynamics, and a failure of spatial smoothness. Five additional stress tests are retained in the Online Appendix.

\section{Relation to existing approaches}
\label{sec:literature}

\subsection{Synthetic control with incomplete pre-treatment information}

Many standard SCM formulations and extensions use treated-unit pre-treatment information to learn donor relationships. Augmented SCM combines SCM balancing with an outcome-model correction \citep{benmichael2021augmented}; partially pooled and staggered-adoption formulations borrow across treated units but retain target-specific pre-treatment outcomes \citep{benmichael2022staggered}; generalized synthetic control imposes an interactive fixed-effect structure on the panel \citep{xu2017generalized}; and synthetic difference-in-differences combines unit and time reweighting \citep{arkhangelsky2021sdid}. Other refinements change how the available pre-treatment trajectory is used. For example, \citet{sun2024temporal} study temporal aggregation of pre-treatment outcomes and propose balancing both aggregated and disaggregated series to trade off noise reduction against loss of signal. These approaches address important limitations of ordinary SCM, but their usual implementations based on pre-treatment outcomes use target PRE information and do not, without an additional cross-location or covariate restriction, determine a site-specific donor representation for a target with no own PRE observations.

In empirical spatial applications, another practical response is to restrict the analysis to locations observed throughout the study period. For example, \citet{kanasugi2018highspeed}, in studying the effect of a high-speed railway on residential land prices, construct a balanced panel for 2008--2015 and explicitly exclude land-price points that newly enter or disappear during the analysis period. This avoids the need to construct counterfactuals for locations without a usable pre-treatment history, but it also changes the population represented by the analysis. The post-only problem considered here instead asks whether such locations can be retained and assigned a counterfactual under an explicit cross-location restriction.

A closely related line of work studies missing pre-treatment outcomes directly. Robust SCM first denoises and completes a noisy panel by singular-value thresholding \citep{amjad2018robust}. \citet{liu2026missing} study conventional SCM when pre-treatment outcomes are missing. Using the vertical-regression representation of SCM, they characterize the bias that can arise when units with incomplete pre-treatment data are excluded and develop a sensitivity-analysis framework for assessing the robustness of the resulting treatment-effect estimates. This is particularly relevant to empirical practice because it formalizes the consequences of restricting the analysis to units with usable pre-treatment histories. Their objective, however, is to assess sensitivity to missing pre-treatment data and the associated sample exclusion, rather than to construct donor weights for a treated target that has no own pre-treatment trajectory. The post-only pattern considered here has no observed PRE outcome for the target, so the target-specific SCM fit term disappears and an additional source of cross-location information is needed to determine the target's donor representation.

The closest causal-identification result to a genuinely post-only target is the proximal synthetic-control approach of \citet{liu2024proximal}. Under proxy and bridge conditions, their framework can identify causal effects using post-intervention surrogates, even in settings with no treated-unit pre-intervention outcomes. The identifying information is fundamentally different from ours. Proximal SCM borrows through post-treatment proxy restrictions. Our estimator does not use the post-treatment outcome of a post-only target to learn its weights; instead, it borrows the donor representation from other treated locations through a graph constructed from geography and pre-treatment or treatment-invariant covariates.

\subsection{Matrix completion and spatial missing-data methods}

Matrix-completion approaches regard unobserved counterfactuals as missing entries of a structured panel. Nuclear-norm methods impose low rank \citep{athey2021matrix}; latent-factor methods allow general missing patterns and provide inferential tools \citep{xiong2023latent}; and causal matrix completion studies non-random observation mechanisms \citep{agarwal2023causal}. These methods borrow through a common low-dimensional outcome surface.

Spatial matrix completion adds another relevant perspective. \citet{vu2022spatial} develop a matrix-completion method for spatially misaligned, high-dimensional air-pollution data, combining low-rank structure with spatial dependence to fill missing pollutant measurements. While their study is developed for spatially misaligned and incomplete pollution data rather than for synthetic-control estimation, it provides an important example of how spatial structure and missing observations can be incorporated jointly within a unified estimation framework. A closer precursor in the synthetic-control literature is the SMaC approach of \citet{grossi2023smac}, which imputes treated-area counterfactual time series as weighted averages of control time series while allowing the weights to vary smoothly over space through a Gaussian-process structure.

Graph-SCM places the structural restriction at a different level. It keeps the usual SCM representation
\begin{equation}
Y_{it}(0) \approx \bm{w}_i^{\top}\bm{Y}_t^C,
\label{eq}
\end{equation}
where $Y_{it}(0)$ denotes the untreated potential outcome for target location $i$ at time $t$, $\bm{Y}_t^C$ denotes the vector of observed outcomes for the untreated donor units at time $t$, and $\bm{w}_i$ is the corresponding vector of donor weights for target location $i$. As in conventional SCM, the elements of $\bm{w}_i$ are nonnegative and sum to one. Collecting the target-specific weight vectors as rows gives the target-by-donor weight matrix $\bm W$. Graph-SCM regularizes this matrix across locations by encouraging geographically connected targets to have similar donor-weight patterns. In simple terms, matrix-completion methods impose structure on the outcome matrix to recover missing entries, whereas Graph-SCM imposes spatial structure on the donor weights used to construct the untreated potential outcomes in Eq.~\eqref{eq}. These modeling structures can both be appropriate for the same data-generating process, and neither method class is uniformly more general than the other.

\begin{table}[!t]
\centering
\caption{Primary structural assumptions and treatment of post-only targets in related approaches.}
\label{tab:method-structure}
\footnotesize
\setlength{\tabcolsep}{4pt}
\renewcommand{\arraystretch}{0.95}
\begin{tabularx}{\textwidth}{@{}p{0.16\textwidth}XXX@{}}
\toprule
Approach & Primary structural assumption & Information used & Post-only target \\
\midrule

Ordinary SCM
& Target outcomes can be represented by a convex combination of donor outcomes
& Own observed PRE outcomes matched to donor PRE outcomes
& Target-specific $\bm{w}_i$ cannot be determined without an additional restriction \\

Matrix completion
& Outcome matrix has a shared low-rank or factor structure
& Observed outcome entries across units and time
& Recovery uses the shared outcome structure; a completely unobserved untreated target row needs additional information to determine its loadings \\

Spatial matrix completion
& Outcome matrix has shared latent structure and spatial dependence
& Observed outcome entries and spatial information
& Missing entries may be recovered using both outcome and spatial structure \\

Proximal SCM
& Proxy/bridge conditions link observed proxy information to the untreated outcome
& Proxy or surrogate information and associated moment restrictions
& Can be handled under the corresponding bridge and proxy conditions \\

Spatial vertical regression
& Control-unit regression coefficients vary smoothly across spatially related treated locations
& Observed PRE outcomes across treated locations, control outcomes, and spatial relationships
& The cited study does not evaluate the post-only setting \\

Graph-SCM
& Target-specific donor weights vary smoothly over a geographic graph
& Own PRE matching where available, plus graph-based sharing of donor weights
& $\bm{w}_i$ is inferred from connected targets with observed PRE outcomes \\

GeoEco-SCM
& Donor weights are graph-smooth, with edge strengths also reflecting PRE economic similarity
& Own PRE matching, geographic connections, and PRE economic similarity
& Same graph-based borrowing, with borrowing strength adjusted by PRE economic similarity \\
\bottomrule
\end{tabularx}
\end{table}

\subsection{Spatially structured donor relationships}

The closest methodological precedent to Graph-SCM is the spatial vertical regression (SVR) approach of \citet{grossi2026svr}. SVR writes the untreated potential outcome of each treated location as a unit-specific linear regression on the control outcomes and places Gaussian-process priors on the coefficient vector associated with each control unit across treated locations. Nearby treated locations are therefore encouraged to have similar relationships with the controls, and the model can also allow spatial dependence in the residuals. This shares the central idea of Graph-SCM that donor or control relationships can themselves vary smoothly across spatially related treated locations, rather than requiring only the outcomes to be spatially smooth.

Our focus is the post-only setting. We connect treated locations through a graph and constrain their donor weights to be nonnegative and sum to one. The resulting convex weight-estimation problem makes the information passed to a post-only location explicit.

A related Bayesian extension by \citet{grossi2025spatialpriors} incorporates spatial information in a different way: its prior penalizes the selection of more distant control units and is embedded in a semiparametric framework intended to account for unobserved spatial confounding. This makes geographic distance informative about which donors should be selected, whereas Graph-SCM uses spatial relationships among treated targets to regularize how their donor-weight vectors vary across locations. Together with the earlier SMaC formulation of \citet{grossi2023smac}, these contributions show that spatial information can enter synthetic-control analysis through several distinct channels: the outcome model, donor selection, or cross-target similarity in donor relationships.

A distinct recent proposal is Graph-Regularized Synthetic Control (Graph-SC) by \citet{zhang2026graph}. Its graph is defined over donors and regularizes coefficients assigned to similar donor trajectories, with the aim of stabilizing counterfactual representations under donor redundancy. The graph in the present paper instead lies on the \emph{target} axis: it couples the weight vectors of different treated locations and supplies an interpolation rule for targets without their own outcome histories.

\subsection{Spatial dependence and spillovers in synthetic control}

Spatial structure has also been incorporated into SCM to address interference between units. \citet{cao2026spillover} develop estimation and inference for direct treatment effects and spillover effects under a specified spillover structure. \citet{sakaguchi2026spillover} extend SCM with a spatial autoregressive panel model so that treatment can affect not only the treated unit but also other units through a spatial dependence structure; they develop Bayesian inference with regularization for both treatment and spillover effects. In that framework, the spatial-weight matrix represents propagation of outcomes and treatment spillovers across units. This use of spatial structure is different from ours: Graph-SCM currently assumes that donor outcomes are not contaminated by treatment and instead uses a graph to regularize similarities in the donor-weight vectors across treated target locations. These approaches are complementary: spillover-aware SCM addresses interference, whereas Graph-SCM addresses counterfactual construction when a target has little or no own pre-treatment outcome information.

\subsection{Outcome-side spatial prediction}

A closely related recent contribution is \citet{hiramoto2026block}, who combine synthetic control with block kriging for point-referenced data. Their work provides a distinct way of incorporating explicit geostatistical structure into SCM. Our focus is different: we ask how donor representations can be constructed for target locations that have no own pre-treatment outcomes.

A different strategy for the post-only problem is to treat the missing history itself as a spatial prediction problem. More generally, \citet{vega2026gp} show how Gaussian-process models can be used for causal inference with spatio-temporal panel data and relate the resulting posterior predictions to weighted combinations of observed outcomes, with weights reflecting spatial and temporal similarity. This is conceptually close to synthetic control but places the primary structure on the outcome process rather than on a target-by-donor weight field. A spatio-temporal model can likewise first predict the missing pre-treatment outcomes and then ordinary SCM can be applied to the completed history. Alternatively, the same spatio-temporal model can predict the untreated post-treatment outcome directly. We use a nonseparable covariance model from the Gneiting class \citep{gneiting2002nonseparable} for both strategies. The economic-covariate benchmarks use the same types of PRE economic variables as \GeoEco, but include them in the outcome mean model rather than in target-graph edge weights. The comparison therefore asks a substantive modeling question: is it more useful to model the outcome surface directly, or to borrow the donor representation across nearby locations?

\section{Problem formulation}
\label{sec:setup}

\subsection{Potential outcomes, donors, and observation regimes}

Throughout the paper, vectors and matrices are written in bold italic type, while scalar quantities are written in ordinary math italic.

Let $i=1,\ldots,N$ index treated target locations, $j=1,\ldots,J$ index untreated donor locations, and $t=1,\ldots,T$ index time. Treatment begins after period $T_0$. Define
\[
\Tpre=\{1,\ldots,T_0\},\qquad
\Tpost=\{T_0+1,\ldots,T\}.
\]
Let $Y_{it}(0)$ and $Y_{it}(1)$ denote the untreated and treated potential outcomes of target $i$. The observed target outcome is
\begin{equation}
Y_{it}=Y_{it}(0)+D_{it}\tau_{it},
\qquad
\tau_{it}=Y_{it}(1)-Y_{it}(0),
\label{eq:potential-outcome}
\end{equation}
where $D_{it}=1$ after treatment for treated targets and zero otherwise. Donor $j$ is untreated throughout the study window, and its observed outcome is denoted $Y^C_{jt}$.

Historical coverage can vary by target. Let
\begin{equation}
\Oi=\{t\in\Tpre:Y_{it}\ \text{is observed}\}
\label{eq:Oi}
\end{equation}
be the set of observed pre-treatment periods for target $i$. We distinguish three regimes:
\begin{align}
\text{complete PRE: }&\Oi=\Tpre,\\
\text{partial PRE: }&0<|\Oi|<|\Tpre|,\\
\text{post-only: }&\Oi=\varnothing.
\end{align}
Define the set of pre-treatment-observed targets
\begin{equation}
\mathcal I_+=\{i:|\Oi|>0\},\qquad N_+=|\mathcal I_+|,
\end{equation}
and the post-only set $\mathcal I_0=\{1,\ldots,N\}\setminus\mathcal I_+$.

Let
\begin{equation}
\bm{Y}_t^C=(Y^C_{1t},\ldots,Y^C_{Jt})^{\top}\in\R^J
\end{equation}
collect donor outcomes at time $t$. Each target receives a donor-weight vector
\begin{equation}
\bm{w}_i=(w_{i1},\ldots,w_{iJ})^{\top}\in\Delta_J,
\qquad
\Delta_J=\{\bm w\in\R^J:\bm w\ge0,\ \1^{\top}\bm w=1\}.
\label{eq:simplex}
\end{equation}
The set $\Delta_J$ is the simplex: all nonnegative donor-weight vectors whose entries sum to one.

\subsection{Why ordinary SCM is underdetermined for a post-only target}

For a target with at least one observed pre-treatment outcome, a simple SCM formulation based on observed pre-treatment outcomes solves
\begin{equation}
\widehat{\bm w}_i^{\mathrm{SCM}}
\in\argmin_{\bm w\in\Delta_J}
\frac{1}{|\Oi|}
\sum_{t\in\Oi}
\left(Y_{it}-\bm w^{\top}\bm{Y}_t^C\right)^2.
\label{eq:ordinary-scm}
\end{equation}
If $\Oi=\varnothing$, define the empty-history extension of this fitting loss to be zero, without dividing by $|\Oi|$. It is then constant over the simplex, so that formally $\argmin_{\bm w\in\Delta_J}0=\Delta_J$. The optimization problem is therefore not undefined; rather, the target-specific representation is completely underdetermined by the target's own PRE outcomes. This is a structural absence of the matching criterion, not a numerical convergence problem.

Discarding post-only targets changes the population over which the average treatment effect on the treated (ATT) is evaluated. If $R_i=1$ denotes having any pre-treatment outcome, then the estimand based only on analyzable targets is
\begin{equation}
ATT_{\mathrm{PRE}}(t)=E[\tau_{it}\mid R_i=1],
\label{eq:attpre}
\end{equation}
whereas the policy estimand for all treated targets is
\begin{equation}
ATT_{\mathrm{ALL}}(t)=E[\tau_{it}].
\label{eq:attall}
\end{equation}
These two quantities can differ when historical coverage is associated with treatment-effect heterogeneity. The target locations are assumed to be definable independently of treatment; the framework does not cover entities whose existence is caused by the intervention.

\section{Graph-regularized synthetic control}
\label{sec:graphscm}

The procedure has four steps. First, construct a graph between target locations from geography and, for GeoEco-SCM, PRE economic attributes. Second, use held-out PRE outcomes to select how strongly neighboring targets share donor weights and how much economic similarity modifies the geographic edges (Section~\ref{sec:tuning}). Third, fit all target weights together using the available PRE outcomes. Finally, combine each target's fitted weights with donor POST outcomes to predict its untreated path. Target POST outcomes enter only afterward, when treatment effects are calculated.

\subsection{Roles of assumptions and support conditions}

The method combines two substantive modeling requirements with two support and design conditions. The donor representation must remain useful from PRE to POST (Assumption~\ref{ass:representation}), and connected targets should have similar donor representations (Assumption~\ref{ass:graphsmooth}). Graph-based borrowing also requires a path from each post-only target to an observed-history target (Condition~\ref{cond:anchor}), while geography and economic covariates must be available independently of target POST outcomes (Condition~\ref{cond:graph-measurable}). These requirements concern donor validity, the plausibility of borrowing, and the information available to the estimator.

For partial histories, the observed PRE periods must be informative about the donor relationship of interest. As in conventional SCM, causal interpretation additionally requires no anticipation and donor outcomes that remain untreated and are not materially contaminated by treatment spillovers.

\subsection{Baseline SCM representation and the additional graph restriction}

The first assumption is inherited from ordinary SCM. Its purpose is to formalize two familiar requirements: the donor pool must provide a useful convex representation of each target, and that representation must remain informative from the pre-treatment period into the post-treatment period.

\begin{assumption}[Baseline SCM donor support and temporal stability]
\label{ass:representation}
For each target location $i$, there exists a donor-weight vector $\bm{w}_i^\star\in\Delta_J$ such that
\begin{equation}
Y_{it}(0)=\bm{w}_i^{\star\top}\bm{Y}_t^C+\xi_{it},
\qquad t\in\Tpre\cup\Tpost,
\label{eq:true-representation}
\end{equation}
where $Y_{it}(0)$ is the untreated potential outcome of target $i$ at time $t$, $\bm{Y}_t^C$ is the $J$-dimensional vector of donor outcomes at time $t$, and $\xi_{it}$ is the approximation error, that is, the part of the untreated outcome not reproduced by the donor combination. The approximation is assumed to be sufficiently accurate for the counterfactual prediction problem of interest.
\end{assumption}

Because $\bm w_i^\star\in\Delta_J$, its elements are nonnegative and sum to one. Thus the synthetic control is a convex combination of donor outcomes rather than an unrestricted linear extrapolation across donors. At the same time, SCM still extrapolates the donor relationship learned before treatment into the post-treatment period. The substantive temporal restriction in Assumption~\ref{ass:representation} is therefore that the same donor-weight vector remains relevant before and after treatment. This requirement applies to ordinary SCM as well as to Graph-SCM.

Assumption~\ref{ass:representation} alone does not solve the post-only problem. If target $i$ has no pre-treatment outcomes, its own data provide no information for learning $\bm{w}_i^\star$. Graph-SCM therefore adds a second restriction whose purpose is to justify borrowing donor-weight information across target locations.

Let $G=(V,E,\bm A)$ denote an undirected weighted graph over the $N$ target locations. Here, $V=\{1,\ldots,N\}$ is the set of target locations, $E$ is the set of undirected edges connecting target pairs, and $\bm A=(a_{ik})\in\mathbb R^{N\times N}$ is the symmetric weighted adjacency matrix. The scalar $a_{ik}=a_{ki}\ge0$ measures the strength of the connection between targets $i$ and $k$; pairs that are not connected have $a_{ik}=0$.

Collect the true donor-weight vectors into
\begin{equation}
\bm{W}^\star=(\bm{w}_1^\star,\ldots,\bm{w}_N^\star)^\top\in\mathbb R^{N\times J},
\label{eq:true-weight-matrix}
\end{equation}
so that the $i$th row of $\bm{W}^\star$ is the donor representation of target $i$.

\begin{assumption}[Similarity of donor representations across connected locations]
\label{ass:graphsmooth}
For the graph used by the estimator,
\begin{equation}
\mathcal E_{\bm A}(\bm{W}^\star)
=\frac{1}{\sum_{(i,k)\in E}a_{ik}}
\sum_{(i,k)\in E}
a_{ik}\|\bm{w}_i^\star-\bm{w}_k^\star\|_2^2
\label{eq:true-energy}
\end{equation}
is sufficiently small for graph-based borrowing to be informative.
\end{assumption}

Here, $\|\cdot\|_2$ denotes the Euclidean norm. Hence $\|\bm{w}_i^\star-\bm{w}_k^\star\|_2^2$ measures how different the donor-weight vectors are at two connected targets, while $a_{ik}$ gives greater importance to differences along stronger graph connections. A small value of $\mathcal E_{\bm A}(\bm{W}^\star)$ therefore means that strongly connected targets tend to use similar donor combinations.

This is the main substantive assumption added by Graph-SCM. It concerns smoothness of the \emph{donor representation}, not necessarily smoothness or stationarity of the observed outcome surface. If connected targets have unrelated donor representations, graph-based borrowing need not help and can introduce bias.

\subsection{A common penalized estimator}
\label{sec:penalized-estimator}

We now translate the two ideas above into an estimator. Recall that $\Oi$ is the set of observed pre-treatment periods for target $i$, $\mathcal I_+=\{i:|\Oi|>0\}$ is the set of targets with at least one observed pre-treatment outcome, and $N_+=|\mathcal I_+|$. For $\bm{W}=(\bm{w}_1,\ldots,\bm{w}_N)^\top\in\mathbb R^{N\times J}$, define the pre-treatment fitting loss
\begin{equation}
\mathcal L_{\mathrm{fit}}(\bm{W})
=\frac{1}{N_+}\sum_{i\in\mathcal I_+}
\frac{1}{|\Oi|}\sum_{t\in\Oi}
\left(Y_{it}-\bm{w}_i^{\top}\bm{Y}_t^C\right)^2.
\label{eq:fit-loss}
\end{equation}
Throughout, at least one target has observed PRE outcomes and the graph has positive total edge weight. The inner average measures pre-treatment fit for one target using only the periods in which that target is observed. The outer average gives each target equal weight, so a complete-history target does not dominate a partial-history target merely because it has more observed periods.

The second part of the estimator penalizes differences in donor weights across connected targets. Recall that $\bm A=(a_{ik})$ is the $N\times N$ symmetric weighted adjacency matrix of $G$, with $a_{ik}>0$ for connected pairs and $a_{ik}=0$ otherwise. Define
\begin{equation}
\mathcal E_{\bm A}(\bm{W})
=\frac{1}{A_+}
\sum_{(i,k)\in E}a_{ik}\|\bm{w}_i-\bm{w}_k\|_2^2,
\qquad
A_+=\sum_{(i,k)\in E}a_{ik},
\label{eq:graph-energy}
\end{equation}
where $A_+$ is the total edge weight. This quantity is small when strongly connected targets have similar donor-weight vectors.

For compact matrix notation, let $\bm D=\operatorname{diag}(d_1,\ldots,d_N)$, where $d_i=\sum_k a_{ik}$ is the total connection strength, or weighted degree, of target $i$. The matrix
\begin{equation}
\bm L=\bm D-\bm A
\label{eq:laplacian-definition}
\end{equation}
is the weighted graph Laplacian. This standard graph matrix summarizes which locations are connected and how strongly. With each undirected edge counted once,
\begin{equation}
\sum_{(i,k)\in E}a_{ik}\|\bm{w}_i-\bm{w}_k\|_2^2
=\operatorname{tr}(\bm W^{\top}\bm L\bm W),
\label{eq:laplacian-trace}
\end{equation}
where $\operatorname{tr}(\cdot)$ denotes the matrix trace. Equation~\eqref{eq:laplacian-trace} is only a compact matrix expression of the pairwise penalty in Eq.~\eqref{eq:graph-energy}; the substantive idea remains the same.

We use a common objective for both proposed estimators. Write $\bm A(\beta)$ for a target graph whose edge strengths reflect geographic proximity and can also account for pre-treatment economic similarity. The parameter $\beta\ge0$ controls this economic adjustment: $\bm A(0)$ uses geography alone. For this geography-only case, Eqs.~\eqref{eq:graph-energy}--\eqref{eq:laplacian-trace} are evaluated with $\bm A=\bm A(0)$; the same definitions apply to the economically adjusted graph with $\bm A=\bm A(\beta)$. Section~\ref{sec:geoeco} gives the graph construction. For a chosen $\beta$, all graph quantities, including $A_+$ and $\bm L$, are computed from $\bm A(\beta)$; we suppress this dependence where it is clear from context.

The full estimator combines pre-treatment fit, graph-based similarity, and a small numerical ridge:
\begin{equation}
\begin{split}
\widehat{\bm W}(\lambda,\beta;\varepsilon)
\in{}&\argmin_{\bm W:\,\bm w_i\in\Delta_J}\\
&\left\{
\mathcal L_{\mathrm{fit}}(\bm W)
+\lambda\mathcal E_{\bm A(\beta)}(\bm W)
+\frac{\varepsilon}{N}\|\bm W\|_F^2
\right\},
\end{split}
\label{eq:graphscm-objective}
\end{equation}
where $\lambda,\beta,\varepsilon\ge0$ and $\|\bm W\|_F^2=\sum_{i=1}^N\sum_{j=1}^J w_{ij}^2$. \GraphSCM fixes $\beta=0$, so borrowing depends on geography alone. \GeoEco uses the same objective but allows $\beta$ to be selected together with $\lambda$, so economic similarity can modify borrowing along the geographic edges. Economic information therefore enters through the graph weights, rather than through a separate fitting-loss term.

The parameter $\lambda$ controls the strength of graph regularization. At locations with observed PRE histories, small values place more emphasis on their own fit, while larger values give greater influence to agreement across connected targets. A post-only target has no own fitting term and receives information through these connections.

Each target still has its own donor-weight vector. Graph regularization trades the flexibility of independent fits for information sharing: it can stabilize estimates from short histories, but can bias weights when connected targets have different donor relationships.

The final term is a \emph{numerical ridge}. Short PRE histories or highly similar donor trajectories can leave several weight matrices with the same or nearly the same fitting and graph-penalty values. A positive $\varepsilon$ makes the objective strictly convex, giving a unique mathematical minimizer while preserving the quadratic-program structure. On the simplex, the ridge favors less concentrated donor weights, toward equal allocation across donors. Conditional on the fitted weights at locations with observed PRE histories, post-only predictions then combine predictions propagated from those locations with the equally weighted average of donor outcomes. The mathematical details are given in the Online Appendix, Section~1.3.

In this study, we fix $\varepsilon=10^{-10}$ throughout tuning and final estimation; only $\lambda$ and, for \GeoEco, $\beta$ are selected by validation. This value keeps the added penalty at most $10^{-10}$ in the normalized objective for any feasible weights. The bound and its implications are given in the Online Appendix, Section~1.3. The following subsections first describe the borrowing mechanism in the special case $\varepsilon=0$, for which the interpolation formulas take a particularly simple form.

\subsection{How donor-weight information reaches a post-only target}

To make the borrowing mechanism explicit, fix a graph $\bm A(\beta)$ and consider $\lambda>0$ and $\varepsilon=0$ in Eq.~\eqref{eq:graphscm-objective}. Separate the targets into two groups. Let $\mathcal A=\mathcal I_+$ denote targets with at least one observed pre-treatment outcome. We call them \emph{anchors} because their donor weights are directly informed by observed PRE outcomes. Let $\mathcal U=\mathcal I_0$ denote post-only targets, which have no observed PRE outcomes.

Let $\bm{W}_{\mathcal A}$ be the matrix obtained by stacking the donor-weight vectors of the anchor locations, and let $\bm{W}_{\mathcal U}$ be the corresponding matrix for post-only locations. Thus, $\bm{W}_{\mathcal A}$ contains weights informed directly by observed PRE outcomes, whereas $\bm{W}_{\mathcal U}$ contains weights that must be inferred through graph borrowing.

Order the targets so that anchors come first and post-only locations second, and partition the graph Laplacian as
\begin{equation}
\bm L=
\begin{pmatrix}
\bm L_{\mathcal A\mathcal A} & \bm L_{\mathcal A\mathcal U}\\
\bm L_{\mathcal U\mathcal A} & \bm L_{\mathcal U\mathcal U}
\end{pmatrix}.
\label{eq:lap-partition}
\end{equation}
For example, $\bm L_{\mathcal U\mathcal A}$ summarizes graph connections from post-only locations to anchors, while $\bm L_{\mathcal U\mathcal U}$ summarizes connections among post-only locations.

For graph borrowing to determine a post-only weight vector, information from at least one anchor must be able to reach that location.

\begin{condition}[Graph support for post-only targets]
\label{cond:anchor}
Every post-only target is connected, directly or through other target locations, to at least one anchor. Equivalently, every connected component of the target graph $G=(V,E,\bm A)$ that contains a post-only target also contains at least one anchor.
\end{condition}

Here, a \emph{connected component} means a group of locations that can be reached from one another through a sequence of graph edges. Condition~\ref{cond:anchor} is directly checkable from the graph and observation pattern. If a group of post-only locations is isolated from all anchors, graph smoothness can make their weights similar to one another.

The next result formalizes what the graph penalty does once the anchor weights are held fixed. Here $\bm W_{\mathcal A}$ can be any fixed matrix of feasible weights, including weights estimated from the observed data.

\begin{proposition}[Donor-weight interpolation for post-only locations]
\label{prop:harmonic}
Fix a graph and simplex-valued anchor weights $\bm{W}_{\mathcal A}$. Consider $\lambda>0$ and $\varepsilon=0$. Under Condition~\ref{cond:anchor}, the post-only weights that minimize the graph penalty satisfy
\begin{equation}
\bm L_{\mathcal U\mathcal U}\bm{W}_{\mathcal U}
+\bm L_{\mathcal U\mathcal A}\bm{W}_{\mathcal A}=0,
\label{eq:dirichlet-system}
\end{equation}
and therefore
\begin{equation}
\bm{W}_{\mathcal U}=\bm H\bm{W}_{\mathcal A},
\qquad
\bm H=-\bm L_{\mathcal U\mathcal U}^{-1}\bm L_{\mathcal U\mathcal A}.
\label{eq:harmonic-solution}
\end{equation}
The entries of $\bm H$ are nonnegative and each row sums to one:
\begin{equation}
\bm H\ge0,
\qquad
\bm H\bm 1_{|\mathcal A|}=\bm 1_{|\mathcal U|},
\label{eq:h-rowstochastic}
\end{equation}
where $\bm 1_m$ denotes an $m$-dimensional vector of ones. Hence each post-only donor-weight vector is a convex combination of anchor donor-weight vectors and remains in $\Delta_J$.
\end{proposition}

\paragraph{What Proposition~\ref{prop:harmonic} means}
The proposition is easiest to read from right to left in Eq.~\eqref{eq:harmonic-solution}. The rows of $\bm{W}_{\mathcal A}$ are donor-weight vectors learned from targets that actually have PRE outcomes. The matrix $\bm H$ tells us how those anchor representations are combined for each post-only target. For example, if one row of $\bm H$ is $(0.6,0.3,0.1)$ for three anchors $a_1,a_2,a_3$, then the corresponding post-only donor vector is $0.6\bm w_{a_1}+0.3\bm w_{a_2}+0.1\bm w_{a_3}$. The graph therefore does not create new outcome information; it determines how existing PRE information is transported from anchors to post-only locations.

The same result can be written locally as
\begin{equation}
\bm{w}_i=\frac{1}{d_i}
\sum_{k:(i,k)\in E}a_{ik}\bm{w}_k,
\label{eq:local-harmonic}
\end{equation}
so a post-only target's donor-weight vector is the edge-weighted average of its neighbors' donor-weight vectors. If some neighbors are themselves post-only, information can pass through them from more distant anchors. In graph terminology this is called a discrete harmonic extension \citep{zhu2003harmonic}. A full proof of Proposition~\ref{prop:harmonic} is provided in the Online Appendix, Section~1.1.

With $\varepsilon=0$, the same identity holds for donor-based predictions at any date:
\begin{equation}
\widehat{\bm Y}_{\mathcal U t}(0)
=\bm H\widehat{\bm W}_{\mathcal A}\bm Y_t^C
=\bm H\widehat{\bm Y}_{\mathcal A t}(0).
\label{eq:harmonic-predictions}
\end{equation}
Thus, conditional on the fitted anchor weights, interpolating donor weights and interpolating the corresponding anchor predictions give exactly the same post-only prediction. The graph penalty also allows the weights at locations with observed PRE histories to adjust in response to one another. Proposition~\ref{prop:kron} below shows how this happens when the interpolation formula for the post-only weights is substituted into the objective.

\subsection{What post-only locations contribute to the graph penalty}

The following algebraic result shows how paths through post-only locations link the anchor weights.

\begin{proposition}[Equivalent graph penalty after post-only interpolation]
\label{prop:kron}
Under Condition~\ref{cond:anchor}, for fixed simplex-valued $\bm{W}_{\mathcal A}$, minimizing the graph penalty over $\bm{W}_{\mathcal U}$ gives the solution in Proposition~\ref{prop:harmonic}, and
\begin{align}
\min_{\bm{W}_{\mathcal U}}\operatorname{tr}(\bm W^\top\bm L\bm W)
&=\operatorname{tr}\left(\bm W_{\mathcal A}^\top
\bm L_{\mathrm{eff}}\bm W_{\mathcal A}\right),\notag\\
\bm L_{\mathrm{eff}}
&=\bm L_{\mathcal A\mathcal A}
-\bm L_{\mathcal A\mathcal U}\bm L_{\mathcal U\mathcal U}^{-1}
\bm L_{\mathcal U\mathcal A}.
\label{eq:kron}
\end{align}
Here, $\bm L_{\mathrm{eff}}$ is the effective graph Laplacian among the anchor locations after accounting for paths that pass through post-only locations.
\end{proposition}

\paragraph{What Proposition~\ref{prop:kron} means}
For fixed anchor weights, Proposition~\ref{prop:harmonic} gives the post-only weights that minimize the graph penalty: $\bm W_{\mathcal U}=\bm H\bm W_{\mathcal A}$. Substituting this expression into the graph penalty gives the right-hand side of Eq.~\eqref{eq:kron}, which depends only on $\bm W_{\mathcal A}$. Thus the optimization can be carried out over the anchor weights alone, and the post-only weights are subsequently obtained from the same interpolation formula. The post-only locations remain part of the model: paths through them determine $\bm L_{\mathrm{eff}}$ and therefore affect how the anchor weights are linked. They change these graph connections without adding PRE outcome observations. This substitution is a Schur-complement reduction, also called Kron reduction in network theory. A full proof is provided in the Online Appendix, Section~1.2.

For $\lambda>0$ and $\varepsilon=0$, substituting $\bm W_{\mathcal U}=\bm H\bm W_{\mathcal A}$ into Eq.~\eqref{eq:graphscm-objective} gives the equivalent optimization over anchor weights:
\begin{equation}
\begin{split}
\widehat{\bm W}_{\mathcal A}\in{}&
\argmin_{\bm w_i\in\Delta_J,\ i\in\mathcal A}\\
&\left\{\mathcal L_{\mathrm{fit}}(\bm W_{\mathcal A})
+\frac{\lambda}{A_+}\operatorname{tr}(\bm W_{\mathcal A}^{\top}
\bm L_{\mathrm{eff}}\bm W_{\mathcal A})\right\}.
\end{split}
\label{eq:reduced-anchor-objective}
\end{equation}
If each location with observed PRE outcomes were fitted separately, only the first term in Eq.~\eqref{eq:reduced-anchor-objective} would be minimized. The second term links the donor weights across these locations through $\bm L_{\mathrm{eff}}$, so their weights are estimated together. The weights at post-only locations are then obtained from Eq.~\eqref{eq:harmonic-solution}. This is how one formulation accommodates complete, partial, and absent PRE histories.

Equation~\eqref{eq:reduced-anchor-objective} omits the ridge because it describes the case $\varepsilon=0$. In computation, we solve the full objective in Eq.~\eqref{eq:graphscm-objective}, with $\varepsilon=10^{-10}$ as specified in Section~\ref{sec:penalized-estimator}. The corresponding conditional interpolation, including the contribution from the equally weighted donor average, is derived in the Online Appendix, Section~1.3.

\subsection{Counterfactuals and causal estimands}

After estimating weights using pre-treatment information, the untreated post-treatment counterfactual is
\begin{equation}
\widehat Y_{it}(0)=\widehat{\bm w}_i^{\top}\bm{Y}_t^C,
\qquad t\in\Tpost.
\label{eq:counterfactual}
\end{equation}
The site- and time-specific treatment effect estimate is
\begin{equation}
\widehat\tau_{it}=Y_{it}-\widehat Y_{it}(0),
\label{eq:tauhat}
\end{equation}
and, for a target set $\mathcal S$,
\begin{equation}
\widehat{ATT}_{\mathcal S}(t)
=\frac{1}{|\mathcal S|}\sum_{i\in\mathcal S}\widehat\tau_{it}.
\label{eq:att-est}
\end{equation}
The graph regularizer addresses counterfactual construction. As in conventional SCM, causal interpretation additionally requires that donor outcomes represent untreated exposure and are not themselves affected by treatment spillovers.

\section{GeoEco-SCM: economically adjusted local borrowing}
\label{sec:geoeco}

Because GeoEco-SCM uses geography and economic covariates to determine how strongly locations borrow from one another, these inputs must be available independently of the post-treatment outcomes being explained. Otherwise, post-treatment information could inadvertently enter the construction of the counterfactual.

\begin{condition}[Graph construction uses only pre-treatment information]
\label{cond:graph-measurable}
Coordinates, graph topology, and economic covariates used to construct the weighted target graph are observed before treatment or are treatment-invariant attributes. They are available for post-only as well as PRE-observed targets, and neither graph construction nor hyperparameter selection uses post-treatment target outcomes or variables that are consequences of treatment.
\end{condition}

Condition~\ref{cond:graph-measurable} is a design requirement rather than a stochastic assumption about the outcome-generating process. Its purpose is to prevent information leakage from the post-treatment period. Because the variables entering the graph are either measured before treatment or treatment invariant, the graph and its edge strengths can be regarded as part of the information set available before the counterfactual is constructed.

\subsection{Geographic candidate graph}

The geographic graph first determines \emph{which locations are allowed to borrow from one another}. Let $\bm s_i=(x_i,y_i)^\top$ be the target coordinate and let $d^g_{ik}=\|\bm s_i-\bm s_k\|_2$. Each target first selects its $k$ nearest target locations. An undirected edge is then retained whenever either member of a pair selects the other; the simulations use $k=8$. This edge set is fixed before economic similarity is considered. GeoEco-SCM therefore does not add new edges between economically similar but geographically non-neighboring locations; economic information only adjusts the strengths of edges that already exist in the geographic graph.

Let $E_d$ denote the directed neighbor list before symmetrization. In accordance with the implementation, define
\begin{equation}
h_g=\operatorname{median}\{d^g_{ik}:(i,k)\in E_d,\ d^g_{ik}>0\}.
\end{equation}
For \GraphSCM, the raw geographic edge strength is
\begin{equation}
\widetilde a^{G}_{ik}=\exp\left(-\frac{d^g_{ik}}{h_g}\right).
\label{eq:graph-geographic}
\end{equation}

\subsection{Pre-treatment economic distance}

Let $\bm x_i\in\R^p$ collect $p$ economic covariates determined before treatment. Here we use continuous quantitative covariates with positive sample variances; in the Monte Carlo study these are proxies for pre-treatment employment density, commercial share, and income. We first standardize each covariate,
\begin{equation}
\bm z_i=\operatorname{diag}(\widehat\sigma_1,\ldots,\widehat\sigma_p)^{-1}(\bm x_i-\widehat{\bm\mu}),
\end{equation}
where $\widehat{\bm\mu}$ and $\widehat\sigma_1,\ldots,\widehat\sigma_p$ are calculated across the target locations. All targets, including post-only targets, have these PRE attributes. Let $\bm S$ be the sample covariance matrix of the standardized vectors $\bm z_i$.

Because highly correlated covariates can make $\bm S^{-1}$ numerically unstable, we slightly shrink the off-diagonal covariances toward zero:
\begin{equation}
\widetilde{\bm S}
=(1-\eta)\bm S+\eta\operatorname{diag}(\bm S)+\delta_{\mathrm{cov}}\bm I_p.
\label{eq:shrink-cov}
\end{equation}
Here $\bm I_p$ is the $p\times p$ identity matrix, $0\le\eta\le1$ controls the amount of covariance shrinkage, and $\delta_{\mathrm{cov}}>0$ ensures invertibility. In this study, we set $\eta=0.15$ and $\delta_{\mathrm{cov}}=10^{-6}$. This covariance adjustment is distinct from the weight ridge $\varepsilon$ in Eq.~\eqref{eq:graphscm-objective}. We then measure economic dissimilarity by
\begin{equation}
d^e_{ik}
=\left\{(\bm z_i-\bm z_k)^{\top}\widetilde{\bm S}^{-1}(\bm z_i-\bm z_k)\right\}^{1/2}.
\label{eq:mahalanobis}
\end{equation}
This is a covariance-adjusted distance: differences along highly variable or strongly correlated economic dimensions are not counted as if all covariates were independent and equally scaled.
Define
\begin{equation}
h_e=\operatorname{median}\{d^e_{ik}:(i,k)\in E_d,\ d^e_{ik}>0\}.
\end{equation}

\subsection{Reweighting the existing geographic edges}

For $\beta\ge0$, \GeoEco assigns
\begin{equation}
\widetilde a_{ik}(\beta)
=\exp\left(
-\frac{d^g_{ik}}{h_g}
-\beta\frac{d^e_{ik}}{h_e}
\right),
\qquad (i,k)\in E.
\label{eq:geoeco-raw}
\end{equation}
Opposite directed edges are collapsed by averaging their raw strengths. After symmetrization, edge weights are normalized to mean one:
\begin{equation}
a_{ik}(\beta)
=\frac{\widetilde a_{ik}(\beta)}{|E|^{-1}\sum_{(r,q)\in E}\widetilde a_{rq}(\beta)}.
\label{eq:edge-normalization}
\end{equation}
Let $\bm A(\beta)=[a_{ik}(\beta)]$ denote the resulting weighted adjacency matrix. GeoEco-SCM substitutes this graph into the common objective in Eq.~\eqref{eq:graphscm-objective}, with $\varepsilon=10^{-10}$ as in Graph-SCM. When $\beta=0$, Eq.~\eqref{eq:geoeco-raw} gives the geography-only graph, so the two estimators coincide for the same $\lambda$. Increasing $\beta$ attenuates economically dissimilar connections more strongly while retaining the geographic edge set. PRE validation selects $\lambda$ and $\beta$, including the option $\beta=0$.

\section{Hyperparameter selection and computation}
\label{sec:tuning}

\subsection{Tuning without post-treatment information}

Neither $\lambda$ nor $\beta$ is tuned with post-treatment target outcomes. We keep $\varepsilon=10^{-10}$ fixed and use two validation exercises, both constructed entirely from the pre-treatment period. For \GraphSCM, $\beta$ is fixed at zero, so only $\lambda$ is selected.

First, a temporal holdout asks whether the fitted donor weights predict years that were temporarily withheld from otherwise observed pre-treatment histories. Specifically, we remove late pre-treatment years $\mathcal H\subset\Tpre$ and evaluate their prediction. Let $\widehat{\bm W}^{-\mathcal H}(\lambda,\beta)$ be fitted without the held-out cells. The temporal score is
\begin{equation}
R_{\mathrm{time}}(\lambda,\beta)
=
\left[
\frac{1}{M_H}
\sum_{(i,t)\in\mathcal V_H}
\left(
Y_{it}-\widehat{\bm w}_i^{(-\mathcal H)\top}\bm{Y}_t^C
\right)^2
\right]^{1/2},
\label{eq:temporal-cv}
\end{equation}
where $\mathcal V_H$ is the set of observed held-out cells and $M_H=|\mathcal V_H|$ is their number. The choice of the holdout window is application dependent and is fixed before tuning; the concrete Monte Carlo setting is reported in the Online Appendix.

Second, we mimic the post-only problem inside the pre-treatment sample. For a small, spatially interspersed subset of complete-history targets, we temporarily hide their entire pre-treatment history, refit the model, and ask how well the hidden values are recovered. We refer to this exercise as \emph{pseudo-post-only validation}. If $\mathcal V_f$ is the set masked in fold $f$, we refit after setting all $Y_{it}$ for $i\in\mathcal V_f$ and $t\in\Tpre$ to missing. The fold score is
\begin{equation}
R_{\mathrm{PO},f}(\lambda,\beta)
=
\left[
\frac{1}{|\mathcal V_f||\Tpre|}
\sum_{i\in\mathcal V_f}\sum_{t\in\Tpre}
\left(Y_{it}-\widehat{\bm w}_{i,-f}^{\top}\bm{Y}_t^C\right)^2
\right]^{1/2}.
\label{eq:pseudo-po}
\end{equation}
The pseudo-post-only targets are selected so that their number of immediate pre-treatment-observed graph neighbors resembles that of actual post-only targets. This matches the local availability of PRE information. It does not ensure that the masked targets represent the economic attributes or selective historical coverage of the actual post-only group; Section~\ref{sec:method-choice} describes observable checks for applications.

Let $\overline R_{\mathrm{PO}}$ be the mean of the fold-specific scores $R_{\mathrm{PO},f}$. Both validation scores measure outcome-prediction RMSE on the same scale; neither measures error in the donor weights. The selection score is
\begin{equation}
S(\lambda,\beta)
=\frac{1}{2}R_{\mathrm{time}}(\lambda,\beta)
+\frac{1}{2}\overline R_{\mathrm{PO}}(\lambda,\beta).
\label{eq:selection-score}
\end{equation}
We use the unstandardized average in Eq.~\eqref{eq:selection-score}, assigning equal weight to an absolute RMSE reduction of the same size in either validation task. This is a prespecified default. The search and practical tie-breaking rule are described below; the fold-confirmation details are given in the Online Appendix, Section~4.

\subsection{Search strategy}

The search compares weak to strong graph regularization and, for GeoEco-SCM, weak to strong economic attenuation. Appropriate numerical ranges depend on the application, so our implementation starts from a scale-adaptive grid and expands a boundary when the best confirmed candidate lies at its edge. The screening, expansion, and fold-confirmation rules are fixed before the search; their numerical settings are given in the Online Appendix, Section~4. Let $\mathcal C$ denote the eligible candidates retained after these steps.

Very similar validation errors provide little reason to impose stronger economic attenuation on the geographic graph. We therefore prefer the smaller $\beta$ when predictive performance is practically tied. In this study, candidates within 0.5\% of the lowest confirmed score are treated as tied. Define
\begin{equation}
\begin{aligned}
S_{\min}&=\min_{(\lambda,\beta)\in\mathcal C}S(\lambda,\beta),\\
\mathcal C_{\mathrm{tol}}
&=\{(\lambda,\beta)\in\mathcal C:
S(\lambda,\beta)\le(1+\delta_{\mathrm{CV}})S_{\min}\},
\end{aligned}
\label{eq:hyperparameter-selection}
\end{equation}
where $\delta_{\mathrm{CV}}=0.005$ is the prespecified practical tolerance. We select the smallest $\beta$ in $\mathcal C_{\mathrm{tol}}$; among candidates with that $\beta$, we select the lowest score, using the smaller $\lambda$ only to resolve any remaining tie. Thus a geography-only specification can be retained when its validation error is sufficiently close to the best score.

In the Monte Carlo implementation, the $\lambda=0$ fit is a diagnostic reference. When post-only targets are present, final selection is restricted to $\lambda>0$, so their representations use graph-based borrowing. Uniform is therefore a separate comparator rather than an automatically selectable no-borrowing fallback.

\subsection{Final numerical optimization}

For fixed $(\lambda,\beta)$, Eq.~\eqref{eq:graphscm-objective} is a convex quadratic program (QP): the objective is quadratic in the unknown weights and the constraints are linear, so any optimum is a global optimum. This form is convenient for reliable numerical solution. Let $r_{it}$ be a residual variable for each observed target--PRE cell, representing the difference between the observed target outcome and its donor-weighted reconstruction. The problem can be written as
\begin{align}
\min_{\bm W,\bm r}\quad
&\frac{1}{N_+}\sum_{i\in\mathcal I_+}\frac{1}{|\Oi|}
\sum_{t\in\Oi}r_{it}^2
+\frac{\lambda}{A_+}\tr(\bm W^{\top}\bm L_{\beta}\bm W)
+\varepsilon\frac{1}{N}\|\bm W\|_F^2,
\label{eq:qp-objective}\\
\text{s.t.}\quad
&r_{it}=Y_{it}-\bm{w}_i^{\top}\bm{Y}_t^C,
\qquad (i,t):t\in\Oi,\label{eq:qp-resid}\\
&\1^{\top}\bm{w}_i=1,\qquad \bm{w}_i\ge0,\qquad i=1,\ldots,N,
\label{eq:qp-simplex}
\end{align}
where $\bm L_{\beta}$ and $A_+$ are computed from the selected graph $\bm A(\beta)$, and the numerical ridge is the same $\varepsilon=10^{-10}$ introduced in Section~\ref{sec:penalized-estimator}. Thus Eq.~\eqref{eq:qp-objective} is a residual-variable form of the full objective, including the ridge, for both estimators; \GraphSCM uses $\beta=0$. The positive coefficient gives a unique mathematical weight solution. The exact conditional interpolation formula is given in the Online Appendix, Section~1.3. At $\lambda=0$, a post-only target has no fitting observations or graph penalty, and the ridge selects equal donor weights.

Candidate configurations are screened with a projected Fast Iterative Shrinkage-Thresholding Algorithm (FISTA; \citealp{beck2009fista}) to limit computation during hyperparameter selection. After selecting $(\widehat\lambda,\widehat\beta)$, the model is refitted on all available PRE outcomes with OSQP (Operator Splitting Quadratic Program; \citealp{stellato2020osqp}), using tighter convergence tolerances to obtain the reported weights. Algorithmic details, solver tolerances, and the scope of numerical checks are reported in the Online Appendix, Sections~4 and~5.

\section{Spatio-temporal and SCM benchmarks}
\label{sec:benchmarks}

\subsection{Ordinary SCM}

Ordinary SCM is fitted independently for targets with at least one pre-treatment outcome. It is informative about conditional performance on analyzable targets, but when $\Oi=\varnothing$ its own fitting criterion does not determine a target-specific donor representation without an additional cross-unit or outcome-model restriction. It is therefore excluded from the principal full-coverage ranking.

\subsection{Uniform donor weights}

Uniform weights, $w_{ij}=1/J$, provide a simple full-coverage reference. This comparator uses no target-specific PRE matching or spatial graph and is particularly informative when true donor weights are unrelated across targets.

\subsection{Covariate-balanced synthetic control}

Observed attributes offer a direct way to choose donor weights even when a target has no PRE outcomes. We therefore add a covariate-only comparator, denoted Cov-SCM. Each target and donor has two planar coordinate columns, measured in meters, and the same three PRE economic proxies used by the economic-covariate methods. Standardize each of these five columns using its pooled target-and-donor mean and sample standard deviation, dropping any constant column. Let $\bm q_i\in\mathbb R^{p_c}$ be the standardized target attributes and let $\bm Q_C\in\mathbb R^{J\times p_c}$ contain the standardized donor attributes, where $p_c$ is the number of retained columns. Cov-SCM solves, separately for each target,
\begin{equation}
\widehat{\bm w}_i^{\mathrm{Cov}}
=\argmin_{\bm w\in\Delta_J}
\left\{\frac{1}{p_c}\|\bm q_i-\bm Q_C^\top\bm w\|_2^2
+\varepsilon_c\|\bm w\|_2^2\right\},
\qquad \varepsilon_c=10^{-10}.
\label{eq:covariate-scm}
\end{equation}
Writing $\bm s_j$ for donor $j$'s planar coordinates, coordinate balance compares the target location $\bm s_i$ with the donor-weighted mean $\sum_j w_{ij}\bm s_j$. Donors may occupy different locations; their weighted mean is balanced jointly with the economic proxies, and balance can be approximate. All standardized attributes receive equal weight, and $\varepsilon_c$ is fixed rather than tuned using outcomes. This positive numerical ridge makes the weight solution unique and contributes at most $10^{-10}$ to the objective on the simplex. No target outcome, true donor weight, or true counterfactual enters this fit; the counterfactual prediction uses the resulting weights and donor POST outcomes. The comparison asks whether this direct covariate-balancing rule can replace borrowing donor representations learned from PRE outcomes. Because Cov-SCM uses both geography and economic covariates, it is not an isolated test of the economic edge adjustment. Exact implementation and data-provenance checks are given in the Online Appendix, Section~8.

\subsection{Nonseparable spatio-temporal prediction benchmarks}

To represent outcome-side spatial prediction, we use a stationary nonseparable spatio-temporal covariance model from the Gneiting class \citep{gneiting2002nonseparable}. ``Nonseparable'' means that spatial dependence is allowed to change with temporal separation rather than being written as a simple product of an independent spatial covariance and an independent temporal covariance. The correlation function is
\begin{equation}
R(h,u)
=\frac{
\exp\left[
-(h/\phi_s)^{a_s}
/\{1+(u/\phi_t)^{a_t}\}^{\gamma a_s/2}
\right]
}{1+(u/\phi_t)^{a_t}},
\label{eq:gneiting}
\end{equation}
where $h$ and $u$ are spatial and temporal lags, respectively. To keep the benchmark parsimonious and computationally feasible, we fix the spatial and temporal shape parameters at $a_s=a_t=1$. The parameters $\phi_s$ and $\phi_t$ control the spatial and temporal correlation ranges, $\gamma$ controls the degree of spatio-temporal interaction, and the nugget ratio is the variance of independent noise relative to the variance of the correlated spatio-temporal process. These remaining covariance parameters are selected using pre-treatment information only. Their candidate ranges are scaled to the spatial and temporal extent of the data and are expanded when necessary to verify that the selected solution is not driven by a search boundary; the specific Monte Carlo grids are reported in the Online Appendix.

We evaluate two basic strategies, each with and without economic covariates. The first strategy \emph{pre-imputes the missing pre-treatment history}: missing target outcomes are predicted from the nonseparable spatio-temporal model and ordinary SCM is then fitted to the completed pre-treatment panel. We refer to this as \emph{spatio-temporal pre-imputation + SCM}; a second version additionally includes PRE economic covariates in the outcome mean model. The spatio-temporal fit uses observed donor and target PRE outcomes, with covariates for both groups in the economic-covariate version. The second strategy bypasses SCM and predicts the untreated post-treatment outcome directly using the target PRE outcomes and donor outcomes available up to the prediction date, again with and without the economic covariates. We refer to these as \emph{direct spatio-temporal prediction}. In all cases the spatio-temporal model uses the Gneiting covariance function. The direct predictors never use target post-treatment outcomes or future donor outcomes when predicting a given post-treatment date.

The result tables abbreviate spatio-temporal PRE imputation followed by SCM as STI and direct prediction as STD; the suffix -X denotes inclusion of the PRE economic covariates. Both versions of each strategy are reported.

\section{Monte Carlo design}
\label{sec:mcdesign}

To make the simulation concrete without tying it to a particular substantive field, we use a stylized place-based intervention. At the treatment date, a new service or accessibility node is introduced at one location, and the treatment effect declines with distance from that site. A new transit station is one intuitive interpretation, but the same geometry could represent the opening of a health-care facility, a public service center, an environmental remediation site, or another spatially localized intervention. This device is used only to generate transparent spatial heterogeneity in treatment effects and in the observation process; neither \GraphSCM nor \GeoEco relies on transport-specific mechanisms.

\subsection{Common spatial and temporal architecture}

\begin{figure}[t]
\centering
\includegraphics[width=0.88\linewidth]{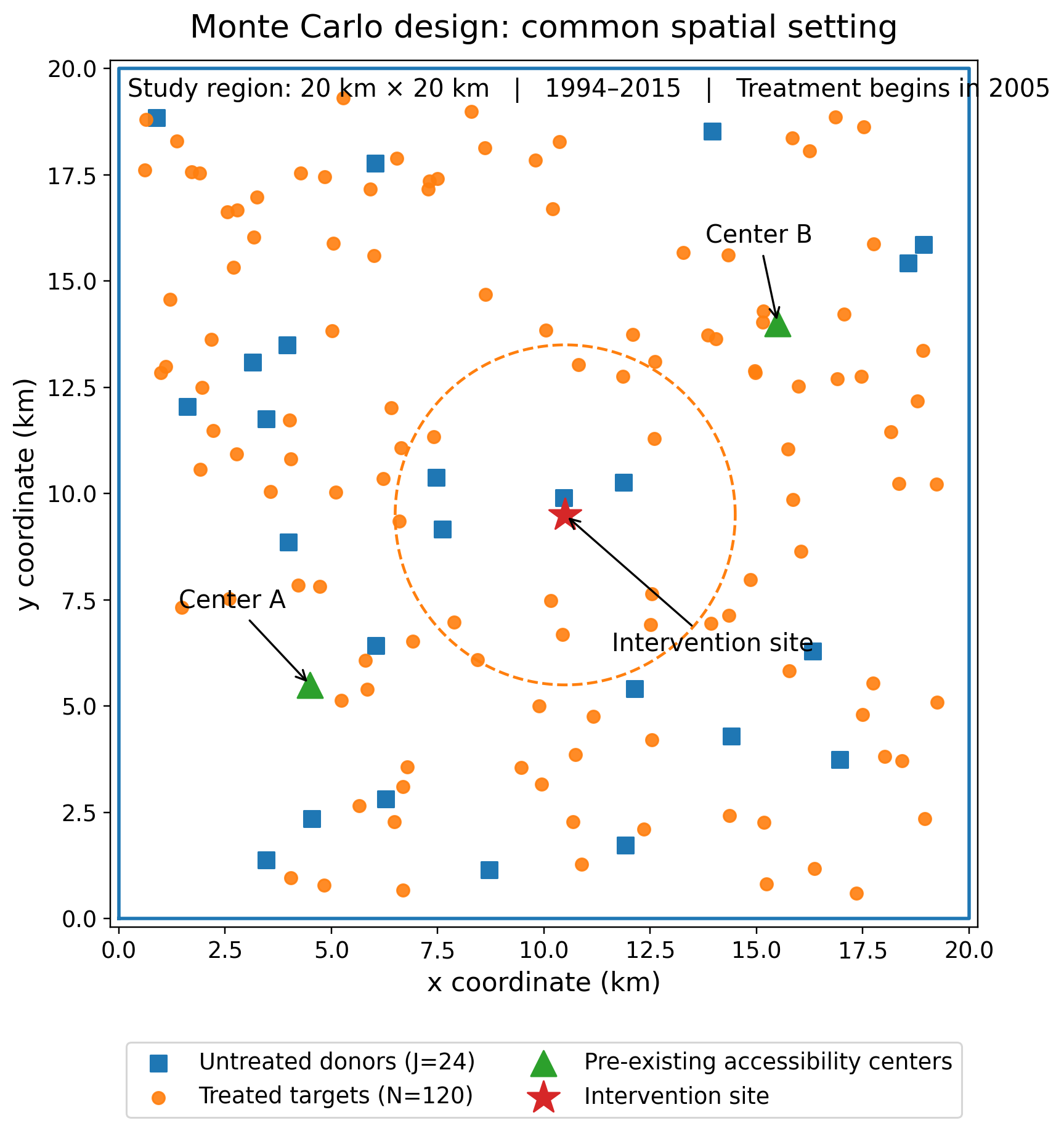}
\caption{Schematic of the common spatial setting used in the Monte Carlo experiments. Untreated donors and treated targets are distributed over the study region, while two pre-existing accessibility centers generate baseline spatial structure and the intervention site generates spatially localized treatment-effect heterogeneity. The dashed circle is included only as a visual guide to the local nature of the intervention and is not a distance cutoff used by the DGP.}
\label{fig:mc-design}
\end{figure}

Each replication contains $J=24$ untreated donors and $N=120$ treated targets observed over 1994--2015. All Monte Carlo outcomes are measured on a logarithmic scale. Treatment begins in 2005, giving 11 pre-treatment and 11 post-treatment years. Except in the separated-donor design, donor and target coordinates are sampled independently and uniformly on a $19\,\mathrm{km}\times19\,\mathrm{km}$ interior of a $20\,\mathrm{km}\times20\,\mathrm{km}$ study region. Two pre-existing accessibility centers, which generate baseline spatial structure, are fixed at $(4.5,5.5)$ and $(15.5,14)$ km, and the newly introduced intervention site is fixed at $(10.5,9.5)$ km. Figure~\ref{fig:mc-design} shows one illustrative realization; the two PRE centers and intervention site are fixed across replications.

Different DGPs use separate seed streams, except for the paired supplementary designs D9--D10.

The five primary DGPs described in Section~\ref{sec:mc-dgps} each have 100 replications, giving 500 primary datasets. Five supplementary DGPs also have 100 replications each. The complete study therefore contains $10\times100=1{,}000$ datasets; Cov-SCM uses the same 500 primary datasets. The replication materials record the design-specific seeds and the common replacement seeds for infeasible D9--D10 observation patterns.

\subsection{Five primary DGPs}
\label{sec:mc-dgps}

The five primary designs ask when donor relationships can be borrowed across targets, what economic information adds, and when borrowing fails. Table~\ref{tab:dgp-purpose} summarizes their roles. The following descriptions introduce the mechanisms before the common equations; the detailed selective-observation and donor-factor formulas are given after their ingredients have been defined.

\paragraph{D1: geographic baseline}
True donor weights vary smoothly with geographic market state, and historical coverage is assigned randomly. Nearby observed-history targets should therefore carry information about a post-only target's donor representation. This is the baseline for sharing PRE-informed representations rather than using uniform weights or balancing observed attributes directly.

\paragraph{D2: informative economic structure}
Donor weights depend on both geography and an economic state with substantial variation beyond geography. Nearby targets can therefore require different donor combinations. This design asks whether economic proxies help identify useful local borrowing connections, and compares that use with direct covariate balancing and outcome-side prediction.

\paragraph{D3: selective post-only coverage}
The donor-weight and market-state architecture is the same as in D2, but targets near the intervention with low latent PRE signals are more likely to lack histories. The number of post-only targets remains 24. This design examines borrowing from an observed-history group that differs systematically from the post-only population, and the effect of excluding that population from ATT estimation.

\paragraph{D4: complex donor dynamics and nonlinear economic loadings}
Donor weights remain constant over time, while donor trajectories use spatially generated coefficients and additional persistent, cyclical, and level-shift factors. Their loadings also contain nonlinear economic structure. This is a joint stress test of outcome structure and temporal dynamics: it asks whether borrowing stable donor representations remains useful when the specified spatio-temporal covariance and linear economic mean are less aligned with the outcome process.

\paragraph{D5: no spatial weight smoothness}
Each target receives an independent symmetric Dirichlet weight draw. Neither its neighbors nor observed attributes reveal that particular draw. Uniform weights are the conditional mean of the unknown weights, making this a negative control for unnecessary spatial or economic borrowing.

\begin{table}[!t]
\centering
\caption{Five primary Monte Carlo DGPs and their roles.}
\label{tab:dgp-purpose}
\small
\begin{tabularx}{\textwidth}{@{}lXX@{}}
\toprule
DGP & Main feature & Role in the comparison \\
\midrule
D1 & Geographically smooth donor weights; random historical coverage. & Baseline for sharing donor representations across nearby targets. \\
D2 & Donor weights depend on geographic and partly nonspatial economic state. & Whether economic information contributes beyond geography. \\
D3 & Geography--economics weights with selective post-only coverage. & Whether observed-history targets provide useful representations for a systematically different post-only group. \\
D4 & Stable donor weights, spatially generated donor trajectories, and additional factors with nonlinear economic loadings. & Borrowing representations under the combined changes in economic outcome structure and temporal dynamics. \\
D5 & Independent donor-weight draws across targets. & Negative control without spatial donor-weight information to borrow. \\
\bottomrule
\end{tabularx}
\end{table}

The Online Appendix completes the coefficient, spatial-function, and factor specifications and gives five supplementary stress tests: geography-only selective coverage (D6), spatially separated donors (D7), abrupt weight boundaries (D8), and a paired local geography--economics experiment (D9--D10). The paired experiment conditions on the feasibility of its post-only selection constraints, as detailed in the Online Appendix. Cov-SCM is evaluated in D1--D5; the original seven full-coverage methods are retained for all ten designs.

\subsection{Common outcome model and information available to estimation}

Table~\ref{tab:mc-information} separates inputs available to the estimators from quantities used to generate and evaluate the simulations. In particular, the observed economic proxies are distinct from the latent states that determine the true donor weights.

\begin{table}[!t]
\centering
\caption{Observed inputs and generating quantities in the Monte Carlo study.}
\label{tab:mc-information}
\small
\begin{tabularx}{\textwidth}{@{}lX@{}}
\toprule
Quantity & Meaning and use \\
\midrule
\multicolumn{2}{@{}l}{\textit{Available inputs for the relevant estimators}}\\
$C_{jt}=Y_{jt}^C$ & Observed untreated-donor outcomes in PRE and POST.\\
$Y_{it}$ & Target PRE outcomes where observed; target POST outcomes are used to calculate effects after fitting.\\
$\bm s_a,\ \bm x_a$ & Coordinates and the three observed PRE economic proxies for donor or target location $a$.\\
\midrule
\multicolumn{2}{@{}l}{\textit{Generator and evaluation quantities, not supplied as estimation inputs}}\\
$g_{a\ell},e_{a\ell},m_{a\ell}$ & Geographic, economic, and combined market states used to construct the DGP.\\
$\bm w_i^\star$ & Time-invariant generating donor weights.\\
$\mu_{it},\tau_{it}$ & Latent untreated signal and true log treatment effect used to evaluate recovery.\\
\bottomrule
\end{tabularx}
\end{table}

The common target-outcome model is
\begin{equation}
\begin{aligned}
\mu_{it}&=\sum_{j=1}^{J}w_{ij}^{\star}C_{jt},
\qquad \bm w_i^{\star}\in\Delta_J,\\
Y_{it}(0)&=\mu_{it}+\epsilon_{it},
\qquad Y_{it}=Y_{it}(0)+\tau_{it},\\
\epsilon_{it}&\stackrel{\mathrm{iid}}{\sim}N(0,0.008^2).
\end{aligned}
\label{eq:mc-signal-model}
\end{equation}
Here $C_{jt}=Y^C_{jt}$ is donor $j$'s log outcome, the true weights are constant over time, and $\tau_{it}=0$ before treatment. Counterfactual RMSE is evaluated against the latent signal $\mu_{it}$.

Donor trajectories combine heterogeneous levels, trends, cycles, and common and donor-specific temporal variation. With calendar year $y_t$ and elapsed time $r_t=y_t-1994$, the baseline trajectory is
\begin{equation}
\begin{split}
C_{jt}^{\mathrm{base}}={}&\alpha_j+b_jr_t+q_jr_t^2\\
&+c_{1j}\sin(r_t/2.8)+c_{2j}\cos(r_t/4.2)
+\ell_j f_t^c+u_{jt}.
\end{split}
\label{eq:mc-donor-base}
\end{equation}
The coefficients $\alpha_j,b_j,q_j,c_{1j},c_{2j}$ determine the donor's level, linear and quadratic trends, and cycle amplitudes; $\ell_j$ loads on the centered common temporal process $f_t^c$. This process combines periodic terms with an AR(1) component of coefficient 0.70 and innovation SD 0.004. The independent donor-specific component $u_{jt}$ is AR(1) with coefficient 0.55 and innovation SD 0.006.

For D1--D3 and D5, $C_{jt}=C_{jt}^{\mathrm{base}}$ and coefficient means depend linearly on the donor market states defined in Section~\ref{sec:mc-states}, with independent Gaussian perturbations. For example, $\alpha_j=\log(220000)+0.18m_{j1}-0.10m_{j2}+\epsilon_{\alpha j}$, where $\epsilon_{\alpha j}\sim N(0,0.10^2)$. D4 changes the baseline coefficient structure and adds temporal factors, as specified below. The Online Appendix gives the remaining coefficient distributions and the complete temporal-process construction. The companion R package supplies the fixed seeds and scripts for reproducing the simulation tables.

\subsection{Latent geographic and economic state}
\label{sec:mc-states}

Let $a$ index a generic donor or target location with coordinates $\bm s_a=(x_a,y_a)^\top$. In this subsection, standardization pools $M=N+J=144$ locations. For a nonconstant vector $\bm v=(v_1,\ldots,v_M)^\top$, define
\[
\begin{aligned}
z[\bm v]_a&=\frac{v_a-\bar v}{s_v},\qquad a=1,\ldots,M,\\
\bar v&=\frac{1}{M}\sum_{b=1}^{M}v_b,\qquad
s_v=\left[\frac{1}{M-1}\sum_{b=1}^{M}(v_b-\bar v)^2\right]^{1/2}.
\end{aligned}
\]
Thus $z[\cdot]$ gives each variable zero sample mean and unit sample standard deviation across the pooled donor and target locations. Each use below applies this operation separately to the variable inside the brackets.

Broad geographic market characteristics are
\begin{equation}
\begin{aligned}
g_{a1}&=0.70x_a^\circ+0.20\sin(\pi y_a^\circ),\\
g_{a2}&=0.65y_a^\circ+0.18\cos(\pi x_a^\circ),
\end{aligned}
\label{eq:mc-geographic-state}
\end{equation}
where $x_a^\circ=(x_a-10000)/9500$ and $y_a^\circ=(y_a-10000)/9500$. To introduce economic variation beyond geography, define
\begin{equation}
\begin{aligned}
e_{a\ell}&=z\!\left[\sqrt{0.20}\,f_\ell(\bm s_a)
+\sqrt{0.80}\,z[v_{a\ell}]\right],\\
m_{a\ell}&=z[g_{a\ell}+\kappa_D e_{a\ell}],
\qquad \ell=1,2,
\end{aligned}
\label{eq:mc-economic-state}
\end{equation}
with independent $v_{a\ell}\sim N(0,1)$. The functions $f_1,f_2$ are standardized smooth sinusoidal fields, specified in the Online Appendix. The mixing puts greater weight on location-specific than spatial economic variation. Set $\kappa_D=0$ in D1 and D5, and $\kappa_D=0.65$ in D2--D4; write $\bm m_a=(m_{a1},m_{a2})^\top$.

Estimators observe three noisy PRE economic proxies, representing employment density, commercial share, and income:
\begin{equation}
\begin{aligned}
x_a^{(1)}&=z[0.85e_{a1}+\nu_{a1}],\\
x_a^{(2)}&=z[-0.25e_{a1}+0.85e_{a2}+\nu_{a2}],\\
x_a^{(3)}&=z[0.55e_{a1}-0.35e_{a2}+\nu_{a3}],\\
\nu_{ap}&\stackrel{\mathrm{iid}}{\sim}N(0,0.18^2),\qquad p=1,2,3.
\end{aligned}
\label{eq:mc-economic-proxies}
\end{equation}
All methods receiving economic covariates use these same observed proxies; the latent states $e_{a\ell}$ and $m_{a\ell}$ are unavailable to estimation.

\subsection{True donor weights}

For D1--D4, the market states determine a target's donor representation through
\begin{equation}
\begin{aligned}
q_{ij}&=-\frac12\left[
\frac{\|\bm m_i-\bm m_j\|_2^2}{0.60^2}
+0.60\left(\frac{a_i-a_j}{0.80}\right)^2\right],\\
w_{ij}^{\star}&=\frac{\exp(q_{ij})}{\sum_{r=1}^{J}\exp(q_{ir})}.
\end{aligned}
\label{eq:mc-donor-weights}
\end{equation}
Here $a_i$ and $a_j$ are standardized negative distances to the nearer of the two PRE accessibility centers, using the pooled donor--target sample. Nearby targets tend to have similar market characteristics and donor weights; target--donor physical distance does not enter this score directly. D1 has geography-only market states, whereas D2--D4 also use the economic component in Eq.~\eqref{eq:mc-economic-state}.

D5 instead draws target-specific weights independently:
\begin{equation}
\bm w_i^{\star}\stackrel{\mathrm{ind}}{\sim}
\operatorname{Dirichlet}(0.2,\ldots,0.2).
\label{eq:mc-independent-weights}
\end{equation}
This retains the convex donor representation while removing spatial and covariate information about a post-only target's particular weight draw.

\subsection{Treatment effect and observation pattern}

Let $d_i^{\mathrm{new}}$ be the target's distance in meters from the intervention site and $\tau_i^\infty$ its long-run log treatment effect. Treatment effects decrease with distance and ramp up over five years:
\begin{equation}
\begin{aligned}
\tau_i^\infty&=0.015+0.105\exp(-d_i^{\mathrm{new}}/5000),\\
\tau_{it}&=
\begin{cases}
0, & y_t<2005,\\
\tau_i^\infty\min\{1,(y_t-2004)/5\}, & y_t\ge2005.
\end{cases}
\end{aligned}
\label{eq:mc-treatment-effect}
\end{equation}
Each replication assigns the 120 targets to 60 complete-PRE, 36 partial-PRE, and 24 post-only locations. Thus 96 targets have at least one PRE observation. Partial histories retain a random 4--8 of the 11 PRE years. Observation groups are randomly assigned in D1, D2, D4, and D5; D3 uses the selective rule below while keeping the same group sizes. The target graph uses $k=8$ throughout; alternative graph topologies and spatially blocked PRE missingness are not evaluated.

\subsection{Selective observation and complex donor trajectories}

\paragraph{Selective observation in D3}
To implement the selective-coverage mechanism introduced above, define $n_i=\exp(-d_i^{\mathrm{new}}/3000)$, $\bar\mu_{i,\mathrm{PRE}}=T_0^{-1}\sum_{t\in\Tpre}\mu_{it}$, and $l_i=-z[\bar\mu_{i,\mathrm{PRE}}]$, where standardization here uses the target sample. The selection score is
\begin{equation}
s_i=z[n_i]+0.90l_i+0.50z[n_i]l_i.
\label{eq:mc-selective-coverage}
\end{equation}
Exactly 24 targets are sampled without replacement using weights proportional to $\exp(s_i)$. Keeping this count fixed separates the composition of the post-only population from the amount of missingness. This design examines whether observed-history locations remain informative for a systematically different group and how excluding that group changes the ATT population.

\paragraph{Donor trajectories in D4}
The baseline trajectory retains the form in Eq.~\eqref{eq:mc-donor-base}, but coefficient means are generated from smooth spatial bases rather than directly from the economic market state. The actual donor outcome is
\begin{equation}
C_{jt}=C_{jt}^{\mathrm{base}}
+\frac{0.060}{\sqrt{3}}\sum_{h=1}^{3}\ell_{jh}^{\mathrm{LF}}F_{ht}.
\label{eq:mc-latent-factors}
\end{equation}
The three factors $F_{ht}$, standardized over time, have persistent AR(1), cyclical, and level-shift dynamics. Their donor loadings $\ell_{jh}^{\mathrm{LF}}$ combine spatial and nonlinear economic structure with donor-specific noise. The nonlinear economic bases are residualized against an intercept and the three observed economic proxies, weakening their direct linear relationship with outcome levels. The target signal still uses the time-invariant weights in Eq.~\eqref{eq:mc-donor-weights}. D4 therefore changes both temporal dynamics and the way economic information enters outcomes relative to D2; it does not isolate a temporal-complexity effect. Exact factor and loading formulas are given in the Online Appendix.

\subsection{Spatio-temporal benchmark tuning in the Monte Carlo experiments}
\label{sec:mc-benchmark-tuning}

The numerical covariance grids, mean specification, and likelihood calculation are detailed in the Online Appendix, Section~6. Covariance parameters are fitted to PRE observations within each replication. To check dependence on the candidate range, the spatial correlation-range grid was extended toward shorter distances. This grid development used known simulation truth to assess changes in average POST prediction errors and rankings, and is distinct from PRE-only parameter fitting within each replication. Section~\ref{sec:numerical-sensitivity} reports the results of this exploratory sensitivity check.

\subsection{Evaluation criteria}

For target group $\mathcal G$, the primary metric is the root mean squared error (RMSE) for prediction of the latent untreated signal,
\begin{equation}
RMSE_{\mathcal G}
=\left[\frac{1}{|\mathcal G||\Tpost|}
\sum_{i\in\mathcal G}\sum_{t\in\Tpost}
\{\widehat Y_{it}(0)-\mu_{it}\}^2\right]^{1/2}.
\label{eq:rmse}
\end{equation}
We evaluate complete-PRE, partial-PRE, post-only, and all-target groups. Tables average replication-specific RMSEs rather than taking a square root after pooling squared errors. Ordinary SCM is evaluated only on its PRE-observed population and is not ranked as an all-target predictor. Every full-coverage comparison uses the same target group within a replication. Post-only performance is the main focus; the all-target average also includes 96 PRE-observed targets and therefore answers a different performance question.

Counterfactual recovery is primary because errors can cancel in aggregate ATT estimates. We also report annual ATT RMSE and signed bias. We report the percentage-scale ATT as
\begin{equation}
\theta_{\mathcal G}(t)=100\left[\exp\left\{
|\mathcal G|^{-1}\sum_{i\in\mathcal G}\tau_{it}\right\}-1\right].
\label{eq:percentage-att}
\end{equation}
This transforms the mean log effect; it is not the arithmetic mean of site-specific percentage effects. The estimate uses the same transformation of the mean estimated log effect. ATT RMSE is computed over POST years within each replication and then averaged; bias is estimated minus true ATT, in percentage points.

Monte Carlo standard errors quantify variability across the 100 replications. For a paired difference $d_r$ between two methods on dataset $r$, let $\overline d$ and $s_d$ be the sample mean and standard deviation over the 100 replications. The reported standard error is $s_d/\sqrt{100}$, and the 95\% interval is $\overline d\pm t_{99,.975}s_d/\sqrt{100}$, where $t_{99,.975}$ is the 97.5th percentile of a $t$ distribution with 99 degrees of freedom. These unadjusted intervals concern mean simulation-performance differences for the specified DGPs and search settings. Only D9--D10 are paired changes to the same underlying ingredients; method comparisons within every DGP are paired by dataset.

\section{Monte Carlo results}
\label{sec:mcresults}

The five primary designs compare eight full-coverage methods on identical datasets and target groups. Cov-SCM balances coordinates and PRE economic covariates by matching their donor-weighted averages to target attributes; Uniform denotes equal donor weights; STI and STD denote spatio-temporal PRE imputation followed by SCM and direct spatio-temporal prediction, respectively; -X adds PRE economic covariates to the outcome mean model. We focus first on post-only targets, then on ATT recovery and the implications of extending population coverage. Tables report means over 100 replications, with Monte Carlo standard errors or paired intervals as indicated.

\subsection{Post-only counterfactual recovery}

Table~\ref{tab:postonly-primary} reports counterfactual signal RMSE for the 24 post-only targets in each replication. STI has the lowest mean in the geographic baseline D1, STI-X in D2 and D3, GeoEco-SCM in D4, and Uniform in D5. Economic reweighting improves on Graph-SCM in D2, D3, and D4; the paired intervals exclude zero in these three designs (Table~\ref{tab:paired-primary}). In D1 and D5, GeoEco-SCM instead has slightly higher signal RMSE than Graph-SCM, and both paired intervals exclude zero. Economic adjustment can therefore be detrimental when the proxies do not reveal relevant differences in donor relationships.

\input{tables/postonly_primary.tex}
\input{tables/paired_primary.tex}

The comparison with outcome-side prediction is also conditional on the DGP. STI-X and STD-X have smaller mean post-only errors than GeoEco-SCM in D2 and D3. In D4, GeoEco-SCM has mean RMSE 0.0601, compared with 0.0675 for STI-X; the paired difference is $-0.00744$, with a 95\% MC interval of $[-0.00960,-0.00528]$. This result favors borrowing stable donor representations in D4's combined setting of spatial donor coefficients, nonlinear economic loadings, and complex temporal factors. Because these features change together, the comparison does not attribute the gain to temporal complexity alone; the linear economic mean used by STI-X is also less aligned with this construction. This advantage concerns counterfactual signal recovery; the corresponding ATT-RMSE difference from STI-X remains uncertain.

Economic edge adjustment improves graph borrowing in D2--D4, but no method dominates across the five designs.

\subsection{Can direct covariate balancing replace graph-based borrowing?}

The Cov-SCM comparison separates direct balancing of observed attributes from borrowing representations informed by other targets\textquotesingle{} PRE histories. Graph-SCM and GeoEco-SCM have lower post-only RMSE than Cov-SCM in D1, D4, and D5, with paired intervals excluding zero. Their improvement shows that this covariate-balancing rule does not recover all the useful information available through PRE-informed donor representations. In D2, Cov-SCM improves on geography-only Graph-SCM and has slightly lower mean RMSE than GeoEco-SCM (0.0977 versus 0.1040), but their paired difference remains uncertain: Cov-minus-GeoEco is $-0.00631$, with interval $[-0.01265,0.00002]$.

\input{tables/covariate_pairs_primary.tex}

D3 provides the important opposite result. Cov-SCM has mean post-only RMSE 0.0926, compared with 0.1385 for GeoEco-SCM and 0.1810 for Graph-SCM. The Cov-minus-GeoEco interval is $[-0.05632,-0.03537]$. Selective historical coverage favors post-only targets with low latent PRE signal near the intervention, making their representations harder to borrow from the remaining observed-history targets. Direct use of their own observed covariates may then be more informative. STI-X remains more accurate still, with mean RMSE 0.0750. Its improvement over STI\textquotesingle{}s 0.1699 may reflect using economic information directly in outcome-level prediction, whereas GeoEco-SCM uses it to adjust existing geographic edges.

Uniform has the lowest mean post-only RMSE in D5: 0.0978, compared with 0.1040 for Graph-SCM, 0.1073 for GeoEco-SCM, and 0.2034 for Cov-SCM. D5 draws independent symmetric Dirichlet weights, whose conditional mean is uniform; unrelated neighbors or attributes do not reveal an individual target\textquotesingle{}s draw. Thus outperforming Cov-SCM in this negative control is not evidence that spatial borrowing itself is useful there.

The usefulness of direct covariate balancing depends on the design: it outperforms graph borrowing in D3 and is less accurate in D1, D4, and D5.

\subsection{ATT recovery and selective historical coverage}

Table~\ref{tab:att-primary} reports post-only ATT RMSE and signed bias. RMSE is calculated over POST years within each replication and then averaged. Pointwise counterfactual error and aggregate ATT error need not rank methods identically because errors can cancel across targets. In D2, Cov-SCM has slightly lower counterfactual RMSE than GeoEco-SCM, but its ATT RMSE is higher: 2.998 versus 2.092 percentage points. The paired Cov-minus-GeoEco ATT-RMSE interval is $[0.364,1.448]$ (Table~\ref{tab:covariate-pairs-primary}).

\input{tables/att_primary.tex}

D3 is a failure case for graph-based inclusion. Its post-only group\textquotesingle{}s average true percentage ATT is about $+4.87\%$. GeoEco-SCM\textquotesingle{}s signed bias is $-8.164$ percentage points, implying an average estimate of about $-3.30\%$ and reversing the sign. Graph-SCM\textquotesingle{}s bias is $-11.838$ points. Cov-SCM has bias $+0.306$ and ATT RMSE 2.961, compared with GeoEco-SCM\textquotesingle{}s 8.199. STI-X has lower ATT RMSE still, at 2.145 points. Matching the number of PRE-observed graph neighbors during validation does not ensure that pseudo-post-only targets resemble the selectively unobserved group in outcome levels or donor relationships.

In practice, comparing the observed PRE covariate distributions and local anchor availability of post-only and observed-history targets helps assess whether borrowing between the groups is plausible.

Population coverage remains a separate issue. In D3, the true mean percentage ATT over POST years is 3.63\% for all targets, 3.32\% for the PRE-observed group, and 4.87\% for post-only targets. Ordinary SCM estimates about 3.32\% for its PRE-observed population. Its proximity to that population\textquotesingle{}s truth does not establish recovery of the all-target effect. The Online Appendix reports the full population comparison, all-target ATT recovery, and group-specific uncertainty summaries.

Including post-only targets extends population coverage, while counterfactual errors in that group can still distort the all-target ATT.

\subsection{Overall coverage and economic adjustment}

All-target counterfactual RMSE is lower for both graph estimators than Cov-SCM in each primary design. This average includes 60 complete-PRE and 36 partial-PRE targets, where the graph estimators can use own PRE outcomes and Cov-SCM deliberately does not. It therefore complements, rather than replaces, the post-only comparisons. The all-target and separate complete- and partial-PRE results are reported in the Online Appendix. Ordinary SCM remains competitive on PRE-observed targets in those comparisons; extending population coverage is a separate benefit from improving their individual predictions.

The selected economic adjustment also responds to the available structure. GeoEco-SCM selects $\beta=0$ in 87/100 D1 replications and 64/100 D5 replications, often selecting the geography-only specification when economic similarity supplies no relevant donor information. Selection frequencies for all designs are given in the Online Appendix. One objective therefore covers all three history regimes, and economic reweighting can be switched off by PRE validation; the D1 and D5 results also show that this selection does not eliminate every cost of unnecessary adjustment.

Secondary diagnostics in the Online Appendix, Section~8.6, compare the fitted donor weights with the generating weights. Economic adjustment reduces weight error relative to Graph-SCM in D2--D4, but GeoEco-SCM does not consistently have lower weight error than Uniform, emphasizing that recovery of the generating weights and counterfactual prediction are distinct evaluation targets.

\subsection{Numerical checks and supplementary designs}
\label{sec:numerical-sensitivity}

Final weight optimization completed for all 1,000 datasets, with OSQP status \texttt{solved}. Three cases triggered an upper-$\beta$ search flag during adaptive screening; the flags and final confirmed choices are reported in the Online Appendix.

In the earlier 20-replication-per-design range-sensitivity check, adding shorter spatial correlation ranges changed mean all-target and post-only RMSE by less than 1\% for STI, STI-X, STD, and STD-X. The 100-replication study uses the resulting expanded grid. The exploratory grid development using known simulation truth is described in Section~\ref{sec:mc-benchmark-tuning}.

The supplementary designs examine selective geography-only coverage, separated donor support, abrupt boundaries, and local economic alignment. Full results and construction details are retained in the Online Appendix.

The local pair favors economic reweighting under both alignment settings. Its point estimate suggests a larger gain under misalignment, but the paired interval still includes zero.

\subsection{Practical recommendations}
\label{sec:method-choice}

The simulations suggest choosing a method according to whether useful information lies primarily in transferable donor relationships or in predictable outcome levels.

Graph-SCM and GeoEco-SCM are attractive when the donor pool provides plausible counterfactual building blocks and nearby locations with PRE histories can inform a post-only target\textquotesingle{}s donor representation. D4 suggests considering GeoEco-SCM when stable donor combinations coexist with nonlinear economic outcome structure and complex temporal dynamics. Its lower post-only signal RMSE than STI-X concerns this combination of features, rather than an isolated advantage under temporal complexity. Within the graph family, economic reweighting is useful when geography alone misses relevant differences in donor relationships, as in D2--D4. Geography-only Graph-SCM is preferable when economic proxies add no useful donor information, as the D1 and D5 comparisons illustrate.

Geographically separated donors are particularly relevant when nearby potential donors are excluded to limit spillover exposure. D7 illustrates this use: GeoEco-SCM has post-only RMSE comparable to STI-X while using the separated donor pool. The local information comes from PRE-observed targets within the treated region; donor validity remains a requirement, and the graph does not itself model spillover effects.

When geographic and economic characteristics are informative about outcome levels, spatio-temporal prediction should be a leading candidate. STI performs well in D1, and the strong STI-X results in D2 and D3 favor considering economic covariates directly in the outcome mean model. Under selective historical coverage, direct covariate balancing also deserves consideration: D3 shows that a post-only target's own attributes can be more informative than donor relationships borrowed from an unrepresentative observed-history group.

Before relying on post-only predictions, compare the actual post-only targets with the pseudo-post-only validation targets in terms of their observed PRE economic attributes, distance to the nearest observed-history target, and number of immediate observed-history neighbors. These checks assess whether the validation exercise covers the population to be predicted. D3 shows why matching neighbor counts alone is insufficient when historical coverage selects targets with different donor relationships. Where observed overlap is poor, assess direct covariate balancing and outcome-side prediction alongside graph borrowing.

When spatial and economic borrowing information is weak, retain uniform donor weights as a separate reference, as illustrated by D5. In applications, Uniform can be evaluated on the same temporal and pseudo-post-only PRE holdouts as the graph candidates; the implemented graph-tuning rule does not select it automatically. These PRE comparisons can guide method choice, while selective coverage driven by unobserved signal differences can remain difficult to detect.

\section{Conclusion}
\label{sec:conclusion}

Graph-SCM constructs synthetic controls for locations with complete, partial, or absent PRE histories within one convex estimation problem. Locations with observed histories inform donor representations, and a geographic graph determines how these representations are shared with post-only locations. GeoEco-SCM retains the geographic topology while adjusting local borrowing strengths using PRE economic similarity. The common formulation provides a coherent constrained approach to heterogeneous historical coverage.

The five primary experiments identify useful conditions and limits. PRE-informed graph borrowing improves on direct covariate balancing in several settings, but selective post-only coverage can favor covariate balancing instead. Spatio-temporal imputation with economic covariates is strong in the economic and selective designs, while borrowing representations is more accurate for the post-only signal in the design combining nonlinear economic loadings with complex donor dynamics. Economic reweighting improves on geography-only borrowing when it carries relevant donor information and can hurt otherwise. Extending coverage is therefore valuable when the borrowed representations are appropriate; it does not itself ensure accurate counterfactuals or treatment effects.

Future research could combine graph-regularized donor representations with explicit models of spatial treatment spillovers. Such an extension would address settings where nearby donor locations are also affected by the intervention. Other extensions include adding economic-neighbor edges, fitting local subgraphs, and examining sensitivity to target density and PRE-history length; the present experiments fix $N=120$ and 11 PRE periods.

\section*{Data and code availability}
All data are synthetic. Replication code and replication-level evaluation data are available on Zenodo, version 0.0.0 \citep{seya2026graphscmcode}, at \url{https://doi.org/10.5281/zenodo.24056020}.

\bibliographystyle{elsarticle-harv}
\bibliography{references,references_additions}

\end{document}


\maketitle

\section*{Reading guide and notation}

D1--D5 are the five main-paper designs; D6--D10 are supplementary stress tests. Section~1 gives the interpolation and graph-reduction proofs, Sections~3--6 specify the graph and numerical procedures, Section~7 defines the DGPs, and Sections~8--9 report additional results.

\paragraph{Panel, outcomes, and weights.}
There are $N$ treated targets, indexed by $i$, $J$ untreated donors, indexed by $j$, and $T$ dates, indexed by $t$. Write $\Tpre=\{1,\ldots,T_0\}$ and $\Tpost=\{T_0+1,\ldots,T\}$. The set $\Oi\subseteq\Tpre$ contains the observed PRE dates for target $i$. Anchors are $\mathcal A=\{i:|\Oi|>0\}$, post-only targets are $\mathcal U=\{i:|\Oi|=0\}$, and $N_+=|\mathcal A|$. The donor-outcome vector is $\bm Y_t^C=(Y_{1t}^C,\ldots,Y_{Jt}^C)^\top$. Each row of $\bm W\in\mathbb R^{N\times J}$ is $\bm w_i^\top$, where
\[
\Delta_J=\{\bm w\in\mathbb R^J:\bm w\ge0,\ \bm1_J^\top\bm w=1\}.
\]
A star denotes generating weights and a hat denotes fitted weights. In the simulations, all outcomes are logarithmic, $C_{jt}=Y_{jt}^C$, $\mu_{it}=\bm w_i^{\star\top}\bm Y_t^C$ is the latent untreated signal, and $\tau_{it}$ is the log treatment effect. The realized untreated outcome also contains observation noise. The settings are $N=120$, $J=24$, $T=22$, and $T_0=11$, with calendar years 1994--2015 and treatment beginning in 2005.

Complete-PRE targets have $\Oi=\Tpre$, partial-PRE targets have $0<|\Oi|<T_0$, and post-only targets have $\Oi=\varnothing$. Their numbers are 60, 36, and 24 per replication. Write $Y_{it}$ for the observed treated-target outcome and $Y_{it}(0)$ for its untreated potential outcome. Counterfactual predictions are $\widehat Y_{it}(0)=\widehat{\bm w}_i^\top\bm Y_t^C$, and estimated log effects are $\widehat\tau_{it}=Y_{it}-\widehat Y_{it}(0)$. Coordinates are $\bm s_i=(x_i,y_i)^\top$ for targets and $\bm s_j=(x_j,y_j)^\top$ for donors; scalar coordinate $x_i$ is distinct from the economic covariate vector $\bm x_i$ used in Section~3.

Coordinates, donor outcomes, available target PRE outcomes, and the three noisy economic proxies are estimation inputs. The states $g_{a\ell},e_{a\ell},m_{a\ell}$, generating weights $\bm w_i^\star$, latent signals $\mu_{it}$, and true effects $\tau_{it}$ are used only for DGP construction or evaluation. Section~7 defines these states and proxies separately.

\paragraph{Graph and matrix notation.}
The graph $G=(V,E,\bm A)$ connects targets, with symmetric adjacency entries $a_{ik}\ge0$, weighted degrees $d_i=\sum_k a_{ik}$, and Laplacian $\bm L=\operatorname{diag}(d_1,\ldots,d_N)-\bm A$. Each undirected edge is counted once, and $A_+=\sum_{(i,k)\in E}a_{ik}>0$ is total edge weight. Subscripts $\mathcal A,\mathcal U$ select anchor and post-only blocks of matrices. The vector $\bm1_m$ has $m$ entries equal to one, $\bm I_m$ is the $m\times m$ identity matrix, $\operatorname{tr}$ denotes trace, $I(\cdot)$ is an indicator, and $\|\cdot\|_2$ and $\|\cdot\|_F$ are the Euclidean and Frobenius norms. The nonnegative parameters $\lambda$ and $\beta$ control graph regularization and economic edge adjustment; $\varepsilon=10^{-10}$ is the numerical ridge in the weight objective, distinct from the DGP observation noise $\epsilon_{it}$.

\section{Proofs and numerical qualifications for the main-paper propositions}
\label{app:proofs}

This section proves the two main-paper propositions: conditional graph interpolation and the equivalent graph penalty after substituting the post-only weights. Both describe the case of positive graph regularization and zero numerical ridge. Section~\ref{app:numerical-ridge} gives the corresponding conditional interpolation with the numerical ridge used in computation.

Throughout the Online Appendix, vectors and matrices are written in bold italic type, while scalar quantities are written in ordinary math italic. Let $G=(V,E,\bm A)$ be the undirected weighted graph defined in the main paper, with weighted adjacency matrix $\bm A=(a_{ik})$, weighted degree matrix $\bm D$, and graph Laplacian $\bm L=\bm D-\bm A$. Let $\mathcal A$ denote anchor targets with observed PRE outcomes and $\mathcal U$ post-only targets. The graph-support condition requires every connected component containing a post-only target to contain at least one anchor. After ordering anchors first and post-only targets second,
\begin{equation}
\bm L=
\begin{pmatrix}
\bm L_{\mathcal A\mathcal A} & \bm L_{\mathcal A\mathcal U}\\
\bm L_{\mathcal U\mathcal A} & \bm L_{\mathcal U\mathcal U}
\end{pmatrix},
\qquad
\bm W=
\begin{pmatrix}
\bm W_{\mathcal A}\\
\bm W_{\mathcal U}
\end{pmatrix}.
\end{equation}
The unnormalized graph penalty is
\begin{equation}
Q(\bm W)
=\sum_{(i,k)\in E}a_{ik}\|\bm w_i-\bm w_k\|_2^2
=\operatorname{tr}(\bm W^\top\bm L\bm W).
\end{equation}
For fixed anchors and zero numerical ridge, dividing the graph penalty by its positive total edge weight does not change its minimizer. With positive ridge, the relative scaling is accounted for by $\rho$ in Section~\ref{app:numerical-ridge}.

\subsection{Proof of Proposition 1: donor-weight interpolation}
\label{app:proof-prop1}

\begin{proof}[Proof of Proposition 1]
Fix simplex-valued $\bm W_{\mathcal A}$. The fitting loss contains no post-only outcomes, so for $\lambda>0$ and $\varepsilon=0$ minimizing the full objective over its post-only rows is equivalent to minimizing $Q$. Expanding the quadratic form gives
\begin{align}
Q(\bm W)
={}&\operatorname{tr}(\bm W_{\mathcal A}^\top\bm L_{\mathcal A\mathcal A}\bm W_{\mathcal A})
+2\operatorname{tr}(\bm W_{\mathcal A}^\top\bm L_{\mathcal A\mathcal U}\bm W_{\mathcal U})\notag\\
&+\operatorname{tr}(\bm W_{\mathcal U}^\top\bm L_{\mathcal U\mathcal U}\bm W_{\mathcal U}).
\end{align}
Differentiating with respect to $\bm W_{\mathcal U}$ yields
\begin{equation}
\frac{\partial Q}{\partial \bm W_{\mathcal U}}
=2\bm L_{\mathcal U\mathcal U}\bm W_{\mathcal U}
+2\bm L_{\mathcal U\mathcal A}\bm W_{\mathcal A}.
\end{equation}
We first show that $\bm L_{\mathcal U\mathcal U}$ is invertible under the graph-support condition. For any nonzero vector $\bm x\in\mathbb R^{|\mathcal U|}$, extend it to all vertices by setting $\bm z_{\mathcal A}=\bm 0$ and $\bm z_{\mathcal U}=\bm x$. Then
\begin{equation}
\bm x^\top\bm L_{\mathcal U\mathcal U}\bm x
=\bm z^\top\bm L\bm z
=\sum_{(i,k)\in E}a_{ik}(z_i-z_k)^2\ge0.
\end{equation}
If this quantity were zero, $\bm z$ would be constant on every connected component. Every connected component containing a post-only target also contains an anchor, and every anchor has $z_i=0$ by construction. Hence $\bm z$ must be zero on every component containing post-only targets, implying $\bm x=\bm0$, a contradiction. Therefore $\bm L_{\mathcal U\mathcal U}$ is positive definite and invertible.

Setting the derivative equal to zero gives the unique minimizer
\begin{equation}
\bm W_{\mathcal U}
=-\bm L_{\mathcal U\mathcal U}^{-1}\bm L_{\mathcal U\mathcal A}\bm W_{\mathcal A}
=\bm H\bm W_{\mathcal A}.
\end{equation}
It remains to show that the rows of $\bm H$ are convex-combination weights. We first show nonnegativity without introducing additional matrix terminology. Consider one column of $\bm H$, corresponding to an anchor $a$. Give that anchor boundary value one, all other anchors boundary value zero, and let the post-only values be the corresponding column of $\bm H$. The first-order condition says that every post-only value equals the edge-weighted average of its neighboring values. Suppose, for contradiction, that some post-only value were negative, and choose a post-only location attaining the minimum value. All of its neighboring post-only values are at least as large, and all anchor boundary values are nonnegative. A weighted average of those neighboring values therefore cannot equal a strictly smaller negative minimum unless every connected neighbor has exactly the same negative value. Repeating this argument along the connected component would force the same negative value to reach an anchor, contradicting the nonnegative anchor boundary values. The graph-support condition ensures that such an anchor is reached. Hence every entry of $\bm H$ is nonnegative.

Finally, $\bm L\bm 1=\bm0$. Taking the rows corresponding to $\mathcal U$ gives
\begin{equation}
\bm L_{\mathcal U\mathcal U}\bm 1_{|\mathcal U|}
+\bm L_{\mathcal U\mathcal A}\bm 1_{|\mathcal A|}=\bm0.
\end{equation}
Multiplying by $-\bm L_{\mathcal U\mathcal U}^{-1}$ yields
\begin{equation}
\bm H\bm 1_{|\mathcal A|}=\bm 1_{|\mathcal U|}.
\end{equation}
Thus each row of $\bm H$ is nonnegative and sums to one. If each anchor donor-weight vector lies in the simplex $\Delta_J$, every row of $\bm W_{\mathcal U}=\bm H\bm W_{\mathcal A}$ is a convex combination of simplex vectors and therefore also belongs to $\Delta_J$.
\end{proof}

The same first-order condition can be read locally. For a post-only target $i$,
\begin{equation}
\bm w_i=\frac{1}{d_i}\sum_{k:(i,k)\in E}a_{ik}\bm w_k,
\end{equation}
where $d_i=\sum_k a_{ik}$. Thus the formal result is exactly the weighted-neighbor interpretation given below Proposition~1 in the main paper.

\subsection{Proof of Proposition 2: equivalent graph penalty after interpolation}
\label{app:proof-kron}

\begin{proof}[Proof of Proposition 2]
Using the block expansion of $Q(\bm W)$ from Section~\ref{app:proof-prop1}, substitute the minimizing post-only weights
\begin{equation}
\bm W_{\mathcal U}
=-\bm L_{\mathcal U\mathcal U}^{-1}\bm L_{\mathcal U\mathcal A}\bm W_{\mathcal A}.
\end{equation}
Because $\bm L$ is symmetric, $\bm L_{\mathcal A\mathcal U}=\bm L_{\mathcal U\mathcal A}^\top$ and $\bm L_{\mathcal U\mathcal U}^{-1}$ is symmetric. Straightforward substitution and collection of the cross and quadratic terms yield
\begin{equation}
\min_{\bm W_{\mathcal U}}Q(\bm W)
=\operatorname{tr}\!\left[\bm W_{\mathcal A}^\top
\left(
\bm L_{\mathcal A\mathcal A}
-\bm L_{\mathcal A\mathcal U}\bm L_{\mathcal U\mathcal U}^{-1}\bm L_{\mathcal U\mathcal A}
\right)
\bm W_{\mathcal A}\right].
\end{equation}
Defining
\begin{equation}
\bm L_{\mathrm{eff}}
=\bm L_{\mathcal A\mathcal A}
-\bm L_{\mathcal A\mathcal U}\bm L_{\mathcal U\mathcal U}^{-1}\bm L_{\mathcal U\mathcal A}
\end{equation}
produces the expression in Proposition~2.
\end{proof}

The matrix $\bm L_{\mathrm{eff}}$ is the Schur complement obtained after eliminating the post-only rows. In network theory the same reduction is often called Kron reduction. The terminology is secondary here: substantively, post-only locations affect the indirect pathways linking anchors but do not add PRE outcome observations.

\subsection{Numerical ridge and conditional interpolation}
\label{app:numerical-ridge}

\paragraph{Purpose and scale of the numerical ridge.}
Short histories, redundant donor trajectories, or weakly determined graph relationships can leave multiple or nearly equivalent donor-weight matrices. The full objective defined in the main text contains $\varepsilon R(\bm W)$, where $R(\bm W)=\|\bm W\|_F^2/N$. In this study, we fix $\varepsilon=10^{-10}$ for both Graph-SCM and GeoEco-SCM throughout tuning and final estimation. This quadratic term has Hessian $(2\varepsilon/N)\bm I_{NJ}$ in the vectorized weights. It therefore makes the weight objective strictly convex, yielding a unique mathematical minimizer on the feasible set, while retaining a quadratic objective and the same linear constraints.

Let $\bm u=\bm1_J/J$ denote uniform donor weights. Because each row belongs to $\Delta_J$,
\[
\|\bm w_i\|_2^2
=\|\bm w_i-\bm u\|_2^2+\frac1J.
\]
Thus minimizing the ridge penalty is equivalent to favoring weights closer to uniform allocation. It is an explicit computational selection rule, not evidence that uniform weights or the selected weights are closer to the true donor representation. For finite $\varepsilon>0$, the penalty can change the fitting and graph terms rather than merely choose between exactly tied solutions.

The simplex constraint gives
\[
\frac1J\le R(\bm W)\le1,
\qquad
\frac{\varepsilon}{J}\le\varepsilon R(\bm W)\le\varepsilon.
\]
Hence $\varepsilon=10^{-10}$ contributes at most $10^{-10}$ to the normalized objective for any feasible weights. More precisely, let $F(\bm W)$ denote the objective without the numerical ridge, let $\bm W_0$ minimize $F$, and let $\bm W_{\varepsilon}$ minimize $F+\varepsilon R$, all for the same fixed graph and tuning parameters. Comparing their objective values gives
\begin{align*}
0\le F(\bm W_{\varepsilon})-F(\bm W_0)
&\le\varepsilon\{R(\bm W_0)-R(\bm W_{\varepsilon})\}\\
&\le\varepsilon(1-1/J).
\end{align*}
This justifies the choice as a positive numerical penalty with a bounded effect on the optimized original objective. It is an exact-optimization statement, not a bound on solver error or on differences in weights and POST predictions. Weakly determined directions can still produce appreciable prediction changes.

\paragraph{Conditional post-only interpolation.}
For $\lambda>0$, define
\begin{equation}
\rho=\frac{\varepsilon A_+}{\lambda N},\qquad
\bm K=-(\bm L_{\mathcal U\mathcal U}+\rho\bm I_{|\mathcal U|})^{-1}
\bm L_{\mathcal U\mathcal A}.
\end{equation}
For fixed simplex-valued anchor weights, the exact conditional minimizer is
\begin{equation}
\widehat{\bm W}_{\mathcal U}
=\bm K\widehat{\bm W}_{\mathcal A}
+(\bm1_{|\mathcal U|}-\bm K\bm1_{|\mathcal A|})\bm u^\top.
\label{eq:ridge-conditional-supp}
\end{equation}
To derive the formula, center all rows at $\bm u$. The centered row-sum constraints are zero, and the conditional quadratic objective has first-order equation
\begin{equation}
(\bm L_{\mathcal U\mathcal U}+\rho\bm I_{|\mathcal U|})
(\widehat{\bm W}_{\mathcal U}-\bm1_{|\mathcal U|}\bm u^\top)
=-\bm L_{\mathcal U\mathcal A}
(\widehat{\bm W}_{\mathcal A}-\bm1_{|\mathcal A|}\bm u^\top).
\end{equation}
The solution is feasible for the simplex. Interpret $\rho$ as an edge from each post-only vertex to an additional fixed boundary with donor weights $\bm u$. The maximum-principle proof of Proposition~1 gives $\bm K\ge0$ and $\bm K\bm1\le\bm1$. Equation~\eqref{eq:ridge-conditional-supp} is therefore a convex mixture of anchor weights and the uniform vector, and is also the constrained conditional minimizer.

At a POST date, write $\widehat{\bm Y}_{\mathcal A t}(0)=\widehat{\bm W}_{\mathcal A}\bm Y_t^C$ and $\overline Y_t^C=\bm u^\top\bm Y_t^C$. Multiplying Eq.~\eqref{eq:ridge-conditional-supp} by $\bm Y_t^C$ yields
\[
\widehat{\bm Y}_{\mathcal U t}(0)
=\bm K\widehat{\bm Y}_{\mathcal A t}(0)
+(\bm1-\bm K\bm1)\overline Y_t^C.
\]
The second term is the contribution from the equally weighted average of donor outcomes. In a general graph, $\bm K$ also changes the relative contributions of different anchors. Thus this expression need not be a mixture of the unchanged no-ridge harmonic prediction and the donor mean. Moreover, it is conditional on the fitted anchor weights; those weights can themselves change when ridge is added to the full problem.

For a simple illustration with only one post-only vertex $i$, let $d_i=\sum_{a\in\mathcal A}a_{ia}$ and $\bm m_i=d_i^{-1}\sum_{a\in\mathcal A}a_{ia}\widehat{\bm w}_a$. Then
\[
\widehat{\bm w}_i
=\frac{d_i}{d_i+\rho}\bm m_i
+\frac{\rho}{d_i+\rho}\bm u.
\]
The graph penalty pulls toward the neighbor-weighted vector $\bm m_i$, and the numerical ridge pulls toward uniform donor weights. Prediction uses the same two mixing coefficients for $\bm m_i^\top\bm Y_t^C$ and $\overline Y_t^C$.

The influence of ridge depends on $\rho$ and the strength of connections to anchors, not solely on $\varepsilon$. The ridge supplies no PRE observations and is not an identification assumption.

At $\lambda=0$, each post-only row is determined by the ridge alone and is uniform. Without the ridge, every simplex vector is a minimizer for that row. A component containing only post-only locations similarly has no anchor boundary information: graph smoothing alone cannot determine its common donor vector, even though the ridge selects the uniform solution. Connectivity remains a graph-support diagnostic.

\section{Relation to post-only and missing-data methods}

\paragraph{Post-intervention-only identification.}
\citet{liu2024proximal} develop a proximal synthetic-control method with post-intervention surrogates. Under bridge and proxy conditions, treatment effects can in some cases be identified using exclusively post-intervention information. Their source of identification is post-treatment surrogate moments. Graph-SCM instead never uses the post-only target's own POST outcome to learn its weights; its representation is propagated from PRE-observed targets through the graph.

\paragraph{Missing PRE outcomes within SCM.}
\citet{liu2026missing} study SCM with missing pre-treatment outcomes and the bias and sample-selection consequences of missingness. Robust SCM uses singular-value thresholding to denoise and complete incomplete panels \citep{amjad2018robust}. These approaches are natural for partial histories. A completely post-only target removes the unit-specific SCM fit criterion and therefore requires either an additional outcome model, a proxy restriction, or a cross-unit restriction such as the graph assumption used here.

\paragraph{Matrix completion and latent factors.}
Nuclear-norm matrix completion \citep{athey2021matrix}, general missing-pattern latent-factor methods \citep{xiong2023latent}, and causal matrix completion under non-random missingness \citep{agarwal2023causal} impose structure on the outcome matrix or latent factors. \citet{vu2022spatial} add spatial structure to matrix completion for spatially misaligned air-pollution measurements. Graph-SCM instead preserves the donor representation and regularizes the weight field. Spatial vertical regression also shares information across treated locations through donor-coefficient fields \citep{grossi2026svr}. The methods therefore encode different extrapolation restrictions.

\section{Exact graph construction and economic metric}

The public implementation uses a symmetrized geographic $k$-nearest-neighbor graph with $k=8$. Let $\bm s_i=(x_i,y_i)^\top$ be the coordinate vector of target $i$, and let $E_d$ denote the directed kNN edges. For each directed edge,
\begin{equation}
d^g_{ik}=\|\bm s_i-\bm s_k\|_2,
\end{equation}
and $h_g$ is the median positive value of $d^g_{ik}$ over the directed neighbor list $E_d$. Symmetrization retains a pair whenever either endpoint selects the other; $E$ is the resulting set of undirected pairs. After computing raw weights, duplicate opposite-direction edges are collapsed by averaging them.

For the economic metric, let $\bm x_i\in\mathbb R^p$ collect the continuous quantitative PRE economic covariates for target $i$, each with positive sample variance in this study; here $p=3$. Each component is centered and scaled using the target sample, producing the standardized vector $\bm z_i$. Let $\bm S$ be the $p\times p$ sample covariance matrix of these standardized covariate vectors, and let $\bm I_p$ be the $p\times p$ identity matrix. The operator $\diag(\bm S)$ retains the diagonal entries and sets the off-diagonal entries to zero.
\begin{equation}
\widetilde{\bm S}=(1-\eta)\bm S+\eta\diag(\bm S)+\delta_{\mathrm{cov}}\bm I_p.
\end{equation}
For this study, we set $\eta=0.15$ and $\delta_{\mathrm{cov}}=10^{-6}$. The covariance adjustment $\delta_{\mathrm{cov}}$ is distinct from the donor-weight ridge $\varepsilon=10^{-10}$ in the full estimation objective.
The economic distance between targets $i$ and $k$ is the Mahalanobis distance
\begin{equation}
d^e_{ik}=\sqrt{(\bm z_i-\bm z_k)^\top \widetilde{\bm S}^{-1}(\bm z_i-\bm z_k)}.
\end{equation}
Let $h_e$ be the median positive economic distance $d^e_{ik}$ over $E_d$, using the same directed neighbor list as for $h_g$. Both bandwidths are determined before economic reweighting. For a fixed nonnegative economic-weighting parameter $\beta$,
\begin{equation}
\widetilde a_{ik}=\exp[-d^g_{ik}/h_g-\beta d^e_{ik}/h_e],
\end{equation}
The final weights are $a_{ik}=\widetilde a_{ik}/\{|E|^{-1}\sum_{(r,q)\in E}\widetilde a_{rq}\}$ for $(i,k)\in E$, with zero entries off $E$. Thus the mean edge weight is one and $A_+=|E|$. This normalization keeps the magnitude of the smoothness penalty comparable when $\beta$ changes.

\section{PRE-only tuning algorithm}

For $N_+>0$, the PRE fitting loss is
\[
\mathcal L_{\rm fit}(\bm W)=\frac{1}{N_+}\sum_{i\in\mathcal A}
\frac{1}{|\Oi|}\sum_{t\in\Oi}(Y_{it}-\bm w_i^\top\bm Y_t^C)^2.
\]
The graph term is $\lambda Q(\bm W)/A_+$ and the numerical ridge is $\varepsilon\|\bm W\|_F^2/N$. The observed-history sets and counts in the fitting loss are recomputed within each validation fit after holding out or masking cells. The final replication code follows the sequence below.
\begin{enumerate}[leftmargin=*]
\item Construct the geography-only $k=8$ graph and remove 2002--2004 from the target PRE panel.
\item Fit the $\lambda=0,\beta=0$ reference model, denoting its weights by $\widehat{\bm W}^{\rm ref}$, and define the scale $\lambda_b=\max\{\mathcal L_{\rm fit}(\widehat{\bm W}^{\rm ref})/2,10^{-8}\}$.
\item Screen $\lambda/\lambda_b\in\{0,0.1,0.3,1,3\}$ for $\beta=0$.
\item Cross every positive screened $\lambda$ with $\beta\in\{0,0.5,1,2,4\}$ and evaluate temporal holdout.
\item For all retained candidates, create spatially interspersed pseudo-post-only folds by masking complete-PRE targets. The selected pseudo targets are chosen so that their immediate PRE-anchor counts resemble those of the actual post-only targets.
\item Compute the combined validation score
\[
S(\lambda,\beta)=0.5R_{\rm time}(\lambda,\beta)+0.5\overline R_{\rm PO}(\lambda,\beta),
\]
where $R_{\rm time}$ is the outcome RMSE on the held-out PRE years and $\overline R_{\rm PO}$ is the average outcome RMSE across the pseudo-post-only validation folds. Both compare predicted with held-out observed target outcomes on the same log scale; neither is a donor-weight error.
When post-only targets are present, only positive-$\lambda$ candidates are eligible for final selection; the zero-$\lambda$ fit remains a diagnostic reference.
\item Adaptively extend the upper $\lambda$ tail by factors of 3 when necessary. The production safety search permits up to 11 tail steps, corresponding to a multiplier as large as 531441. If an economic candidate remains at the upper $\beta$ boundary, expand by factors of 2, with at most three expansion steps and a maximum $\beta=32$; reaching a safety limit is recorded.
\item Re-evaluate the leading configurations using three pseudo-post-only folds of eight targets each. Candidates within 0.5\% of the best confirmed score are treated as practically tied. Prefer the smallest $\beta$, since nearly equal predictive errors offer little support for stronger economic attenuation. Among candidates with that $\beta$, choose the lowest score, and then the smaller $\lambda$ if a tie remains. This is the practical-tolerance rule defined in the main paper.
\item Refit the selected model on the full available PRE panel using the sparse OSQP quadratic program with tight convergence tolerances.
\end{enumerate}
The equal coefficients in the validation score assign the same value to an absolute RMSE reduction in either task; they do not standardize the variability of the two scores. Applications with a different substantive balance can prespecify $\omega R_{\rm time}+(1-\omega)\overline R_{\rm PO}$, with $0\le\omega\le1$. This study uses $\omega=0.5$; fold confirmation and the preference for smaller $\beta$ among practically tied candidates are specified above.

Projected FISTA (Fast Iterative Shrinkage-Thresholding Algorithm; \citealp{beck2009fista}) is used as an accelerated proximal-gradient method for candidate screening. Each iteration takes a gradient step for the smooth quadratic objective, projects each target's weight vector onto the simplex, and applies the acceleration step. This scheme limits computation when many related fitting problems are solved during hyperparameter selection.

Two pseudo-post-only folds of eight targets are used initially. Screening permits 1500 FISTA iterations at weight-change tolerance $2\times10^{-5}$; intermediate confirmation permits 3000 at $5\times10^{-6}$. All simulation candidate fits use cold starts, so candidate comparisons do not use different warm-start histories. After selection, the model is refitted on the full available PRE panel using OSQP (Operator Splitting Quadratic Program; \citealp{stellato2020osqp}) with tighter convergence tolerances, as detailed in Section~5. FISTA is used for screening, whereas OSQP determines the reported final weights.

\section{Sparse quadratic program for the final fit}

Let $\mathcal M=\{(i,t):t\in\Oi\}$ be the observed target--PRE cells and let $M=|\mathcal M|$. Introduce residual vector $\bm r\in\mathbb R^M$. Vectorize $\bm W$ rowwise as $\bm w=\operatorname{vec}_r(\bm W)$. If $\bm B$ is the weighted graph incidence matrix, each row corresponds to an edge $e=(i,k)$ and is $\sqrt{a_{ik}}(\bm e_i-\bm e_k)^\top$, where $\bm e_i$ is the unit vector that selects target $i$. Let $\bm I_J$ be the $J\times J$ identity matrix. Then
\begin{equation}
\tr(\bm W^\top\bm L\bm W)=\|(\bm B\otimes\bm I_J)\bm w\|_2^2.
\end{equation}
The symbol $\otimes$ denotes the Kronecker product. Enumerate the observed cells by $m=1,\ldots,M$, and let $i(m)$ and $t(m)$ be the target and date of cell $m$. The fit residual constraints are sparse because each observed cell involves only the $J$ weights of one target:
\begin{equation}
r_m+\bm Y_{t(m)}^{C\top}\bm w_{i(m)}=Y_{i(m),t(m)}.
\end{equation}
The Hessian is block sparse and positive semidefinite. The ridge in the objective is $\varepsilon\|\bm W\|_F^2/N$ with $\varepsilon=10^{-10}$, so its weight-block Hessian contribution is $(2\varepsilon/N)\bm I_{NJ}$ under OSQP\textquotesingle{}s $\tfrac12\bm x^\top\bm P\bm x$ convention, where $\bm x$ stacks the weights and residual variables and $\bm P$ is the quadratic-coefficient matrix. The ridge adds nothing to the residual block. Nonnegativity and row-sum-one constraints are explicit. OSQP uses absolute and relative tolerances $10^{-8}$. The final QP status in all 1,000 datasets was \texttt{solved}. No $\lambda$ search retained a boundary flag. Adaptive $\beta$ screening retained flags in D4/index 99, D8/index 87 and D10/index 23; the final confirmed $\beta$ values were 8, 0 and 16, respectively. These screening flags are reported rather than treated as evidence that the search is insensitive to further expansion.

\section{Spatio-temporal benchmark details (Gneiting covariance)}

Let $h\ge0$ be geographic separation in meters and $u\ge0$ temporal separation in years. The positive parameters $\phi_s$ and $\phi_t$ are spatial and temporal correlation ranges, and $0\le\gamma\le1$ controls their interaction. The correlation model is
\begin{equation}
R(h,u)=\frac{\exp\left[-(h/\phi_s)/(1+u/\phi_t)^{\gamma/2}\right]}{1+u/\phi_t},
\end{equation}
which is the chosen fixed-shape member of the nonseparable class in \citet{gneiting2002nonseparable}. Spatial ranges are measured in meters and temporal ranges in years. The reference distance $\phi_{s,0}$ is the median of the positive, finite distances obtained by pooling, over distinct sites in the PRE training panel, the distances from each site to its first five nearest other sites. If fewer than six distinct sites are available, all available other sites are used; a fallback of 2000 m is used if no positive finite reference distance can be obtained. This scale is computed before subsampling PRE cells for covariance tuning. Using PRE data only, a profiled Gaussian marginal likelihood selects covariance parameters from
\begin{align}
\phi_t&\in\{0.5,1.5,4,8,12,24,48,96,192,384\},\\
\phi_s/\phi_{s,0}&\in\{0.03125,0.0625,0.125,0.25,0.5,1.5,4\},\\
\gamma&\in\{0,0.5,1\},\\
\delta_{\rm nugget}&\in\{0,0.01,0.15\}.
\end{align}
Here $\delta_{\rm nugget}$ is the ratio of independent-noise variance to correlated-process variance. For process variance $\sigma^2>0$, the correlated covariance is $\sigma^2R(h,u)$, and $\sigma^2\delta_{\rm nugget}$ is added to the diagonal of the observation covariance matrix. This yields $10\times7\times3\times3=630$ covariance candidates for each version, without or with economic covariates. At most 350 PRE cells are used in covariance tuning. In the economic-covariate version, the same PRE covariates used by GeoEco-SCM enter the mean model for spatio-temporal prediction. The final short-range sensitivity check reduced the minimum spatial-range candidate from $0.125\phi_{s,0}$ to $0.03125\phi_{s,0}$ by adding shorter-range candidates.

The direct spatio-temporal predictor for a post-treatment time $t$ uses only untreated-control outcomes observed no later than $t$ plus the target's observed PRE outcomes. Target POST outcomes and future donor outcomes are prohibited. A programmatic information-set audit checks these restrictions in every replication.

The mean model contains year effects and linear coordinates; the -X versions add the three PRE economic proxies. PRE OLS residuals enter the profiled covariance likelihood. Grid expansion used known simulation truth to examine mean POST errors and rankings; within-replication PRE-only fitting should therefore be distinguished from independent study-level validation. CF and ATT evaluations are taken from the same completed case and matched by design, seed, method, and observation group. The provenance audit has no unmatched results.

\section{Exact common Monte Carlo equations and design-specific construction}

\subsection{Overview of the primary and supplementary DGPs}
\label{app:all-dgps}

The ten DGPs are organized into three conceptual families. The first family evaluates when spatial smoothing of donor weights is appropriate or inappropriate: \emph{spatially smooth donor weights based on geography only} is a positive control, \emph{no spatial smoothness in donor weights} removes cross-location smoothness in the true donor weights, and \emph{abrupt donor-weight change at a spatial boundary} introduces a discontinuity that can make smoothing across nearby locations harmful. The second family adds economic or latent structure: \emph{donor weights that depend on geography and economics} tests whether economic similarity contributes beyond geography, \emph{latent temporal factors added} makes the outcome process harder to represent by a single global stationary covariance model while preserving stable donor weights, and the paired \emph{local geography--economics alignment/misalignment} designs change the local correspondence between geography and economics while holding the remaining paired ingredients fixed. The third family changes spatial support or the observation process: the two \emph{selective post-only} DGPs make the absence of PRE outcomes nonrandom, while \emph{donors spatially separated from targets} creates a spatial extrapolation problem in which relevant donor support lies away from the target region.

In D4 and D7, baseline donor levels, trends, and temporal loadings are generated from smooth functions of location instead of their direct linear dependence on latent market state in the standard construction. This additional change means that comparisons with D2 do not isolate only added latent factors in D4 or spatial separation in D7. Table~\ref{tab:dgp-purpose-all} states the role of each DGP. The designs organize controlled comparisons of borrowing mechanisms: whether the true weight field is spatially smooth, whether economic similarity contains nongeographic information, whether missingness selects high-effect locations, how donor-outcome structure affects the competing predictors, and whether local graph smoothness breaks at boundaries.

\small
\setlength{\tabcolsep}{4pt}
\begin{longtable}{@{}p{0.07\linewidth}p{0.38\linewidth}p{0.50\linewidth}@{}}
\caption{Sequential DGP labels and the purpose of each experiment.}\label{tab:dgp-purpose-all}\\
\toprule DGP & Defining feature & Intended comparison \\\midrule
\endfirsthead
\toprule DGP & Defining feature & Intended comparison \\\midrule
\endhead
\bottomrule\endfoot
D1 & Geographic weight smoothness; random coverage. & Baseline for borrowing PRE-informed donor representations from nearby targets. \\
D2 & Geographic and partly nonspatial economic weight structure. & Contribution of economic information beyond geography. \\
D3 & Economic structure with selective post-only coverage. & Transfer to a systematically different post-only population, and the consequence of excluding that population. \\
D4 & Spatial donor trajectories with nonlinear latent temporal factors. & Borrowing representations when the specified stationary covariance and linear mean are less aligned with the outcome process. \\
D5 & Independent target-specific Dirichlet weights. & Negative control without spatial weight information; comparison with the uniform conditional mean. \\
D6 & Geography-only weights with selective post-only coverage. & Whether the selective-coverage difficulty also arises without nonspatial economic structure. \\
D7 & Target disk surrounded by a separated donor annulus. & Predicting targets away from donor locations while retaining a valid donor dictionary. \\
D8 & Abrupt change in donor weights across a geographic boundary. & Risk of borrowing across nearby targets from different regimes. \\
D9 & Local economic alignment; common paired ingredients. & Reference member for the paired local-borrowing comparison. \\
D10 & Joint permutation of target economic states and proxies. & Weakening local alignment while holding the paired geography, coverage, effects, and donor trajectories fixed. \\
\end{longtable}
\normalsize

The remaining subsections specify the generating equations and fixed constants. D1--D5 are the primary designs, and D6--D10 are supplementary stress tests. The sequential labels are shared by the paper and public code; the catalog also records legacy labels and the unchanged random seeds.

D9--D10 are constructed to examine local borrowing. Their common observation pattern is selected using the misaligned design, where each post-only location must have nearby PRE-observed neighbors that are both economically similar and dissimilar. Using the aligned design as a reference, donor trajectories are also modified to weaken a linear relationship between target outcome levels and their observed economic covariates. The same modified donor trajectories are used in both designs. The pair preserves coordinates, observation groups, effects, noise, and the collection of economic covariate rows; the assignment of those rows to target locations changes.

\label{app:dgp}
The following equations give the full construction behind the main-paper simulation model. The latent untreated signal is
\begin{equation}
\mu_{it}=\sum_{j=1}^Jw_{ij}^{\star}C_{jt},\qquad
Y_{it}(0)=\mu_{it}+\epsilon_{it},\qquad
\epsilon_{it}\sim N(0,0.008^2).
\label{eq:dgp-exact-scm}
\end{equation}
Only $\mu_{it}$ has the exact convex donor representation; the realized untreated potential outcome additionally contains noise.
\subsection{Why the supplementary designs are included}
D6 complements D3 by applying selective historical coverage to geography-only donor weights. D6 can be read alongside the random-coverage geographic baseline D1, just as D3 can be read alongside D2. Together they examine whether a change in the population with observed histories is difficult even when geography is the only source of weight structure. These are separate random-seed designs, not paired interventions on one dataset.

D7 separates donor support from target geography: targets occupy a central disk and donors a surrounding annulus. Donor outcomes can still span the target signal, but prediction at target coordinates requires geographic extrapolation from donor locations. Like D4, this design also changes baseline donor trajectories, so it is a composite stress test rather than an isolated geometry change.

D8 introduces an east/west discontinuity in donor relationships. It checks the cost of treating geographic proximity as evidence for similar weights when nearby targets can belong to different regimes. Economic reweighting may reduce some inappropriate borrowing, but the design does not assume that economic proxies perfectly label the boundary.

D9--D10 form the only paired change of DGP ingredients. The same locations and observation groups are used, and target economic rows are reassigned jointly in D10. This tests sensitivity to local geography--economics alignment without changing the marginal collection of economic characteristics. The shared construction below deliberately retains mixed economic support and weakens a linear outcome-mean channel; it should therefore be read as a controlled local-borrowing experiment.

\subsection{Geometry and standardization}
Except in D7, donors and targets are drawn uniformly on $[500,19500]^2$ meters. D7 places targets uniformly in the disk of radius 5 km centered at $(10,10)$ km and donors in an annulus of radii 7--13 km, truncated to the study square. Two PRE accessibility centers are $(4500,5500)$ and $(15500,14000)$ m; the intervention is $(10500,9500)$ m. Standardized accessibility is negative distance to the nearer PRE center. For a vector $\bm v=(v_1,\ldots,v_M)^\top$, write $z[\bm v]_m=(v_m-\bar v)/s_v$, where $\bar v=M^{-1}\sum_m v_m$ and $s_v$ is its sample standard deviation; a constant vector is mapped to zero. Latent market variables and economic proxies are standardized jointly over donors and targets. Donor-only bases use the donor sample, target-selection quantities use the target sample, and time factors use the $T$ dates. All AR processes below begin with the first innovation rather than a stationary initialization.

\subsection{Latent geographic and economic state}

Scale coordinates around the center of the region by
\begin{equation}
x_i^\circ=(x_i-10{,}000)/9500,\qquad
y_i^\circ=(y_i-10{,}000)/9500.
\end{equation}
Two smooth geographic market coordinates are
\begin{align}
g_{i1}&=0.70x_i^\circ+0.20\sin(\pi y_i^\circ),\\
g_{i2}&=0.65y_i^\circ+0.18\cos(\pi x_i^\circ).
\label{eq:geo-latent}
\end{align}

At each donor or target location, let $q\in[0,1]$ control the mixing of the spatial economic component and define two latent economic states
\begin{equation}
e_{i\ell}
=\operatorname{z}\left[
\sqrt{q}\,f_{\ell}(\bm s_i)+\sqrt{1-q}\,\operatorname{z}[u_{i\ell}]
\right],
\qquad \ell\in\{1,2\},
\label{eq:econ-latent}
\end{equation}
where the raw $u_{i\ell}$ are independent $N(0,1)$ draws and $\operatorname{z}[\cdot]$ denotes standardization across donors and targets. Both the spatial field and the location-specific Gaussian component are standardized before mixing, as in the simulation code. D1--D8 use $q=0.20$; the local pair D9--D10 uses $q=0.60$ as specified below. The spatial components are
\begin{align}
f_1(\bm s_i)&=\operatorname{z}\left[
\sin(2\pi x_i/8500)\cos(2\pi y_i/11000)
\right],\\
f_2(\bm s_i)&=\operatorname{z}\left[
\cos(2\pi x_i/10500+y_i/6000)
\right].
\end{align}
When economic similarity is active, the latent market coordinates are
\begin{equation}
m_{i\ell}=\operatorname{z}[g_{i\ell}+0.65e_{i\ell}],
\label{eq:market-econ}
\end{equation}
and otherwise $m_{i\ell}=\operatorname{z}[g_{i\ell}]$.

Observed pre-treatment economic proxies are noisy functions of the latent state:
\begin{align}
x^{(1)}_i&=\operatorname{z}[0.85e_{i1}+\nu_{i1}],\\
x^{(2)}_i&=\operatorname{z}[-0.25e_{i1}+0.85e_{i2}+\nu_{i2}],\\
x^{(3)}_i&=\operatorname{z}[0.55e_{i1}-0.35e_{i2}+\nu_{i3}],
\label{eq:econ-proxies}
\end{align}
with $\nu_{ip}\sim N(0,0.18^2)$. The estimators see these proxies but never the latent economic states.

\subsection{True donor weights}

Let $a_i$ be standardized pre-treatment accessibility, defined as negative distance to the closest pre-existing accessibility center. For the baseline smooth-weight DGPs, the target--donor matching score is
\begin{equation}
q_{ij}
=-\frac{1}{2}
\left[
\left(\frac{m_{j1}-m_{i1}}{0.60}\right)^2
+\left(\frac{m_{j2}-m_{i2}}{0.60}\right)^2
+0.60\left(\frac{a_j-a_i}{0.80}\right)^2
\right].
\label{eq:true-score}
\end{equation}
The true donor weights are the softmax
\begin{equation}
w_{ij}^{\star}
=\frac{\exp(q_{ij})}{\sum_{r=1}^{J}\exp(q_{ir})}.
\label{eq:true-softmax}
\end{equation}
Here, target--donor \emph{physical distance does not enter Eq.~\eqref{eq:true-score}}. Geography matters because nearby targets tend to have similar latent market states and hence similar donor representations. The construction therefore evaluates smoothness of donor representations, not a nearest-donor rule.

The no-spatial-smoothness negative control instead draws
\begin{equation}
\bm w_i^{\star}=(w_{i1}^{\star},\ldots,w_{iJ}^{\star})^\top
\stackrel{ind}{\sim}\operatorname{Dirichlet}(0.2,\ldots,0.2),
\label{eq:no-spatial-weights}
\end{equation}
so that target weights are concentrated but unrelated across nearby locations.

\subsection{Treatment effect and observation pattern}

Let $d_i^{\mathrm{new}}$ be distance in meters from target $i$ to the newly introduced intervention site. The long-run log treatment effect is
\begin{equation}
\tau_i^{\infty}=0.015+0.105\exp(-d_i^{\mathrm{new}}/5000).
\label{eq:longrun-tau}
\end{equation}
For calendar year $y_t$, the treatment path is
\begin{equation}
\tau_{it}=
\begin{cases}
0, & y_t<2005,\\
\tau_i^{\infty}\min\{1,(y_t-2004)/5\}, & y_t\ge2005.
\end{cases}
\label{eq:tau-dgp}
\end{equation}
The observed target outcome is
\begin{equation}
Y_{it}=\mu_{it}+\tau_{it}+\epsilon_{it},
\qquad
\epsilon_{it}\sim N(0,0.008^2).
\label{eq:target-observed-dgp}
\end{equation}

Targets are assigned to 50\% complete PRE, 30\% partial PRE, and 20\% post-only groups. A partial-PRE target retains a random 4--8 of the 11 pre-treatment years. In the random-entry DGPs, group membership is random. In the selective post-only designs, the post-only set has a fixed size but is sampled with higher probability for sites that are both close to the intervention site and have low latent pre-treatment counterfactual levels. This creates a deliberate difference between $ATT_{\mathrm{PRE}}$ and $ATT_{\mathrm{ALL}}$ without changing the number of missing histories.

\subsection{Donor outcome trajectories in the standard architecture}

Let $r_t=y_t-1994$. Define a centered common factor
\begin{align}
\widetilde f_t^{c}&=0.012\sin(r_t/3.1)+0.008\cos(r_t/5.0)+u_t^c,\\
f_t^{c}&=\widetilde f_t^{c}-T^{-1}\sum_{s=1}^{T}\widetilde f_s^{c},
\end{align}
where
\begin{equation}
u_t^c=0.70u_{t-1}^c+\eta_t^c,\qquad \eta_t^c\sim N(0,0.004^2).
\end{equation}
For ordinary DGPs, donor $j$ has
\begin{align}
\alpha_j&=\log(220000)+0.18m_{j1}-0.10m_{j2}+\epsilon_{\alpha j},
&\epsilon_{\alpha j}&\sim N(0,0.10^2),\\
b_j&=0.004+0.003m_{j1}+0.0015m_{j2}+\epsilon_{bj},
&\epsilon_{bj}&\sim N(0,0.0045^2),\\
q_j&=-0.00005+0.00003m_{j2}+\epsilon_{qj},
&\epsilon_{qj}&\sim N(0,0.00008^2),\\
c_{1j}&=0.010m_{j1}+\epsilon_{1j},
&\epsilon_{1j}&\sim N(0,0.014^2),\\
c_{2j}&=-0.008m_{j2}+\epsilon_{2j},
&\epsilon_{2j}&\sim N(0,0.011^2),\\
\ell_j&=1+0.18m_{j1}-0.10m_{j2}+\epsilon_{\ell j},
&\epsilon_{\ell j}&\sim N(0,0.10^2).
\end{align}
The idiosyncratic temporal component follows
\begin{equation}
u_{jt}=0.55u_{j,t-1}+\eta_{jt},\qquad \eta_{jt}\sim N(0,0.006^2).
\end{equation}
The baseline donor log trajectory is
\begin{equation}
C_{jt}^{\mathrm{base}}=\alpha_j+b_jr_t+q_jr_t^2
+c_{1j}\sin(r_t/2.8)+c_{2j}\cos(r_t/4.2)
+\ell_j f_t^c+u_{jt}.
\label{eq:donor-standard-app}
\end{equation}
For the unmodified standard architecture, $C_{jt}=C_{jt}^{\mathrm{base}}$. The spatial-coefficient, additional-factor, and paired-local modifications below define the corresponding donor outcomes in D4, D7, and D9--D10.

\subsection{Alternative outcome architecture for added-latent-factor and spatially-separated-donor designs}

For D4 and D7, baseline donor-outcome coefficients change from the standard architecture to smooth spatial bases, removing their direct linear dependence on economic market state. Thus D4 and D7 are composite designs; their differences from D2 cannot be attributed solely to added factors or separated geometry.

Define standardized spatial bases
\begin{align}
B_{1j}&=z[\sin(\pi x_j/9000)+0.7\cos(\pi y_j/11000)],\\
B_{2j}&=z[\cos\{\pi(x_j+y_j)/17000\}],\\
B_{3j}&=z[\sin(2\pi x_j/20000)\cos(\pi y_j/20000)].
\end{align}
Each normal-distribution term below denotes an independent draw. Then
\begin{align}
\alpha_j&=\log(220000)+0.10B_{1j}+N(0,0.10^2),\\
b_j&=0.004+0.002B_{2j}+N(0,0.0045^2),\\
q_j&=-0.00005+N(0,0.00008^2),\\
c_{1j}&=0.008B_{3j}+N(0,0.012^2),\\
c_{2j}&=-0.006B_{1j}+N(0,0.010^2).
\end{align}
The common-factor loading is $1+0.12z[\sin(\pi x_j/10000)+0.6\cos(\pi y_j/12000)]+N(0,0.10^2)$.

The latent-factor DGP additionally adds
\begin{equation}
C^{LF}_{jt}=C_{jt}^{\mathrm{base}}+\frac{0.060}{\sqrt{3}}\sum_{r=1}^{3}\ell_{jr}^{LF}F_{rt},
\end{equation}
For D4, the donor outcome entering the target signal is $C_{jt}=C^{LF}_{jt}$. Each loading combines a weighted spatial basis and a nonlinear economic-archetype basis plus donor-specific noise; their exact coefficients are given below. The economic archetype bases are nonlinear radial-basis or sinusoidal functions of the latent economic state, residualized against the intercept and three observed economic proxies. The temporal factors include persistent AR(1), cyclical, and level-shift dynamics. This construction makes economic proximity informative about donor composition without making the observed economic proxies simple linear regressors for the outcome level.

\subsection{Design-specific modifications}

\subsubsection{Abrupt boundary (D8)}

For the spatial-boundary design, donor regime labels are defined by whether
\begin{equation}
m_{j1}+0.30m_{j2}
\end{equation}
is above or below its donor median. A target is assigned an east/west regime according to whether $x_i\ge10000$ m. Write $R_i$ and $R_j$ for these target and donor regime labels within this construction. The baseline score in Eq.~\eqref{eq:true-score} is modified to
\begin{equation}
q_{ij}^{B}=q_{ij}+2.50\,I(R_i=R_j).
\end{equation}
Thus adjacent targets on opposite sides of the $x=10$ km boundary can have abruptly different donor representations even though their geographic coordinates are close.

\subsubsection{Selective post-only sampling (D3 and D6)}

Let
\begin{equation}
n_i=\exp(-d_i^{\rm new}/3000),
\end{equation}
standardized across targets as $z(n_i)$. Let $\bar\mu_{i,PRE}=|\Tpre|^{-1}\sum_{t\in\Tpre}\mu_{it}$ be the mean latent untreated PRE signal and define $l_i=-z(\bar\mu_{i,PRE})$. The selection score is
\begin{equation}
s_i=z(n_i)+0.90l_i+0.50z(n_i)l_i.
\end{equation}
Exactly 20\% of targets are selected as post-only without replacement using sampling weights proportional to $\exp(s_i)$. The fixed count isolates composition effects from changes in the amount of missingness.

\subsubsection{Paired local construction (D9--D10)}

The paired local design increases the spatial share of the latent economic field from $q=0.20$ to $q=0.60$, making the aligned member a genuinely aligned baseline. The complete target economic row
\begin{equation}
(e_{i1},e_{i2},x_i^{(1)},x_i^{(2)},x_i^{(3)})
\end{equation}
is then permuted jointly across target locations for the misaligned member. Joint permutation preserves the complete empirical distribution and all within-row relationships exactly.

The mismatch permutation is selected using only pre-specified DGP diagnostics. For each target, the ten nearest geographic neighbors are identified and their Euclidean distances in the three-dimensional observed economic-proxy space are calculated. Edges below the 35th percentile of all local economic distances are labeled economically similar and those above the 65th percentile dissimilar. A location is mixed-support eligible when it has at least two similar and two dissimilar local neighbors. Candidate deterministic interleavings are scored by
\begin{equation}
\mathcal D
=\Delta\overline d_{econ}^{local}
+0.20\,p_{mixed}
+0.02\,\overline b_{mixed},
\end{equation}
To define these quantities, let $n_i^s$ and $n_i^d$ be the counts of similar and dissimilar neighbors among the ten local neighbors. Then $p_{mixed}=N^{-1}\sum_i I(n_i^s\ge2,n_i^d\ge2)$ and $\overline b_{mixed}=N^{-1}\sum_i\min(n_i^s,n_i^d)$. The first term is the candidate-minus-aligned change in the mean of all local economic distances. The thresholds and counts are recomputed for each candidate. No fitted Graph/GeoEco estimate, RMSE, treatment effect, or counterfactual truth enters this choice.

Both paired members use the same post-only selection set, determined from the mismatch-design feasible candidate set but a common geographic/station-proximity priority. The same complete, partial, and post-only IDs are therefore used in the two members. The paired audit additionally verifies equality of coordinates, retained PRE counts, treatment effects, target noise, donor trajectories, and the multiset of economic covariate rows.

Sampling in D9--D10 is conditional on the existence of a feasible set of 24 post-only targets, each retaining at least two economically similar and two dissimilar geographic neighbors with PRE observations. Any seed with no such set is excluded for both members and replaced by the next common seed, retaining the fixed candidate threshold and neighborhood constraints; the common indices are 1--102 excluding 50 and 70. Those two draws admit at most 23 post-only targets, and indices 101 and 102 replace them. Excluded and replacement seeds are recorded in the replication code.

To reduce direct linear prediction of untreated outcome levels from the observed economic proxies in the paired experiment, donor outcome vectors are partially orthogonalized to the target-level economic direction implied by the \emph{aligned} pair. The realized increase in PRE $R^2$ from adding economic proxies to a regression on year effects and coordinates is recorded as a diagnostic; it does not exclude replications. Let $\bm E_{res}\in\mathbb R^{N\times3}$ contain the residuals from separate regressions of the three aligned-target economic proxies on an intercept and the two coordinates, and let $\bm W_{ref}\in\mathbb R^{N\times J}$ be the true aligned-design weight matrix. Define
\begin{equation}
\bm A_E=\bm E_{res}^{\top}\bm W_{ref}.
\end{equation}
The matrix $\bm A_E\bm A_E^\top$ is $3\times3$. For each donor vector $\bm c_t=(C_{1t},\ldots,C_{Jt})^\top$, calculate the ridge-stabilized projection
\begin{equation}
\bm p_t=\bm A_E^{\top}(\bm A_E\bm A_E^{\top}+\kappa\bm I_3)^{-1}\bm A_E\bm c_t,
\end{equation}
and replace
\begin{equation}
\bm c_t\leftarrow \bm c_t-0.75\bm p_t,
\end{equation}
with $\kappa=10^{-10}\max\{\operatorname{tr}(\bm A_E\bm A_E^\top)/3,1\}$ for the three economic proxies. The same transformed donor outcomes are used for both paired members. Here $\bm\mu$ denotes the matrix of latent untreated target signals and $\bm C$ the donor-outcome matrix. This truth-side construction preserves the exact SCM identity $\bm\mu=\bm W^\star\bm C$ but reduces the possibility that a spatio-temporal mean model with economic covariates performs well simply because the DGP placed the target economic proxies directly in the outcome level.

\subsection{Details of the additional latent factors (D4)}
For donor coordinates $(x_j,y_j)$, the three standardized spatial bases are
\begin{align}
S_{1j}&=z[\sin(2\pi x_j/20000)\cos(\pi y_j/20000)],\\
S_{2j}&=z[\cos(\pi x_j/10000+y_j/14000)],\\
S_{3j}&=z[\sin\{\pi(x_j+y_j)/18000\}].
\end{align}
Let $b_j(c_1,c_2)=\exp[-\{(e_{j1}-c_1)^2+(e_{j2}-c_2)^2\}/(2\times0.70^2)]$. Form $b_j(-0.9,-0.5)-b_j(0.8,0.4)$, $b_j(0.9,-0.7)-b_j(-0.4,0.9)$, and $\sin(1.6e_{j1})\cos(1.4e_{j2})$. Regress each of these three nonlinear donor vectors separately on an intercept and the three observed donor economic proxies. Standardizing the residual vectors over donors gives $E_{1j},E_{2j},E_{3j}$. Then
\begin{equation}
\ell_{jr}^{LF}=z[\sqrt{0.35}S_{rj}+\sqrt{0.65}E_{rj}+v_{jr}],
\qquad v_{jr}\sim N(0,0.25^2).
\end{equation}
Write $A_t(\phi,\sigma)$ for an AR(1) sequence with coefficient $\phi$ and innovation SD $\sigma$, where $A_t(\phi,\sigma)$ starts with $A_1\sim N(0,\sigma^2)$ and then follows $A_t=\phi A_{t-1}+\eta_t$ with independent $\eta_t\sim N(0,\sigma^2)$. Independently generated temporal factors are
\begin{align}
F_{1t}&=z[A_t(0.88,0.018)],\\
F_{2t}&=z[A_t(0.25,0.018)+0.60\sin(r_t/1.9)],\\
F_{3t}&=z[0.55\cos(r_t/3.1)+0.45I(y_t\ge2001)+A_t(0.55,0.010)].
\end{align}
The donor increment is $0.060\sum_{r=1}^3\ell_{jr}^{LF}F_{rt}/\sqrt3$. The residualization deliberately weakens a linear economic outcome-mean channel while retaining nonlinear economic information. It is part of the positive-control construction and does not establish a general advantage over covariate-rich outcome models.

\clearpage
\section{Additional Monte Carlo results}
\label{app:additional-results}

\subsection{All-target results and the added covariate baseline}
The five primary designs use eight full-coverage methods on the same 100 datasets. Table~\ref{tab:all-primary} complements the main-paper post-only comparisons with all-target RMSE. Table~\ref{tab:covariate-groups} reports Cov-SCM separately for each observation group, including its Monte Carlo standard errors and signed ATT bias.

For a target group $\mathcal G$, the signal RMSE within one replication is
\[
\left[\frac{1}{|\mathcal G||\Tpost|}\sum_{i\in\mathcal G}\sum_{t\in\Tpost}
\{\widehat Y_{it}(0)-\mu_{it}\}^2\right]^{1/2}.
\]
Tables average these replication-specific values. An MCSE is the sample standard deviation across replications divided by the square root of the replication count. Paired intervals use $\overline d\pm t_{n-1,.975}s_d/\sqrt n$, where $d$ is the within-dataset method difference, $s_d$ its sample standard deviation, and $t_{n-1,.975}$ the Student-$t$ quantile; here $n=100$.
\input{tables/all_primary.tex}
\input{tables/covariate_groups.tex}

\subsection{Cov-SCM implementation and provenance}
The covariate fit uses only the two observed planar coordinate columns in meters and three PRE economic proxy columns. Balance concerns donor-weighted attribute averages, so individual donors need not share the target location. Let $\bm q_i\in\mathbb R^{p_c}$ be a standardized target attribute vector and $\bm Q_C\in\mathbb R^{J\times p_c}$ the donor attribute matrix, after dropping constant columns. Cov-SCM minimizes $p_c^{-1}\|\bm q_i-\bm Q_C^\top\bm w\|_2^2+10^{-10}\|\bm w\|_2^2$ over $\bm w\in\Delta_J$ independently for each target. Columns are standardized jointly across the 120 targets and 24 donors. The mean squared standardized imbalance has equal feature coefficients and a fixed ridge of $10^{-10}$. The simplex-constrained quadratic program is solved by a base-R active-set algorithm with an absolute duality-gap tolerance of $10^{-13}$.

Cov-SCM was evaluated on the same 500 primary datasets and target groups as the other methods. The accompanying replication materials document data alignment and implementation checks.

\subsection{Paired comparisons with Cov-SCM}
These comparisons use the same design, seed, and target group for both methods. Differences are Cov-SCM minus the reference method; negative values favor Cov-SCM. The full paired CSVs include all seven original full-coverage comparators and all four observation groups. The tables below focus on post-only targets.
\input{tables/covariate_pairs_cf_supp.tex}
\input{tables/covariate_pairs_att_supp.tex}

\subsection{Original ten-design counterfactual comparisons}
The seven original full-coverage methods were evaluated in all ten designs. Cov-SCM is evaluated only in the five primary designs, so the tables in this subsection retain the seven-method comparison throughout. All cells use 100 replications. Complete- and partial-PRE comparisons include ordinary SCM on the same observed-history group.
\input{tables/all_existing.tex}
\input{tables/postonly_existing.tex}
\input{tables/paired_existing.tex}
\input{tables/cf_complete_existing.tex}
\input{tables/cf_partial_existing.tex}
\input{tables/ordinary_covered_existing.tex}
\input{tables/paired_all_existing.tex}
\input{tables/beta_existing.tex}

\begin{figure}[!htbp]
\centering
\includegraphics[width=0.94\linewidth]{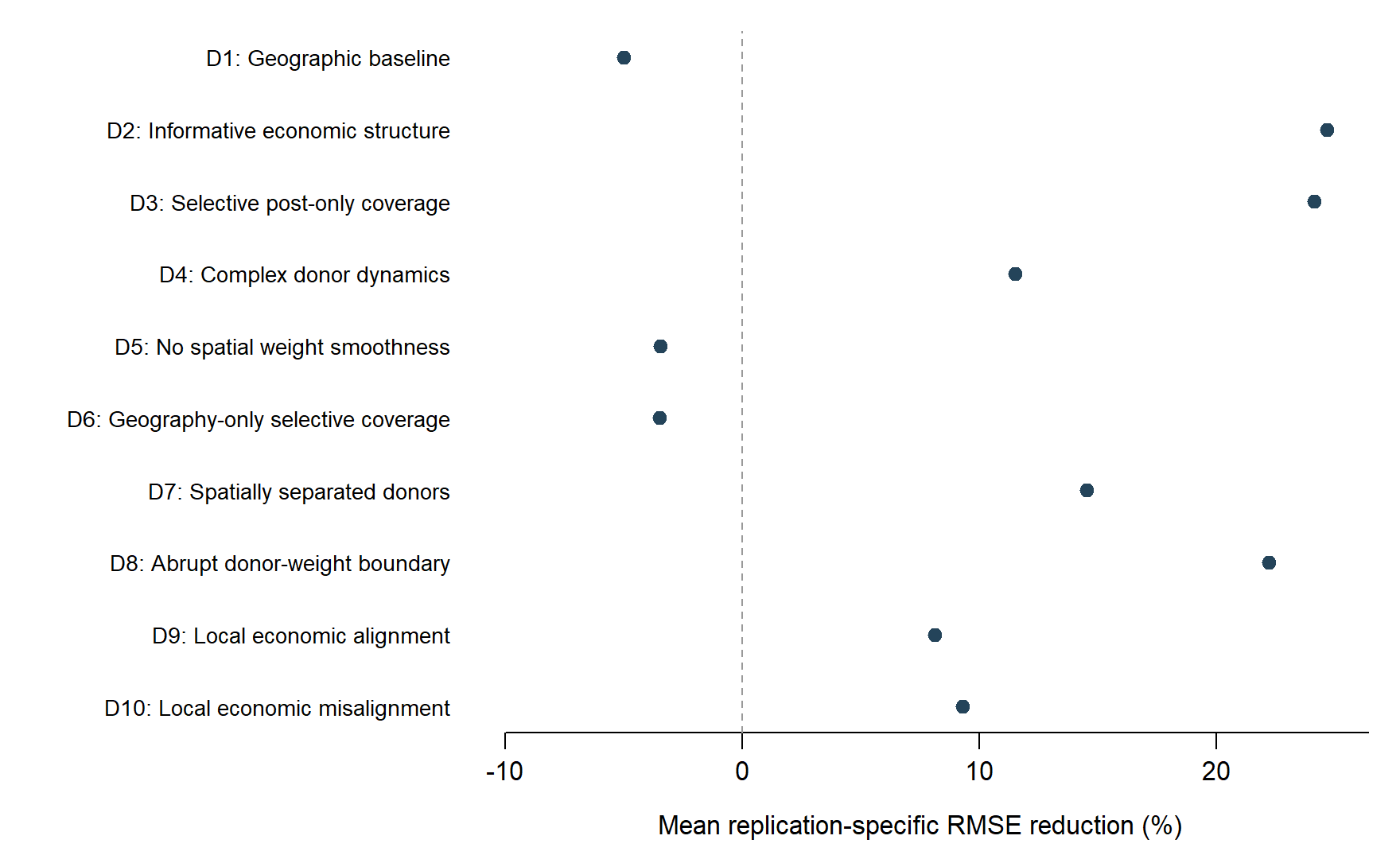}
\caption{Mean replication-specific percentage reduction in post-only RMSE of \GeoEco relative to \GraphSCM. Positive values favor \GeoEco. This average of within-replication percentages differs from the percentage change calculated from the two mean RMSEs. Paired intervals for absolute RMSE differences are reported in Table~\ref{tab:paired-existing}.}
\label{fig:improvement}
\end{figure}

\clearpage
\subsection{Paired local geography--economics alignment experiment}

The final local pair holds geography, membership in the post-only, partial-PRE, and complete-PRE groups, treatment effects, target noise, donor outcomes, and the multiset of economic covariate rows fixed. Only the spatial assignment of the target economic state changes between the two members. As described in Section~\ref{app:all-dgps}, the common observation pattern is selected to retain both economically similar and dissimilar observed neighbors in the misaligned design. Donor trajectories are modified in the same way in both members to weaken a linear economic outcome relationship. These choices provide economically informative observed neighbors for edge reweighting, while weakening the linear outcome-mean channel that can benefit STI-X. Section~7 gives the construction.

Economic misalignment makes prediction harder for both methods. Post-only RMSE rises from 0.0496 to 0.0602 for \GraphSCM and from 0.0454 to 0.0543 for \GeoEco. The paired contrast
\begin{align}
\Delta_{\mathrm{pair}}
={}&\left(RMSE_{\mathrm{GeoEco}}-RMSE_{\mathrm{Graph}}\right)_{\mathrm{misalign}}\notag\\
&-\left(RMSE_{\mathrm{GeoEco}}-RMSE_{\mathrm{Graph}}\right)_{\mathrm{align}}
\label{eq:paired-contrast}
\end{align}
is $-0.00169$ for post-only targets, with a Monte Carlo interval of $[-0.00346,0.00008]$. Local misalignment degrades both estimators, while \GeoEco remains more accurate in both post-only samples. The point estimate suggests a larger GeoEco advantage under misalignment, but the interval includes zero, so the experiment does not establish an increase in that advantage.

\input{tables/local_existing.tex}

\subsection{Donor-weight recovery and concentration}
\label{app:weight-recovery}

The generating weights $\bm w_i^\star$ are known in these simulations. For a target group $\mathcal G$, we calculate the mean of the target-specific weight RMSEs within each replication,
\[
 E_{\mathcal G}^{W}
 =\frac{1}{|\mathcal G|}\sum_{i\in\mathcal G}
 \left[\frac{1}{J}\sum_{j=1}^{J}
 (\widehat w_{ij}-w_{ij}^\star)^2\right]^{1/2}.
\]
This averages target-specific RMSEs, rather than pooling squared weight errors before taking a square root. We also calculate the effective number of donors, $D_i^{\mathrm{eff}}=1/\sum_{j=1}^{J}\widehat w_{ij}^{\,2}$; its range is 1 to $J$, and Uniform has $D_i^{\mathrm{eff}}=24$. This describes concentration across donors. Because locations and generating weights change across the original replications, their raw between-replication variation is not a fixed-design estimate of sampling variance.

Table~\ref{tab:weight-recovery-postonly} reports post-only results from all 1,000 saved datasets for Graph-SCM, GeoEco-SCM and Uniform. It includes paired GeoEco-minus-Graph and GeoEco-minus-Uniform weight-RMSE differences. Ordinary SCM has no post-only weights; its PRE-observed weight diagnostics, together with group-specific summaries for the other methods, are included in the replication materials.

\input{tables/weight_recovery_postonly.tex}

Economic adjustment reduces post-only weight RMSE relative to Graph-SCM in D2--D4, but GeoEco-SCM does not consistently improve recovery of the generating weights relative to Uniform. Correlated donor trajectories can produce similar counterfactual predictions under substantially different weights, so weight-error rankings need not follow the counterfactual-error rankings. These diagnostics therefore provide supplementary information about the fitted representations, while counterfactual prediction remains the primary evaluation target.

\clearpage
\section{ATT recovery and observation selection}
\label{app:att-results}

The percentage ATT transforms each year's mean log effect before evaluating errors. RMSE is first calculated over POST years within each replication; means and Monte Carlo standard errors are then taken across replications. The first two tables include Cov-SCM for the five primary designs. The remaining tables preserve the original seven-method comparison across all ten designs.

For group $\mathcal G$, write $\theta_{\mathcal G}(t)=100[\exp\{|\mathcal G|^{-1}\sum_{i\in\mathcal G}\tau_{it}\}-1]$, and define $\widehat\theta_{\mathcal G}(t)$ by replacing $\tau_{it}$ with $\widehat\tau_{it}$. With $e_{\mathcal Gt}=\widehat\theta_{\mathcal G}(t)-\theta_{\mathcal G}(t)$, a replication's ATT RMSE is $\{\sum_{t\in\Tpost}e_{\mathcal Gt}^2/|\Tpost|\}^{1/2}$ and its signed bias is $\sum_{t\in\Tpost}e_{\mathcal Gt}/|\Tpost|$, both in percentage points.
\input{tables/att_all_primary.tex}
\input{tables/att_mcse_primary.tex}
\input{tables/selection_existing.tex}
\input{tables/att_existing_main.tex}
\input{tables/att_post_existing.tex}
\input{tables/att_mcse_all_existing.tex}
\input{tables/att_mcse_post_existing.tex}

\clearpage
\bibliographystyle{plainnat}
\bibliography{references,references_additions}

%% file: tables/postonly_primary.tex
\begin{table}[!t]\centering
\caption{Post-only counterfactual signal RMSE: five primary designs.}\label{tab:postonly-primary}
\begin{threeparttable}\footnotesize\setlength{\tabcolsep}{3pt}
\begin{tabular}{@{}lrrrrrrrr@{}}\toprule
DGP & Graph & GeoEco & Cov-SCM & Uniform & STI & STI-X & STD & STD-X \\\midrule
D1 & \shortstack[r]{0.0372\\(0.0011)} & \shortstack[r]{0.0388\\(0.0012)} & \shortstack[r]{0.1071\\(0.0025)} & \shortstack[r]{0.1867\\(0.0036)} & \shortstack[r]{0.0338\\(0.0010)} & \shortstack[r]{0.0370\\(0.0009)} & \shortstack[r]{0.0359\\(0.0011)} & \shortstack[r]{0.0399\\(0.0011)} \\
D2 & \shortstack[r]{0.1395\\(0.0032)} & \shortstack[r]{0.1040\\(0.0029)} & \shortstack[r]{0.0977\\(0.0023)} & \shortstack[r]{0.1635\\(0.0035)} & \shortstack[r]{0.1442\\(0.0032)} & \shortstack[r]{0.0756\\(0.0019)} & \shortstack[r]{0.1453\\(0.0032)} & \shortstack[r]{0.0769\\(0.0018)} \\
D3 & \shortstack[r]{0.1810\\(0.0042)} & \shortstack[r]{0.1385\\(0.0047)} & \shortstack[r]{0.0926\\(0.0020)} & \shortstack[r]{0.1751\\(0.0047)} & \shortstack[r]{0.1699\\(0.0040)} & \shortstack[r]{0.0750\\(0.0018)} & \shortstack[r]{0.1648\\(0.0042)} & \shortstack[r]{0.0765\\(0.0018)} \\
D4 & \shortstack[r]{0.0684\\(0.0017)} & \shortstack[r]{0.0601\\(0.0015)} & \shortstack[r]{0.1056\\(0.0020)} & \shortstack[r]{0.0765\\(0.0020)} & \shortstack[r]{0.0719\\(0.0016)} & \shortstack[r]{0.0675\\(0.0016)} & \shortstack[r]{0.0738\\(0.0015)} & \shortstack[r]{0.0698\\(0.0015)} \\
D5 & \shortstack[r]{0.1040\\(0.0019)} & \shortstack[r]{0.1073\\(0.0020)} & \shortstack[r]{0.2034\\(0.0041)} & \shortstack[r]{0.0978\\(0.0018)} & \shortstack[r]{0.1104\\(0.0022)} & \shortstack[r]{0.1111\\(0.0021)} & \shortstack[r]{0.1172\\(0.0024)} & \shortstack[r]{0.1179\\(0.0023)} \\
\bottomrule\end{tabular}
\begin{tablenotes}[flushleft]\scriptsize\item Means of replication-specific latent-signal RMSEs on the log scale over 100 replications. Parentheses contain Monte Carlo standard errors; lower is better. All methods evaluate the same target group. Graph/GeoEco: graph estimators; Cov-SCM: coordinate and PRE economic-covariate balancing; Uniform: equal donor weights; STI: spatio-temporal PRE imputation followed by SCM; STD: direct spatio-temporal prediction; -X: economic covariates in the outcome mean model.
\end{tablenotes}\end{threeparttable}\end{table}

%% file: tables/paired_primary.tex
\begin{table}[!t]\centering
\caption{Paired post-only signal-RMSE differences for the graph estimators.}\label{tab:paired-primary}
\begin{threeparttable}\footnotesize\setlength{\tabcolsep}{3pt}
\begin{tabular}{@{}lrrrr@{}}\toprule
 & \multicolumn{2}{c}{GeoEco $-$ Graph} & \multicolumn{2}{c}{GeoEco $-$ STI-X} \\
DGP & Difference & 95\% MC interval & Difference & 95\% MC interval \\\midrule
D1 & 0.0016 & [0.0004, 0.0029] & 0.0018 & [-0.0012, 0.0048] \\
D2 & -0.0354 & [-0.0402, -0.0307] & 0.0285 & [0.0228, 0.0341] \\
D3 & -0.0425 & [-0.0482, -0.0368] & 0.0635 & [0.0545, 0.0726] \\
D4 & -0.0083 & [-0.0100, -0.0066] & -0.0074 & [-0.0096, -0.0053] \\
D5 & 0.0034 & [0.0016, 0.0051] & -0.0038 & [-0.0067, -0.0008] \\
\bottomrule\end{tabular}
\begin{tablenotes}[flushleft]\scriptsize\item Negative differences favor GeoEco. Intervals are unadjusted paired $t_{99}$ Monte Carlo intervals over the same 100 datasets.
\end{tablenotes}\end{threeparttable}\end{table}

%% file: tables/covariate_pairs_primary.tex
\begin{table}[!t]\centering
\caption{Cov-SCM minus GeoEco: paired post-only performance differences.}\label{tab:covariate-pairs-primary}
\begin{threeparttable}\footnotesize\setlength{\tabcolsep}{3pt}
\begin{tabular}{@{}lrrrr@{}}\toprule
 & \multicolumn{2}{c}{Signal RMSE (log)} & \multicolumn{2}{c}{ATT RMSE (pp)} \\
DGP & Difference & 95\% MC interval & Difference & 95\% MC interval \\\midrule
D1 & 0.06828 & [0.06334, 0.07323] & 1.786 & [1.356, 2.217] \\
D2 & -0.00631 & [-0.01265, 0.00002] & 0.906 & [0.364, 1.448] \\
D3 & -0.04585 & [-0.05632, -0.03537] & -5.239 & [-6.168, -4.309] \\
D4 & 0.04548 & [0.04165, 0.04931] & 2.579 & [2.105, 3.053] \\
D5 & 0.09605 & [0.08890, 0.10321] & 3.075 & [2.268, 3.882] \\
\bottomrule\end{tabular}
\begin{tablenotes}[flushleft]\scriptsize\item Differences are Cov-SCM minus GeoEco, paired by dataset. Negative values favor Cov-SCM. Intervals quantify Monte Carlo uncertainty in mean performance differences. Each design has 100 replications.
\end{tablenotes}\end{threeparttable}\end{table}

%% file: tables/att_primary.tex
\begin{table}[!t]\centering
\caption{ATT recovery: post-only targets, five primary designs.}\label{tab:att-primary}
\begin{minipage}{\linewidth}\centering\footnotesize\setlength{\tabcolsep}{3pt}
\begin{tabular}{@{}lrrrrrrrr@{}}\toprule
\multicolumn{9}{@{}l}{\textit{A. Graph, covariate, and uniform-weight methods}}\\
DGP & \multicolumn{2}{c}{Graph} & \multicolumn{2}{c}{GeoEco} & \multicolumn{2}{c}{Cov-SCM} & \multicolumn{2}{c}{Uniform} \\
 & RMSE & Bias & RMSE & Bias & RMSE & Bias & RMSE & Bias \\\midrule
D1 & 0.926 & 0.292 & 0.971 & 0.285 & 2.757 & -0.437 & 4.002 & 0.302 \\
D2 & 2.769 & -0.109 & 2.092 & -0.012 & 2.998 & 0.030 & 3.877 & 0.364 \\
D3 & 11.873 & -11.838 & 8.199 & -8.164 & 2.961 & 0.306 & 9.910 & -9.856 \\
D4 & 1.508 & -0.174 & 1.337 & -0.117 & 3.917 & -2.891 & 1.845 & -0.917 \\
D5 & 2.047 & -0.147 & 2.145 & -0.268 & 5.220 & 0.473 & 1.629 & 0.087 \\
\bottomrule\end{tabular}\par\medskip
\begin{tabular}{@{}lrrrrrrrr@{}}\toprule
\multicolumn{9}{@{}l}{\textit{B. Spatio-temporal methods}}\\
DGP & \multicolumn{2}{c}{STI} & \multicolumn{2}{c}{STI-X} & \multicolumn{2}{c}{STD} & \multicolumn{2}{c}{STD-X} \\
 & RMSE & Bias & RMSE & Bias & RMSE & Bias & RMSE & Bias \\\midrule
D1 & 0.730 & 0.012 & 0.805 & 0.058 & 0.888 & -0.100 & 0.943 & -0.053 \\
D2 & 2.763 & -0.051 & 1.519 & 0.163 & 3.220 & 0.277 & 1.771 & 0.420 \\
D3 & 10.186 & -10.156 & 2.145 & -1.430 & 9.136 & -9.005 & 2.292 & -0.930 \\
D4 & 1.513 & -0.281 & 1.415 & -0.182 & 1.771 & -0.679 & 1.772 & -0.605 \\
D5 & 2.186 & 0.231 & 2.244 & 0.174 & 2.364 & 0.325 & 2.465 & 0.350 \\
\bottomrule\end{tabular}\par\medskip
\par\medskip{\raggedright\scriptsize\noindent Percentage points, averaged over 100 replication-specific annual errors. ATT is $100\{\exp(\overline\tau_t)-1\}$, the percentage transformation of the mean log effect. Bias is estimated minus true ATT. All methods evaluate the same indicated population. Monte Carlo standard errors are reported in the Online Appendix. Graph/GeoEco: graph estimators; Cov-SCM: coordinate and PRE economic-covariate balancing; Uniform: equal donor weights; STI: spatio-temporal PRE imputation followed by SCM; STD: direct spatio-temporal prediction; -X: economic covariates in the outcome mean model.
\par}\end{minipage}\end{table}

%% file: tables/all_primary.tex
\begin{table}[!htbp]\centering
\caption{All-target counterfactual signal RMSE: five primary designs.}\label{tab:all-primary}
\begin{threeparttable}\footnotesize\setlength{\tabcolsep}{3pt}
\begin{tabular}{@{}lrrrrrrrr@{}}\toprule
DGP & Graph & GeoEco & Cov-SCM & Uniform & STI & STI-X & STD & STD-X \\\midrule
D1 & \shortstack[r]{0.0236\\(0.0005)} & \shortstack[r]{0.0241\\(0.0005)} & \shortstack[r]{0.1073\\(0.0018)} & \shortstack[r]{0.1867\\(0.0031)} & \shortstack[r]{0.0214\\(0.0005)} & \shortstack[r]{0.0224\\(0.0005)} & \shortstack[r]{0.0239\\(0.0006)} & \shortstack[r]{0.0253\\(0.0006)} \\
D2 & \shortstack[r]{0.0663\\(0.0014)} & \shortstack[r]{0.0518\\(0.0012)} & \shortstack[r]{0.0994\\(0.0019)} & \shortstack[r]{0.1654\\(0.0028)} & \shortstack[r]{0.0669\\(0.0014)} & \shortstack[r]{0.0377\\(0.0008)} & \shortstack[r]{0.0736\\(0.0015)} & \shortstack[r]{0.0454\\(0.0009)} \\
D3 & \shortstack[r]{0.0839\\(0.0018)} & \shortstack[r]{0.0657\\(0.0020)} & \shortstack[r]{0.1018\\(0.0017)} & \shortstack[r]{0.1683\\(0.0025)} & \shortstack[r]{0.0780\\(0.0017)} & \shortstack[r]{0.0376\\(0.0007)} & \shortstack[r]{0.0823\\(0.0017)} & \shortstack[r]{0.0460\\(0.0008)} \\
D4 & \shortstack[r]{0.0406\\(0.0007)} & \shortstack[r]{0.0373\\(0.0006)} & \shortstack[r]{0.1070\\(0.0019)} & \shortstack[r]{0.0757\\(0.0016)} & \shortstack[r]{0.0423\\(0.0007)} & \shortstack[r]{0.0408\\(0.0007)} & \shortstack[r]{0.0553\\(0.0008)} & \shortstack[r]{0.0536\\(0.0008)} \\
D5 & \shortstack[r]{0.0488\\(0.0008)} & \shortstack[r]{0.0501\\(0.0008)} & \shortstack[r]{0.2061\\(0.0030)} & \shortstack[r]{0.0999\\(0.0014)} & \shortstack[r]{0.0514\\(0.0009)} & \shortstack[r]{0.0517\\(0.0009)} & \shortstack[r]{0.0583\\(0.0010)} & \shortstack[r]{0.0585\\(0.0010)} \\
\bottomrule\end{tabular}
\begin{tablenotes}[flushleft]\scriptsize\item Means of replication-specific latent-signal RMSEs on the log scale over 100 replications. Parentheses contain Monte Carlo standard errors; lower is better. All methods evaluate the same target group. Graph/GeoEco: graph estimators; Cov-SCM: coordinate and PRE economic-covariate balancing; Uniform: equal donor weights; STI: spatio-temporal PRE imputation followed by SCM; STD: direct spatio-temporal prediction; -X: economic covariates in the outcome mean model.
\end{tablenotes}\end{threeparttable}\end{table}

%% file: tables/covariate_groups.tex
\begin{table}[!htbp]\centering
\caption{Cov-SCM performance by observation group in the five primary designs.}\label{tab:covariate-groups}
\begin{threeparttable}\footnotesize\setlength{\tabcolsep}{3pt}
\begin{tabular}{@{}llrrr@{}}\toprule
DGP & Group & Signal RMSE (log) & ATT RMSE (pp) & ATT bias (pp) \\\midrule
D1 & complete PRE & \shortstack[r]{0.1071\\(0.0019)} & \shortstack[r]{2.657\\(0.173)} & \shortstack[r]{-0.483\\(0.312)} \\
D1 & partial PRE & \shortstack[r]{0.1057\\(0.0021)} & \shortstack[r]{2.566\\(0.175)} & \shortstack[r]{-0.474\\(0.305)} \\
D1 & post-only & \shortstack[r]{0.1071\\(0.0025)} & \shortstack[r]{2.757\\(0.204)} & \shortstack[r]{-0.437\\(0.338)} \\
D1 & all targets & \shortstack[r]{0.1073\\(0.0018)} & \shortstack[r]{2.374\\(0.152)} & \shortstack[r]{-0.482\\(0.276)} \\
D2 & complete PRE & \shortstack[r]{0.0997\\(0.0022)} & \shortstack[r]{2.797\\(0.227)} & \shortstack[r]{0.282\\(0.358)} \\
D2 & partial PRE & \shortstack[r]{0.0982\\(0.0021)} & \shortstack[r]{2.970\\(0.216)} & \shortstack[r]{0.200\\(0.364)} \\
D2 & post-only & \shortstack[r]{0.0977\\(0.0023)} & \shortstack[r]{2.998\\(0.226)} & \shortstack[r]{0.030\\(0.373)} \\
D2 & all targets & \shortstack[r]{0.0994\\(0.0019)} & \shortstack[r]{2.710\\(0.208)} & \shortstack[r]{0.199\\(0.339)} \\
D3 & complete PRE & \shortstack[r]{0.1040\\(0.0019)} & \shortstack[r]{3.062\\(0.202)} & \shortstack[r]{-0.431\\(0.362)} \\
D3 & partial PRE & \shortstack[r]{0.1023\\(0.0023)} & \shortstack[r]{3.131\\(0.224)} & \shortstack[r]{-0.282\\(0.381)} \\
D3 & post-only & \shortstack[r]{0.0926\\(0.0020)} & \shortstack[r]{2.961\\(0.212)} & \shortstack[r]{0.306\\(0.360)} \\
D3 & all targets & \shortstack[r]{0.1018\\(0.0017)} & \shortstack[r]{2.759\\(0.198)} & \shortstack[r]{-0.252\\(0.337)} \\
D4 & complete PRE & \shortstack[r]{0.1082\\(0.0022)} & \shortstack[r]{3.976\\(0.227)} & \shortstack[r]{-3.160\\(0.308)} \\
D4 & partial PRE & \shortstack[r]{0.1047\\(0.0020)} & \shortstack[r]{4.046\\(0.205)} & \shortstack[r]{-3.170\\(0.300)} \\
D4 & post-only & \shortstack[r]{0.1056\\(0.0020)} & \shortstack[r]{3.917\\(0.215)} & \shortstack[r]{-2.891\\(0.317)} \\
D4 & all targets & \shortstack[r]{0.1070\\(0.0019)} & \shortstack[r]{3.873\\(0.201)} & \shortstack[r]{-3.118\\(0.283)} \\
D5 & complete PRE & \shortstack[r]{0.2043\\(0.0033)} & \shortstack[r]{3.572\\(0.246)} & \shortstack[r]{0.033\\(0.432)} \\
D5 & partial PRE & \shortstack[r]{0.2082\\(0.0033)} & \shortstack[r]{4.503\\(0.340)} & \shortstack[r]{-0.500\\(0.560)} \\
D5 & post-only & \shortstack[r]{0.2034\\(0.0041)} & \shortstack[r]{5.220\\(0.405)} & \shortstack[r]{0.473\\(0.656)} \\
D5 & all targets & \shortstack[r]{0.2061\\(0.0030)} & \shortstack[r]{3.513\\(0.234)} & \shortstack[r]{-0.086\\(0.420)} \\
\bottomrule\end{tabular}
\begin{tablenotes}[flushleft]\scriptsize\item Means over 100 replications; parentheses contain Monte Carlo standard errors. The covariate-fitting rule is identical in all groups and uses no target outcomes. Population sizes are 60 complete-PRE, 36 partial-PRE, 24 post-only, and 120 all-target sites per replication.
\end{tablenotes}\end{threeparttable}\end{table}

%% file: tables/covariate_pairs_cf_supp.tex
\begin{table}[!htbp]\centering
\caption{Cov-SCM paired comparisons: post-only signal RMSE.}\label{tab:covariate-pairs-cf}
\begin{threeparttable}\footnotesize\setlength{\tabcolsep}{3pt}
\begin{tabular}{@{}llrrr@{}}\toprule
DGP & Reference & Difference (log) & MCSE & 95\% MC interval \\\midrule
D1 & Graph & 0.06992 & 0.00228 & [0.06540, 0.07444] \\
D1 & GeoEco & 0.06828 & 0.00249 & [0.06334, 0.07323] \\
D1 & Uniform & -0.07954 & 0.00360 & [-0.08668, -0.07241] \\
D1 & STI-X & 0.07007 & 0.00249 & [0.06513, 0.07500] \\
D2 & Graph & -0.04176 & 0.00333 & [-0.04836, -0.03516] \\
D2 & GeoEco & -0.00631 & 0.00319 & [-0.01265, 0.00002] \\
D2 & Uniform & -0.06578 & 0.00358 & [-0.07287, -0.05868] \\
D2 & STI-X & 0.02216 & 0.00219 & [0.01781, 0.02650] \\
D3 & Graph & -0.08833 & 0.00473 & [-0.09772, -0.07895] \\
D3 & GeoEco & -0.04585 & 0.00528 & [-0.05632, -0.03537] \\
D3 & Uniform & -0.08251 & 0.00507 & [-0.09257, -0.07245] \\
D3 & STI-X & 0.01767 & 0.00221 & [0.01330, 0.02205] \\
D4 & Graph & 0.03718 & 0.00200 & [0.03322, 0.04115] \\
D4 & GeoEco & 0.04548 & 0.00193 & [0.04165, 0.04931] \\
D4 & Uniform & 0.02905 & 0.00213 & [0.02482, 0.03327] \\
D4 & STI-X & 0.03804 & 0.00190 & [0.03427, 0.04182] \\
D5 & Graph & 0.09941 & 0.00335 & [0.09275, 0.10606] \\
D5 & GeoEco & 0.09605 & 0.00361 & [0.08890, 0.10321] \\
D5 & Uniform & 0.10554 & 0.00341 & [0.09878, 0.11230] \\
D5 & STI-X & 0.09226 & 0.00296 & [0.08638, 0.09814] \\
\bottomrule\end{tabular}
\begin{tablenotes}[flushleft]\scriptsize\item Cov-SCM minus reference; negative differences favor Cov-SCM. Paired $t_{99}$ Monte Carlo intervals use 100 replications.
\end{tablenotes}\end{threeparttable}\end{table}

%% file: tables/covariate_pairs_att_supp.tex
\begin{table}[!htbp]\centering
\caption{Cov-SCM paired comparisons: post-only ATT RMSE.}\label{tab:covariate-pairs-att}
\begin{threeparttable}\footnotesize\setlength{\tabcolsep}{3pt}
\begin{tabular}{@{}llrrr@{}}\toprule
DGP & Reference & Difference (pp) & MCSE & 95\% MC interval \\\midrule
D1 & Graph & 1.831 & 0.213 & [1.408, 2.254] \\
D1 & GeoEco & 1.786 & 0.217 & [1.356, 2.217] \\
D1 & Uniform & -1.246 & 0.325 & [-1.890, -0.601] \\
D1 & STI-X & 1.952 & 0.205 & [1.545, 2.359] \\
D2 & Graph & 0.229 & 0.277 & [-0.319, 0.778] \\
D2 & GeoEco & 0.906 & 0.273 & [0.364, 1.448] \\
D2 & Uniform & -0.878 & 0.349 & [-1.572, -0.185] \\
D2 & STI-X & 1.480 & 0.245 & [0.993, 1.967] \\
D3 & Graph & -8.912 & 0.443 & [-9.791, -8.034] \\
D3 & GeoEco & -5.239 & 0.468 & [-6.168, -4.309] \\
D3 & Uniform & -6.949 & 0.525 & [-7.991, -5.907] \\
D3 & STI-X & 0.815 & 0.281 & [0.258, 1.373] \\
D4 & Graph & 2.409 & 0.227 & [1.958, 2.860] \\
D4 & GeoEco & 2.579 & 0.239 & [2.105, 3.053] \\
D4 & Uniform & 2.071 & 0.227 & [1.620, 2.523] \\
D4 & STI-X & 2.501 & 0.227 & [2.050, 2.953] \\
D5 & Graph & 3.173 & 0.402 & [2.375, 3.970] \\
D5 & GeoEco & 3.075 & 0.407 & [2.268, 3.882] \\
D5 & Uniform & 3.591 & 0.403 & [2.791, 4.391] \\
D5 & STI-X & 2.976 & 0.374 & [2.235, 3.717] \\
\bottomrule\end{tabular}
\begin{tablenotes}[flushleft]\scriptsize\item Cov-SCM minus reference; negative differences favor Cov-SCM. Paired $t_{99}$ Monte Carlo intervals use 100 replications.
\end{tablenotes}\end{threeparttable}\end{table}

%% file: tables/all_existing.tex
\begin{table}[!t]\centering
\caption{All-target counterfactual signal RMSE and Monte Carlo standard errors.}\label{tab:s_all}
\begin{threeparttable}\footnotesize\setlength{\tabcolsep}{3pt}
\begin{tabular}{@{}lrrrrrrr@{}}\toprule
DGP & Graph & GeoEco & Uniform & STI & STI-X & STD & STD-X \\\midrule
D1 & \shortstack[r]{0.0236\\(0.0005)} & \shortstack[r]{0.0241\\(0.0005)} & \shortstack[r]{0.1867\\(0.0031)} & \shortstack[r]{0.0214\\(0.0005)} & \shortstack[r]{0.0224\\(0.0005)} & \shortstack[r]{0.0239\\(0.0006)} & \shortstack[r]{0.0253\\(0.0006)} \\
D2 & \shortstack[r]{0.0663\\(0.0014)} & \shortstack[r]{0.0518\\(0.0012)} & \shortstack[r]{0.1654\\(0.0028)} & \shortstack[r]{0.0669\\(0.0014)} & \shortstack[r]{0.0377\\(0.0008)} & \shortstack[r]{0.0736\\(0.0015)} & \shortstack[r]{0.0454\\(0.0009)} \\
D3 & \shortstack[r]{0.0839\\(0.0018)} & \shortstack[r]{0.0657\\(0.0020)} & \shortstack[r]{0.1683\\(0.0025)} & \shortstack[r]{0.0780\\(0.0017)} & \shortstack[r]{0.0376\\(0.0007)} & \shortstack[r]{0.0823\\(0.0017)} & \shortstack[r]{0.0460\\(0.0008)} \\
D4 & \shortstack[r]{0.0406\\(0.0007)} & \shortstack[r]{0.0373\\(0.0006)} & \shortstack[r]{0.0757\\(0.0016)} & \shortstack[r]{0.0423\\(0.0007)} & \shortstack[r]{0.0408\\(0.0007)} & \shortstack[r]{0.0553\\(0.0008)} & \shortstack[r]{0.0536\\(0.0008)} \\
D5 & \shortstack[r]{0.0488\\(0.0008)} & \shortstack[r]{0.0501\\(0.0008)} & \shortstack[r]{0.0999\\(0.0014)} & \shortstack[r]{0.0514\\(0.0009)} & \shortstack[r]{0.0517\\(0.0009)} & \shortstack[r]{0.0583\\(0.0010)} & \shortstack[r]{0.0585\\(0.0010)} \\
D6 & \shortstack[r]{0.0250\\(0.0006)} & \shortstack[r]{0.0255\\(0.0006)} & \shortstack[r]{0.1886\\(0.0029)} & \shortstack[r]{0.0222\\(0.0005)} & \shortstack[r]{0.0231\\(0.0005)} & \shortstack[r]{0.0244\\(0.0005)} & \shortstack[r]{0.0257\\(0.0005)} \\
D7 & \shortstack[r]{0.0380\\(0.0009)} & \shortstack[r]{0.0336\\(0.0009)} & \shortstack[r]{0.0814\\(0.0024)} & \shortstack[r]{0.0386\\(0.0009)} & \shortstack[r]{0.0345\\(0.0008)} & \shortstack[r]{0.0468\\(0.0014)} & \shortstack[r]{0.0430\\(0.0012)} \\
D8 & \shortstack[r]{0.0636\\(0.0013)} & \shortstack[r]{0.0512\\(0.0012)} & \shortstack[r]{0.1778\\(0.0030)} & \shortstack[r]{0.0646\\(0.0013)} & \shortstack[r]{0.0415\\(0.0008)} & \shortstack[r]{0.0709\\(0.0015)} & \shortstack[r]{0.0494\\(0.0009)} \\
D9 & \shortstack[r]{0.0308\\(0.0006)} & \shortstack[r]{0.0292\\(0.0006)} & \shortstack[r]{0.0598\\(0.0015)} & \shortstack[r]{0.0314\\(0.0006)} & \shortstack[r]{0.0320\\(0.0006)} & \shortstack[r]{0.0403\\(0.0008)} & \shortstack[r]{0.0409\\(0.0008)} \\
D10 & \shortstack[r]{0.0344\\(0.0007)} & \shortstack[r]{0.0321\\(0.0007)} & \shortstack[r]{0.0600\\(0.0015)} & \shortstack[r]{0.0337\\(0.0008)} & \shortstack[r]{0.0338\\(0.0008)} & \shortstack[r]{0.0419\\(0.0009)} & \shortstack[r]{0.0421\\(0.0009)} \\
\bottomrule\end{tabular}
\begin{tablenotes}[flushleft]\scriptsize\item Means of replication-specific latent-signal RMSEs across 100 replications. Parentheses contain Monte Carlo standard errors. Lower is better. Graph/GeoEco: graph estimators; Uniform: equal donor weights; STI: spatio-temporal PRE imputation followed by SCM; STD: direct spatio-temporal prediction; -X: PRE economic covariates.
\end{tablenotes}\end{threeparttable}\end{table}

%% file: tables/postonly_existing.tex
\begin{table}[!t]\centering
\caption{Post-only counterfactual signal RMSE and Monte Carlo standard errors.}\label{tab:postonly}
\begin{threeparttable}\footnotesize\setlength{\tabcolsep}{3pt}
\begin{tabular}{@{}lrrrrrrr@{}}\toprule
DGP & Graph & GeoEco & Uniform & STI & STI-X & STD & STD-X \\\midrule
D1 & \shortstack[r]{0.0372\\(0.0011)} & \shortstack[r]{0.0388\\(0.0012)} & \shortstack[r]{0.1867\\(0.0036)} & \shortstack[r]{0.0338\\(0.0010)} & \shortstack[r]{0.0370\\(0.0009)} & \shortstack[r]{0.0359\\(0.0011)} & \shortstack[r]{0.0399\\(0.0011)} \\
D2 & \shortstack[r]{0.1395\\(0.0032)} & \shortstack[r]{0.1040\\(0.0029)} & \shortstack[r]{0.1635\\(0.0035)} & \shortstack[r]{0.1442\\(0.0032)} & \shortstack[r]{0.0756\\(0.0019)} & \shortstack[r]{0.1453\\(0.0032)} & \shortstack[r]{0.0769\\(0.0018)} \\
D3 & \shortstack[r]{0.1810\\(0.0042)} & \shortstack[r]{0.1385\\(0.0047)} & \shortstack[r]{0.1751\\(0.0047)} & \shortstack[r]{0.1699\\(0.0040)} & \shortstack[r]{0.0750\\(0.0018)} & \shortstack[r]{0.1648\\(0.0042)} & \shortstack[r]{0.0765\\(0.0018)} \\
D4 & \shortstack[r]{0.0684\\(0.0017)} & \shortstack[r]{0.0601\\(0.0015)} & \shortstack[r]{0.0765\\(0.0020)} & \shortstack[r]{0.0719\\(0.0016)} & \shortstack[r]{0.0675\\(0.0016)} & \shortstack[r]{0.0738\\(0.0015)} & \shortstack[r]{0.0698\\(0.0015)} \\
D5 & \shortstack[r]{0.1040\\(0.0019)} & \shortstack[r]{0.1073\\(0.0020)} & \shortstack[r]{0.0978\\(0.0018)} & \shortstack[r]{0.1104\\(0.0022)} & \shortstack[r]{0.1111\\(0.0021)} & \shortstack[r]{0.1172\\(0.0024)} & \shortstack[r]{0.1179\\(0.0023)} \\
D6 & \shortstack[r]{0.0398\\(0.0012)} & \shortstack[r]{0.0414\\(0.0014)} & \shortstack[r]{0.1459\\(0.0034)} & \shortstack[r]{0.0361\\(0.0012)} & \shortstack[r]{0.0392\\(0.0011)} & \shortstack[r]{0.0387\\(0.0013)} & \shortstack[r]{0.0423\\(0.0012)} \\
D7 & \shortstack[r]{0.0737\\(0.0022)} & \shortstack[r]{0.0629\\(0.0021)} & \shortstack[r]{0.0813\\(0.0025)} & \shortstack[r]{0.0766\\(0.0022)} & \shortstack[r]{0.0650\\(0.0018)} & \shortstack[r]{0.0813\\(0.0024)} & \shortstack[r]{0.0703\\(0.0020)} \\
D8 & \shortstack[r]{0.1334\\(0.0029)} & \shortstack[r]{0.1036\\(0.0029)} & \shortstack[r]{0.1768\\(0.0035)} & \shortstack[r]{0.1393\\(0.0031)} & \shortstack[r]{0.0839\\(0.0019)} & \shortstack[r]{0.1410\\(0.0032)} & \shortstack[r]{0.0874\\(0.0019)} \\
D9 & \shortstack[r]{0.0496\\(0.0013)} & \shortstack[r]{0.0454\\(0.0013)} & \shortstack[r]{0.0585\\(0.0017)} & \shortstack[r]{0.0548\\(0.0014)} & \shortstack[r]{0.0563\\(0.0015)} & \shortstack[r]{0.0569\\(0.0015)} & \shortstack[r]{0.0588\\(0.0015)} \\
D10 & \shortstack[r]{0.0602\\(0.0017)} & \shortstack[r]{0.0543\\(0.0016)} & \shortstack[r]{0.0615\\(0.0018)} & \shortstack[r]{0.0610\\(0.0018)} & \shortstack[r]{0.0613\\(0.0018)} & \shortstack[r]{0.0621\\(0.0018)} & \shortstack[r]{0.0628\\(0.0018)} \\
\bottomrule\end{tabular}
\begin{tablenotes}[flushleft]\scriptsize\item Means of replication-specific latent-signal RMSEs across 100 replications. Parentheses contain Monte Carlo standard errors. Lower is better. Graph/GeoEco: graph estimators; Uniform: equal donor weights; STI: spatio-temporal PRE imputation followed by SCM; STD: direct spatio-temporal prediction; -X: PRE economic covariates.
\end{tablenotes}\end{threeparttable}\end{table}

%% file: tables/paired_existing.tex
\begin{table}[!t]\centering
\caption{Paired differences in mean signal RMSE: post only.}\label{tab:paired-existing}
\begin{threeparttable}\footnotesize\setlength{\tabcolsep}{5pt}
\begin{tabular}{@{}lrrrr@{}}\toprule
 & \multicolumn{2}{c}{GeoEco $-$ Graph} & \multicolumn{2}{c}{GeoEco $-$ STI-X} \\
DGP & Mean difference & 95\% MC interval & Mean difference & 95\% MC interval \\\midrule
D1 & 0.0016 & [0.0004, 0.0029] & 0.0018 & [-0.0012, 0.0048] \\
D2 & -0.0354 & [-0.0402, -0.0307] & 0.0285 & [0.0228, 0.0341] \\
D3 & -0.0425 & [-0.0482, -0.0368] & 0.0635 & [0.0545, 0.0726] \\
D4 & -0.0083 & [-0.0100, -0.0066] & -0.0074 & [-0.0096, -0.0053] \\
D5 & 0.0034 & [0.0016, 0.0051] & -0.0038 & [-0.0067, -0.0008] \\
D6 & 0.0016 & [0.0005, 0.0027] & 0.0022 & [-0.0012, 0.0055] \\
D7 & -0.0108 & [-0.0127, -0.0088] & -0.0021 & [-0.0048, 0.0007] \\
D8 & -0.0298 & [-0.0339, -0.0258] & 0.0197 & [0.0144, 0.0250] \\
D9 & -0.0043 & [-0.0055, -0.0030] & -0.0110 & [-0.0130, -0.0090] \\
D10 & -0.0059 & [-0.0074, -0.0045] & -0.0070 & [-0.0088, -0.0051] \\
\bottomrule\end{tabular}\begin{tablenotes}[flushleft]\scriptsize\item Negative differences favor GeoEco. Intervals are unadjusted paired $t_{99}$ Monte Carlo intervals over 100 datasets. They quantify uncertainty in average simulation performance and are not treatment-effect confidence intervals.\end{tablenotes}\end{threeparttable}\end{table}

%% file: tables/cf_complete_existing.tex
\begin{table}[!htbp]\centering
\caption{Complete-PRE counterfactual signal RMSE and Monte Carlo standard errors.}\label{tab:cf-complete-existing}
\begin{threeparttable}\footnotesize\setlength{\tabcolsep}{3pt}
\begin{tabular}{@{}lrrrrrrrr@{}}\toprule
DGP & Graph & GeoEco & Uniform & STI & STI-X & STD & STD-X & Ordinary SCM \\\midrule
D1 & \shortstack[r]{0.0176\\(0.0005)} & \shortstack[r]{0.0175\\(0.0005)} & \shortstack[r]{0.1874\\(0.0034)} & \shortstack[r]{0.0160\\(0.0005)} & \shortstack[r]{0.0159\\(0.0004)} & \shortstack[r]{0.0193\\(0.0005)} & \shortstack[r]{0.0196\\(0.0005)} & \shortstack[r]{0.0164\\(0.0005)} \\
D2 & \shortstack[r]{0.0222\\(0.0009)} & \shortstack[r]{0.0225\\(0.0009)} & \shortstack[r]{0.1663\\(0.0031)} & \shortstack[r]{0.0175\\(0.0005)} & \shortstack[r]{0.0164\\(0.0005)} & \shortstack[r]{0.0367\\(0.0011)} & \shortstack[r]{0.0318\\(0.0007)} & \shortstack[r]{0.0170\\(0.0005)} \\
D3 & \shortstack[r]{0.0217\\(0.0009)} & \shortstack[r]{0.0212\\(0.0009)} & \shortstack[r]{0.1646\\(0.0027)} & \shortstack[r]{0.0172\\(0.0005)} & \shortstack[r]{0.0168\\(0.0005)} & \shortstack[r]{0.0377\\(0.0011)} & \shortstack[r]{0.0331\\(0.0007)} & \shortstack[r]{0.0173\\(0.0005)} \\
D4 & \shortstack[r]{0.0279\\(0.0006)} & \shortstack[r]{0.0269\\(0.0006)} & \shortstack[r]{0.0754\\(0.0016)} & \shortstack[r]{0.0286\\(0.0007)} & \shortstack[r]{0.0285\\(0.0006)} & \shortstack[r]{0.0492\\(0.0009)} & \shortstack[r]{0.0483\\(0.0008)} & \shortstack[r]{0.0287\\(0.0006)} \\
D5 & \shortstack[r]{0.0140\\(0.0004)} & \shortstack[r]{0.0139\\(0.0003)} & \shortstack[r]{0.1006\\(0.0017)} & \shortstack[r]{0.0140\\(0.0003)} & \shortstack[r]{0.0139\\(0.0003)} & \shortstack[r]{0.0269\\(0.0006)} & \shortstack[r]{0.0268\\(0.0006)} & \shortstack[r]{0.0143\\(0.0003)} \\
D6 & \shortstack[r]{0.0184\\(0.0006)} & \shortstack[r]{0.0183\\(0.0006)} & \shortstack[r]{0.1971\\(0.0032)} & \shortstack[r]{0.0161\\(0.0004)} & \shortstack[r]{0.0157\\(0.0004)} & \shortstack[r]{0.0183\\(0.0003)} & \shortstack[r]{0.0185\\(0.0003)} & \shortstack[r]{0.0163\\(0.0004)} \\
D7 & \shortstack[r]{0.0191\\(0.0007)} & \shortstack[r]{0.0185\\(0.0007)} & \shortstack[r]{0.0816\\(0.0024)} & \shortstack[r]{0.0182\\(0.0005)} & \shortstack[r]{0.0189\\(0.0006)} & \shortstack[r]{0.0315\\(0.0013)} & \shortstack[r]{0.0314\\(0.0012)} & \shortstack[r]{0.0182\\(0.0005)} \\
D8 & \shortstack[r]{0.0226\\(0.0008)} & \shortstack[r]{0.0219\\(0.0008)} & \shortstack[r]{0.1770\\(0.0034)} & \shortstack[r]{0.0170\\(0.0005)} & \shortstack[r]{0.0179\\(0.0006)} & \shortstack[r]{0.0345\\(0.0010)} & \shortstack[r]{0.0325\\(0.0008)} & \shortstack[r]{0.0175\\(0.0005)} \\
D9 & \shortstack[r]{0.0226\\(0.0006)} & \shortstack[r]{0.0221\\(0.0005)} & \shortstack[r]{0.0597\\(0.0016)} & \shortstack[r]{0.0209\\(0.0005)} & \shortstack[r]{0.0209\\(0.0005)} & \shortstack[r]{0.0346\\(0.0008)} & \shortstack[r]{0.0345\\(0.0008)} & \shortstack[r]{0.0209\\(0.0005)} \\
D10 & \shortstack[r]{0.0222\\(0.0005)} & \shortstack[r]{0.0220\\(0.0005)} & \shortstack[r]{0.0590\\(0.0015)} & \shortstack[r]{0.0205\\(0.0005)} & \shortstack[r]{0.0206\\(0.0005)} & \shortstack[r]{0.0346\\(0.0009)} & \shortstack[r]{0.0346\\(0.0009)} & \shortstack[r]{0.0204\\(0.0005)} \\
\bottomrule\end{tabular}
\begin{tablenotes}[flushleft]\scriptsize\item Means of replication-specific latent-signal RMSEs across 100 replications. Parentheses contain Monte Carlo standard errors. Lower is better. Graph/GeoEco: graph estimators; Uniform: equal donor weights; STI: spatio-temporal PRE imputation followed by SCM; STD: direct spatio-temporal prediction; -X: PRE economic covariates. Ordinary SCM is evaluated on the same PRE-observed group only.
\end{tablenotes}\end{threeparttable}\end{table}

%% file: tables/cf_partial_existing.tex
\begin{table}[!htbp]\centering
\caption{Partial-PRE counterfactual signal RMSE and Monte Carlo standard errors.}\label{tab:cf-partial-existing}
\begin{threeparttable}\footnotesize\setlength{\tabcolsep}{3pt}
\begin{tabular}{@{}lrrrrrrrr@{}}\toprule
DGP & Graph & GeoEco & Uniform & STI & STI-X & STD & STD-X & Ordinary SCM \\\midrule
D1 & \shortstack[r]{0.0189\\(0.0005)} & \shortstack[r]{0.0187\\(0.0005)} & \shortstack[r]{0.1838\\(0.0032)} & \shortstack[r]{0.0170\\(0.0005)} & \shortstack[r]{0.0172\\(0.0005)} & \shortstack[r]{0.0197\\(0.0005)} & \shortstack[r]{0.0199\\(0.0005)} & \shortstack[r]{0.0195\\(0.0006)} \\
D2 & \shortstack[r]{0.0245\\(0.0008)} & \shortstack[r]{0.0245\\(0.0008)} & \shortstack[r]{0.1627\\(0.0033)} & \shortstack[r]{0.0202\\(0.0005)} & \shortstack[r]{0.0197\\(0.0005)} & \shortstack[r]{0.0380\\(0.0011)} & \shortstack[r]{0.0335\\(0.0008)} & \shortstack[r]{0.0197\\(0.0006)} \\
D3 & \shortstack[r]{0.0239\\(0.0008)} & \shortstack[r]{0.0229\\(0.0008)} & \shortstack[r]{0.1650\\(0.0033)} & \shortstack[r]{0.0201\\(0.0005)} & \shortstack[r]{0.0199\\(0.0005)} & \shortstack[r]{0.0393\\(0.0011)} & \shortstack[r]{0.0343\\(0.0008)} & \shortstack[r]{0.0198\\(0.0005)} \\
D4 & \shortstack[r]{0.0312\\(0.0006)} & \shortstack[r]{0.0303\\(0.0006)} & \shortstack[r]{0.0746\\(0.0017)} & \shortstack[r]{0.0322\\(0.0007)} & \shortstack[r]{0.0322\\(0.0007)} & \shortstack[r]{0.0488\\(0.0008)} & \shortstack[r]{0.0479\\(0.0008)} & \shortstack[r]{0.0308\\(0.0007)} \\
D5 & \shortstack[r]{0.0181\\(0.0004)} & \shortstack[r]{0.0179\\(0.0003)} & \shortstack[r]{0.0989\\(0.0017)} & \shortstack[r]{0.0176\\(0.0003)} & \shortstack[r]{0.0173\\(0.0003)} & \shortstack[r]{0.0283\\(0.0006)} & \shortstack[r]{0.0283\\(0.0006)} & \shortstack[r]{0.0172\\(0.0003)} \\
D6 & \shortstack[r]{0.0193\\(0.0006)} & \shortstack[r]{0.0192\\(0.0006)} & \shortstack[r]{0.1974\\(0.0033)} & \shortstack[r]{0.0172\\(0.0003)} & \shortstack[r]{0.0171\\(0.0004)} & \shortstack[r]{0.0193\\(0.0004)} & \shortstack[r]{0.0195\\(0.0004)} & \shortstack[r]{0.0195\\(0.0004)} \\
D7 & \shortstack[r]{0.0209\\(0.0006)} & \shortstack[r]{0.0202\\(0.0006)} & \shortstack[r]{0.0802\\(0.0025)} & \shortstack[r]{0.0202\\(0.0005)} & \shortstack[r]{0.0209\\(0.0006)} & \shortstack[r]{0.0325\\(0.0013)} & \shortstack[r]{0.0324\\(0.0012)} & \shortstack[r]{0.0203\\(0.0006)} \\
D8 & \shortstack[r]{0.0238\\(0.0007)} & \shortstack[r]{0.0231\\(0.0007)} & \shortstack[r]{0.1777\\(0.0032)} & \shortstack[r]{0.0197\\(0.0005)} & \shortstack[r]{0.0205\\(0.0006)} & \shortstack[r]{0.0361\\(0.0010)} & \shortstack[r]{0.0338\\(0.0008)} & \shortstack[r]{0.0195\\(0.0005)} \\
D9 & \shortstack[r]{0.0240\\(0.0006)} & \shortstack[r]{0.0240\\(0.0005)} & \shortstack[r]{0.0598\\(0.0016)} & \shortstack[r]{0.0222\\(0.0005)} & \shortstack[r]{0.0222\\(0.0005)} & \shortstack[r]{0.0341\\(0.0008)} & \shortstack[r]{0.0341\\(0.0008)} & \shortstack[r]{0.0221\\(0.0005)} \\
D10 & \shortstack[r]{0.0244\\(0.0005)} & \shortstack[r]{0.0240\\(0.0005)} & \shortstack[r]{0.0593\\(0.0016)} & \shortstack[r]{0.0224\\(0.0005)} & \shortstack[r]{0.0225\\(0.0005)} & \shortstack[r]{0.0340\\(0.0009)} & \shortstack[r]{0.0340\\(0.0009)} & \shortstack[r]{0.0220\\(0.0005)} \\
\bottomrule\end{tabular}
\begin{tablenotes}[flushleft]\scriptsize\item Means of replication-specific latent-signal RMSEs across 100 replications. Parentheses contain Monte Carlo standard errors. Lower is better. Graph/GeoEco: graph estimators; Uniform: equal donor weights; STI: spatio-temporal PRE imputation followed by SCM; STD: direct spatio-temporal prediction; -X: PRE economic covariates. Ordinary SCM is evaluated on the same PRE-observed group only.
\end{tablenotes}\end{threeparttable}\end{table}

%% file: tables/ordinary_covered_existing.tex
\begin{table}[!htbp]\centering\caption{Ordinary SCM on its PRE-observed population.}\label{tab:ordinary-covered}
\begin{threeparttable}\footnotesize\begin{tabular}{@{}lrrr@{}}\toprule
DGP & CF RMSE & ATT RMSE (pp) & ATT bias (pp) \\\midrule
D1 & \shortstack[r]{0.0177\\(0.0005)} & \shortstack[r]{0.320\\(0.024)} & \shortstack[r]{0.086\\(0.035)} \\
D2 & \shortstack[r]{0.0181\\(0.0005)} & \shortstack[r]{0.424\\(0.026)} & \shortstack[r]{0.030\\(0.044)} \\
D3 & \shortstack[r]{0.0184\\(0.0005)} & \shortstack[r]{0.347\\(0.020)} & \shortstack[r]{0.001\\(0.034)} \\
D4 & \shortstack[r]{0.0296\\(0.0006)} & \shortstack[r]{0.556\\(0.032)} & \shortstack[r]{-0.112\\(0.050)} \\
D5 & \shortstack[r]{0.0155\\(0.0003)} & \shortstack[r]{0.166\\(0.011)} & \shortstack[r]{0.004\\(0.017)} \\
D6 & \shortstack[r]{0.0176\\(0.0004)} & \shortstack[r]{0.361\\(0.026)} & \shortstack[r]{-0.038\\(0.039)} \\
D7 & \shortstack[r]{0.0191\\(0.0005)} & \shortstack[r]{0.720\\(0.042)} & \shortstack[r]{0.186\\(0.063)} \\
D8 & \shortstack[r]{0.0183\\(0.0005)} & \shortstack[r]{0.374\\(0.024)} & \shortstack[r]{-0.020\\(0.039)} \\
D9 & \shortstack[r]{0.0214\\(0.0005)} & \shortstack[r]{0.445\\(0.034)} & \shortstack[r]{-0.043\\(0.048)} \\
D10 & \shortstack[r]{0.0211\\(0.0005)} & \shortstack[r]{0.454\\(0.035)} & \shortstack[r]{-0.069\\(0.049)} \\
\bottomrule\end{tabular}\begin{tablenotes}[flushleft]\scriptsize\item Each evaluation covers the 96 targets with any PRE history, not all 120 targets. Means over 100 replications; parentheses contain Monte Carlo standard errors. These entries are not an all-target ranking. ATT errors refer to the PRE-observed estimand.\end{tablenotes}\end{threeparttable}\end{table}

%% file: tables/paired_all_existing.tex
\begin{table}[!htbp]\centering
\caption{Paired differences in mean signal RMSE: ALL.}\label{tab:paired-all-existing}
\begin{threeparttable}\footnotesize\setlength{\tabcolsep}{5pt}
\begin{tabular}{@{}lrrrr@{}}\toprule
 & \multicolumn{2}{c}{GeoEco $-$ Graph} & \multicolumn{2}{c}{GeoEco $-$ STI-X} \\
DGP & Mean difference & 95\% MC interval & Mean difference & 95\% MC interval \\\midrule
D1 & 0.0005 & [-0.0000, 0.0010] & 0.0017 & [0.0004, 0.0030] \\
D2 & -0.0145 & [-0.0166, -0.0125] & 0.0141 & [0.0118, 0.0164] \\
D3 & -0.0181 & [-0.0206, -0.0157] & 0.0282 & [0.0243, 0.0320] \\
D4 & -0.0033 & [-0.0040, -0.0026] & -0.0035 & [-0.0044, -0.0026] \\
D5 & 0.0014 & [0.0006, 0.0021] & -0.0015 & [-0.0028, -0.0002] \\
D6 & 0.0005 & [0.0000, 0.0010] & 0.0024 & [0.0008, 0.0039] \\
D7 & -0.0045 & [-0.0053, -0.0036] & -0.0009 & [-0.0022, 0.0003] \\
D8 & -0.0124 & [-0.0142, -0.0106] & 0.0096 & [0.0074, 0.0118] \\
D9 & -0.0016 & [-0.0022, -0.0010] & -0.0028 & [-0.0037, -0.0020] \\
D10 & -0.0023 & [-0.0029, -0.0016] & -0.0017 & [-0.0026, -0.0008] \\
\bottomrule\end{tabular}\begin{tablenotes}[flushleft]\scriptsize\item Negative differences favor GeoEco. Intervals are unadjusted paired $t_{99}$ Monte Carlo intervals over 100 datasets. They quantify uncertainty in average simulation performance and are not treatment-effect confidence intervals.\end{tablenotes}\end{threeparttable}\end{table}

%% file: tables/beta_existing.tex
\begin{table}[!htbp]\centering\caption{Selected economic attenuation across 100 replications.}\label{tab:beta-existing}
\begin{tabular}{@{}lrrr@{}}\toprule
DGP & $\beta=0$ count & Median $\beta$ & Range \\\midrule
D1 & 87/100 & 0 & 0--8 \\
D2 & 19/100 & 4 & 0--8 \\
D3 & 18/100 & 4 & 0--16 \\
D4 & 30/100 & 4 & 0--8 \\
D5 & 64/100 & 0 & 0--16 \\
D6 & 91/100 & 0 & 0--8 \\
D7 & 19/100 & 4 & 0--16 \\
D8 & 18/100 & 4 & 0--8 \\
D9 & 29/100 & 4 & 0--16 \\
D10 & 32/100 & 4 & 0--16 \\
\bottomrule\end{tabular}\end{table}

%% file: tables/local_existing.tex
\begin{table}[!htbp]\centering\caption{Change in the GeoEco-minus-Graph RMSE difference under local misalignment.}\label{tab:local-existing}
\begin{threeparttable}\footnotesize\begin{tabular}{@{}lrrr@{}}\toprule
Group & Mean change & MCSE & 95\% MC interval \\\midrule
ALL & -0.00064 & 0.00040 & [-0.00143, 0.00014] \\
complete pre & 0.00032 & 0.00036 & [-0.00039, 0.00103] \\
partial pre & -0.00032 & 0.00042 & [-0.00116, 0.00051] \\
post only & -0.00169 & 0.00089 & [-0.00346, 0.00008] \\
\bottomrule\end{tabular}\begin{tablenotes}[flushleft]\scriptsize\item D10 minus D9, paired by seed. Intervals concern mean performance changes and are not treatment-effect intervals.\end{tablenotes}\end{threeparttable}\end{table}

%% file: tables/weight_recovery_postonly.tex
\begin{table}[!htbp]\centering
\caption{Post-only donor-weight recovery and concentration.}\label{tab:weight-recovery-postonly}
\begin{minipage}{\linewidth}\centering\footnotesize\setlength{\tabcolsep}{4pt}
\begin{tabular}{@{}lrrrr@{}}\toprule
\multicolumn{5}{@{}l}{\textit{A. Weight recovery and concentration}}\\
DGP & \shortstack{Weight RMSE\\Graph} & \shortstack{Weight RMSE\\GeoEco} & \shortstack{Weight RMSE\\Uniform} & \shortstack{Effective donors\\GeoEco} \\\midrule
D1 & \shortstack[r]{0.0784\\(0.0007)} & \shortstack[r]{0.0784\\(0.0007)} & \shortstack[r]{0.0824\\(0.0005)} & \shortstack[r]{10.42\\(0.14)} \\
D2 & \shortstack[r]{0.0897\\(0.0008)} & \shortstack[r]{0.0875\\(0.0008)} & \shortstack[r]{0.0817\\(0.0007)} & \shortstack[r]{11.79\\(0.18)} \\
D3 & \shortstack[r]{0.0892\\(0.0008)} & \shortstack[r]{0.0862\\(0.0009)} & \shortstack[r]{0.0789\\(0.0009)} & \shortstack[r]{11.99\\(0.19)} \\
D4 & \shortstack[r]{0.0813\\(0.0006)} & \shortstack[r]{0.0790\\(0.0006)} & \shortstack[r]{0.0811\\(0.0007)} & \shortstack[r]{13.74\\(0.20)} \\
D5 & \shortstack[r]{0.0859\\(0.0004)} & \shortstack[r]{0.0871\\(0.0006)} & \shortstack[r]{0.0804\\(0.0004)} & \shortstack[r]{15.53\\(0.25)} \\
\midrule
D6 & \shortstack[r]{0.0766\\(0.0009)} & \shortstack[r]{0.0766\\(0.0009)} & \shortstack[r]{0.0701\\(0.0006)} & \shortstack[r]{11.29\\(0.18)} \\
D7 & \shortstack[r]{0.0965\\(0.0008)} & \shortstack[r]{0.0950\\(0.0008)} & \shortstack[r]{0.0963\\(0.0007)} & \shortstack[r]{12.73\\(0.23)} \\
D8 & \shortstack[r]{0.0980\\(0.0009)} & \shortstack[r]{0.0958\\(0.0009)} & \shortstack[r]{0.0914\\(0.0007)} & \shortstack[r]{10.56\\(0.17)} \\
D9 & \shortstack[r]{0.0836\\(0.0006)} & \shortstack[r]{0.0830\\(0.0007)} & \shortstack[r]{0.0811\\(0.0006)} & \shortstack[r]{14.73\\(0.21)} \\
D10 & \shortstack[r]{0.0867\\(0.0006)} & \shortstack[r]{0.0867\\(0.0007)} & \shortstack[r]{0.0834\\(0.0006)} & \shortstack[r]{14.49\\(0.27)} \\
\bottomrule\end{tabular}\par\medskip
\begin{tabular}{@{}lrrrr@{}}\toprule
\multicolumn{5}{@{}l}{\textit{B. Paired weight-RMSE differences}}\\
 & \multicolumn{2}{c}{GeoEco $-$ Graph} & \multicolumn{2}{c}{GeoEco $-$ Uniform} \\
DGP & Difference & 95\% MC interval & Difference & 95\% MC interval \\\midrule
D1 & 0.0000 & [-0.0002, 0.0002] & -0.0040 & [-0.0054, -0.0026] \\
D2 & -0.0022 & [-0.0029, -0.0015] & 0.0058 & [0.0045, 0.0071] \\
D3 & -0.0029 & [-0.0036, -0.0022] & 0.0073 & [0.0058, 0.0088] \\
D4 & -0.0022 & [-0.0027, -0.0018] & -0.0021 & [-0.0030, -0.0012] \\
D5 & 0.0012 & [0.0006, 0.0018] & 0.0067 & [0.0058, 0.0075] \\
\midrule
D6 & 0.0001 & [-0.0001, 0.0002] & 0.0065 & [0.0047, 0.0083] \\
D7 & -0.0015 & [-0.0022, -0.0008] & -0.0013 & [-0.0023, -0.0002] \\
D8 & -0.0022 & [-0.0027, -0.0017] & 0.0043 & [0.0030, 0.0056] \\
D9 & -0.0006 & [-0.0013, 0.0000] & 0.0019 & [0.0010, 0.0027] \\
D10 & -0.0000 & [-0.0008, 0.0007] & 0.0034 & [0.0025, 0.0042] \\
\bottomrule\end{tabular}
\par\medskip{\raggedright\scriptsize\noindent Each DGP uses 100 replications and 24 post-only targets per replication. Parentheses in panel A contain Monte Carlo standard errors. Panel B gives paired means and 95\% MC intervals; negative differences favor GeoEco. Effective donors are averaged over targets and replications; Uniform has 24. D1--D5 are primary and D6--D10 supplementary.
\par}\end{minipage}\end{table}

%% file: tables/att_all_primary.tex
\begin{table}[!htbp]\centering
\caption{ATT recovery: all targets, five primary designs.}\label{tab:att-all-primary}
\begin{minipage}{\linewidth}\centering\footnotesize\setlength{\tabcolsep}{3pt}
\begin{tabular}{@{}lrrrrrrrr@{}}\toprule
\multicolumn{9}{@{}l}{\textit{A. Graph, covariate, and uniform-weight methods}}\\
DGP & \multicolumn{2}{c}{Graph} & \multicolumn{2}{c}{GeoEco} & \multicolumn{2}{c}{Cov-SCM} & \multicolumn{2}{c}{Uniform} \\
 & RMSE & Bias & RMSE & Bias & RMSE & Bias & RMSE & Bias \\\midrule
D1 & 0.488 & 0.173 & 0.489 & 0.165 & 2.374 & -0.482 & 3.283 & 0.112 \\
D2 & 0.773 & -0.032 & 0.676 & -0.001 & 2.710 & 0.199 & 3.152 & 0.609 \\
D3 & 2.500 & -2.463 & 1.774 & -1.692 & 2.759 & -0.252 & 2.816 & -0.569 \\
D4 & 0.634 & -0.142 & 0.578 & -0.124 & 3.873 & -3.118 & 1.547 & -0.877 \\
D5 & 0.481 & -0.051 & 0.494 & -0.076 & 3.513 & -0.086 & 0.727 & 0.033 \\
\bottomrule\end{tabular}\par\medskip
\begin{tabular}{@{}lrrrrrrrr@{}}\toprule
\multicolumn{9}{@{}l}{\textit{B. Spatio-temporal methods}}\\
DGP & \multicolumn{2}{c}{STI} & \multicolumn{2}{c}{STI-X} & \multicolumn{2}{c}{STD} & \multicolumn{2}{c}{STD-X} \\
 & RMSE & Bias & RMSE & Bias & RMSE & Bias & RMSE & Bias \\\midrule
D1 & 0.328 & 0.088 & 0.338 & 0.098 & 0.521 & -0.043 & 0.546 & -0.047 \\
D2 & 0.734 & 0.013 & 0.512 & 0.038 & 2.048 & 0.351 & 1.246 & 0.279 \\
D3 & 2.092 & -2.076 & 0.550 & -0.306 & 2.251 & -0.866 & 1.244 & 0.085 \\
D4 & 0.611 & -0.139 & 0.589 & -0.116 & 1.259 & -0.606 & 1.206 & -0.613 \\
D5 & 0.470 & 0.042 & 0.479 & 0.031 & 1.135 & 0.062 & 1.120 & 0.134 \\
\bottomrule\end{tabular}\par\medskip
\par\medskip{\raggedright\scriptsize\noindent Percentage points, averaged over 100 replication-specific annual errors. ATT is $100\{\exp(\overline\tau_t)-1\}$, the percentage transformation of the mean log effect. Bias is estimated minus true ATT. All methods evaluate the same indicated population. Monte Carlo standard errors are reported in the Online Appendix. Graph/GeoEco: graph estimators; Cov-SCM: coordinate and PRE economic-covariate balancing; Uniform: equal donor weights; STI: spatio-temporal PRE imputation followed by SCM; STD: direct spatio-temporal prediction; -X: economic covariates in the outcome mean model.
\par}\end{minipage}\end{table}

%% file: tables/att_mcse_primary.tex
\begin{table}[!htbp]\centering\caption{ATT RMSE and Monte Carlo standard errors in the primary designs.}\label{tab:att-mcse-primary}
\begin{minipage}{\linewidth}\centering\footnotesize\setlength{\tabcolsep}{3pt}
\begin{tabular}{@{}lrrrrrrrr@{}}\toprule
\multicolumn{9}{@{}l}{\textit{A. Post-only targets}}\\
DGP & Graph & GeoEco & Cov-SCM & Uniform & STI & STI-X & STD & STD-X \\\midrule
D1 & \shortstack[r]{0.926\\(0.069)} & \shortstack[r]{0.971\\(0.073)} & \shortstack[r]{2.757\\(0.204)} & \shortstack[r]{4.002\\(0.293)} & \shortstack[r]{0.730\\(0.049)} & \shortstack[r]{0.805\\(0.054)} & \shortstack[r]{0.888\\(0.065)} & \shortstack[r]{0.943\\(0.070)} \\
D2 & \shortstack[r]{2.769\\(0.208)} & \shortstack[r]{2.092\\(0.172)} & \shortstack[r]{2.998\\(0.226)} & \shortstack[r]{3.877\\(0.312)} & \shortstack[r]{2.763\\(0.225)} & \shortstack[r]{1.519\\(0.104)} & \shortstack[r]{3.220\\(0.249)} & \shortstack[r]{1.771\\(0.130)} \\
D3 & \shortstack[r]{11.873\\(0.360)} & \shortstack[r]{8.199\\(0.396)} & \shortstack[r]{2.961\\(0.212)} & \shortstack[r]{9.910\\(0.486)} & \shortstack[r]{10.186\\(0.360)} & \shortstack[r]{2.145\\(0.160)} & \shortstack[r]{9.136\\(0.427)} & \shortstack[r]{2.292\\(0.170)} \\
D4 & \shortstack[r]{1.508\\(0.096)} & \shortstack[r]{1.337\\(0.092)} & \shortstack[r]{3.917\\(0.215)} & \shortstack[r]{1.845\\(0.115)} & \shortstack[r]{1.513\\(0.085)} & \shortstack[r]{1.415\\(0.081)} & \shortstack[r]{1.771\\(0.112)} & \shortstack[r]{1.772\\(0.106)} \\
D5 & \shortstack[r]{2.047\\(0.148)} & \shortstack[r]{2.145\\(0.157)} & \shortstack[r]{5.220\\(0.405)} & \shortstack[r]{1.629\\(0.125)} & \shortstack[r]{2.186\\(0.168)} & \shortstack[r]{2.244\\(0.170)} & \shortstack[r]{2.364\\(0.181)} & \shortstack[r]{2.465\\(0.185)} \\
\bottomrule\end{tabular}\par\medskip
\begin{tabular}{@{}lrrrrrrrr@{}}\toprule
\multicolumn{9}{@{}l}{\textit{B. All targets}}\\
DGP & Graph & GeoEco & Cov-SCM & Uniform & STI & STI-X & STD & STD-X \\\midrule
D1 & \shortstack[r]{0.488\\(0.036)} & \shortstack[r]{0.489\\(0.036)} & \shortstack[r]{2.374\\(0.152)} & \shortstack[r]{3.283\\(0.250)} & \shortstack[r]{0.328\\(0.022)} & \shortstack[r]{0.338\\(0.022)} & \shortstack[r]{0.521\\(0.046)} & \shortstack[r]{0.546\\(0.048)} \\
D2 & \shortstack[r]{0.773\\(0.053)} & \shortstack[r]{0.676\\(0.041)} & \shortstack[r]{2.710\\(0.208)} & \shortstack[r]{3.152\\(0.239)} & \shortstack[r]{0.734\\(0.052)} & \shortstack[r]{0.512\\(0.040)} & \shortstack[r]{2.048\\(0.147)} & \shortstack[r]{1.246\\(0.082)} \\
D3 & \shortstack[r]{2.500\\(0.092)} & \shortstack[r]{1.774\\(0.091)} & \shortstack[r]{2.759\\(0.198)} & \shortstack[r]{2.816\\(0.206)} & \shortstack[r]{2.092\\(0.085)} & \shortstack[r]{0.550\\(0.038)} & \shortstack[r]{2.251\\(0.150)} & \shortstack[r]{1.244\\(0.086)} \\
D4 & \shortstack[r]{0.634\\(0.041)} & \shortstack[r]{0.578\\(0.039)} & \shortstack[r]{3.873\\(0.201)} & \shortstack[r]{1.547\\(0.089)} & \shortstack[r]{0.611\\(0.029)} & \shortstack[r]{0.589\\(0.028)} & \shortstack[r]{1.259\\(0.069)} & \shortstack[r]{1.206\\(0.066)} \\
D5 & \shortstack[r]{0.481\\(0.031)} & \shortstack[r]{0.494\\(0.032)} & \shortstack[r]{3.513\\(0.234)} & \shortstack[r]{0.727\\(0.051)} & \shortstack[r]{0.470\\(0.033)} & \shortstack[r]{0.479\\(0.033)} & \shortstack[r]{1.135\\(0.080)} & \shortstack[r]{1.120\\(0.084)} \\
\bottomrule\end{tabular}\par\medskip
\par\medskip{\raggedright\scriptsize\noindent Percentage points. Parentheses contain Monte Carlo standard errors across 100 replication-specific annual ATT RMSEs; these are not treatment-effect standard errors. Graph/GeoEco: graph estimators; Cov-SCM: coordinate and PRE economic-covariate balancing; Uniform: equal donor weights; STI: spatio-temporal PRE imputation followed by SCM; STD: direct spatio-temporal prediction; -X: economic covariates in the outcome mean model.
\par}\end{minipage}\end{table}

%% file: tables/selection_existing.tex
\begin{table}[!t]\centering\caption{How selective post-only coverage changes the estimand.}\label{tab:selection}
\begin{threeparttable}\footnotesize\setlength{\tabcolsep}{3pt}
\begin{tabular}{@{}lrrrrrr@{}}\toprule
DGP & \shortstack{True ALL\\ATT} & \shortstack{True PRE\\ATT} & \shortstack{True post-only\\ATT} & \shortstack{Ordinary SCM\\estimate} & \shortstack{Bias vs. PRE\\estimand} & \shortstack{Bias vs. ALL\\estimand} \\\midrule
D6 & 3.63 & 3.27 & 5.10 & 3.23 & -0.04 & -0.40 \\
D3 & 3.63 & 3.32 & 4.87 & 3.32 & 0.00 & -0.31 \\
\bottomrule\end{tabular}\begin{tablenotes}[flushleft]\scriptsize\item Percentage points, averaged over POST years and 100 replications. Ordinary SCM estimates the effect for the 96 PRE-observed targets (60 complete-PRE plus 36 partial-PRE). Comparing that estimate with the all-target truth mixes estimation error with a change of population. As a separate summary for the later POST years 2010--2015, after the five-year treatment-effect ramp-up, the corresponding true-ATT population gaps are 10.0\% in D6 and 8.5\% in D3.\end{tablenotes}\end{threeparttable}\end{table}

%% file: tables/att_existing_main.tex
\begin{table}[!t]\centering
\caption{ATT recovery: all targets.}\label{tab:att-existing}
\begin{minipage}{\linewidth}\centering\footnotesize\setlength{\tabcolsep}{3pt}
\begin{tabular}{@{}lrrrrrr@{}}\toprule
\multicolumn{7}{@{}l}{\textit{A. Graph methods and uniform weights}}\\
DGP & \multicolumn{2}{c}{Graph} & \multicolumn{2}{c}{GeoEco} & \multicolumn{2}{c}{Uniform} \\
 & RMSE & Bias & RMSE & Bias & RMSE & Bias \\\midrule
D1 & 0.488 & 0.173 & 0.489 & 0.165 & 3.283 & 0.112 \\
D2 & 0.773 & -0.032 & 0.676 & -0.001 & 3.152 & 0.609 \\
D3 & 2.500 & -2.463 & 1.774 & -1.692 & 2.816 & -0.569 \\
D4 & 0.634 & -0.142 & 0.578 & -0.124 & 1.547 & -0.877 \\
D5 & 0.481 & -0.051 & 0.494 & -0.076 & 0.727 & 0.033 \\
D6 & 0.576 & -0.309 & 0.590 & -0.319 & 3.379 & -0.017 \\
D7 & 0.623 & 0.146 & 0.572 & 0.154 & 2.743 & -0.990 \\
D8 & 0.721 & 0.043 & 0.623 & 0.051 & 2.422 & 0.201 \\
D9 & 0.666 & 0.060 & 0.627 & 0.061 & 1.533 & -1.024 \\
D10 & 0.599 & 0.079 & 0.567 & 0.078 & 1.518 & -1.097 \\
\bottomrule\end{tabular}
\par\medskip
\begin{tabular}{@{}lrrrrrrrr@{}}\toprule
\multicolumn{9}{@{}l}{\textit{B. Spatio-temporal benchmarks}}\\
DGP & \multicolumn{2}{c}{STI} & \multicolumn{2}{c}{STI-X} & \multicolumn{2}{c}{STD} & \multicolumn{2}{c}{STD-X} \\
 & RMSE & Bias & RMSE & Bias & RMSE & Bias & RMSE & Bias \\\midrule
D1 & 0.328 & 0.088 & 0.338 & 0.098 & 0.521 & -0.043 & 0.546 & -0.047 \\
D2 & 0.734 & 0.013 & 0.512 & 0.038 & 2.048 & 0.351 & 1.246 & 0.279 \\
D3 & 2.092 & -2.076 & 0.550 & -0.306 & 2.251 & -0.866 & 1.244 & 0.085 \\
D4 & 0.611 & -0.139 & 0.589 & -0.116 & 1.259 & -0.606 & 1.206 & -0.613 \\
D5 & 0.470 & 0.042 & 0.479 & 0.031 & 1.135 & 0.062 & 1.120 & 0.134 \\
D6 & 0.412 & -0.088 & 0.434 & -0.137 & 0.567 & -0.003 & 0.585 & -0.047 \\
D7 & 0.804 & 0.141 & 0.800 & 0.118 & 2.317 & -0.699 & 2.309 & -0.604 \\
D8 & 0.618 & 0.010 & 0.468 & 0.034 & 1.800 & 0.331 & 1.314 & 0.129 \\
D9 & 0.491 & -0.054 & 0.501 & -0.057 & 1.117 & -0.661 & 1.109 & -0.700 \\
D10 & 0.491 & -0.029 & 0.482 & -0.036 & 1.096 & -0.749 & 1.125 & -0.779 \\
\bottomrule\end{tabular}
\par\medskip
\par\medskip{\raggedright\scriptsize\noindent Percentage points, averaged across 100 replications. Each replication evaluates annual errors over 2005--2015. ATT is $100\{\exp(\overline\tau_t)-1\}$, the percentage transformation of the mean log effect. Bias is estimated minus true ATT. All methods evaluate the same indicated target group. Monte Carlo standard errors are given in the Online Appendix. Graph/GeoEco: graph estimators; Uniform: equal donor weights; STI: spatio-temporal PRE imputation followed by SCM; STD: direct spatio-temporal prediction; -X: PRE economic covariates.
\par}\end{minipage}\end{table}

%% file: tables/att_post_existing.tex
\begin{table}[!htbp]\centering
\caption{ATT recovery: post-only targets.}\label{tab:att-post-existing}
\begin{minipage}{\linewidth}\centering\footnotesize\setlength{\tabcolsep}{3pt}
\begin{tabular}{@{}lrrrrrr@{}}\toprule
\multicolumn{7}{@{}l}{\textit{A. Graph methods and uniform weights}}\\
DGP & \multicolumn{2}{c}{Graph} & \multicolumn{2}{c}{GeoEco} & \multicolumn{2}{c}{Uniform} \\
 & RMSE & Bias & RMSE & Bias & RMSE & Bias \\\midrule
D1 & 0.926 & 0.292 & 0.971 & 0.285 & 4.002 & 0.302 \\
D2 & 2.769 & -0.109 & 2.092 & -0.012 & 3.877 & 0.364 \\
D3 & 11.873 & -11.838 & 8.199 & -8.164 & 9.910 & -9.856 \\
D4 & 1.508 & -0.174 & 1.337 & -0.117 & 1.845 & -0.917 \\
D5 & 2.047 & -0.147 & 2.145 & -0.268 & 1.629 & 0.087 \\
D6 & 1.941 & -1.664 & 2.025 & -1.687 & 7.669 & -7.329 \\
D7 & 1.572 & 0.159 & 1.441 & 0.258 & 2.926 & -0.954 \\
D8 & 2.527 & 0.290 & 1.958 & 0.363 & 3.951 & 0.232 \\
D9 & 1.202 & 0.171 & 1.116 & 0.091 & 1.759 & -1.049 \\
D10 & 1.273 & 0.457 & 1.243 & 0.366 & 1.558 & -0.879 \\
\bottomrule\end{tabular}
\par\medskip
\begin{tabular}{@{}lrrrrrrrr@{}}\toprule
\multicolumn{9}{@{}l}{\textit{B. Spatio-temporal benchmarks}}\\
DGP & \multicolumn{2}{c}{STI} & \multicolumn{2}{c}{STI-X} & \multicolumn{2}{c}{STD} & \multicolumn{2}{c}{STD-X} \\
 & RMSE & Bias & RMSE & Bias & RMSE & Bias & RMSE & Bias \\\midrule
D1 & 0.730 & 0.012 & 0.805 & 0.058 & 0.888 & -0.100 & 0.943 & -0.053 \\
D2 & 2.763 & -0.051 & 1.519 & 0.163 & 3.220 & 0.277 & 1.771 & 0.420 \\
D3 & 10.186 & -10.156 & 2.145 & -1.430 & 9.136 & -9.005 & 2.292 & -0.930 \\
D4 & 1.513 & -0.281 & 1.415 & -0.182 & 1.771 & -0.679 & 1.772 & -0.605 \\
D5 & 2.186 & 0.231 & 2.244 & 0.174 & 2.364 & 0.325 & 2.465 & 0.350 \\
D6 & 1.274 & -0.274 & 1.381 & -0.444 & 1.404 & -0.293 & 1.477 & -0.475 \\
D7 & 1.738 & -0.040 & 1.627 & -0.097 & 2.841 & -0.877 & 2.698 & -0.814 \\
D8 & 2.584 & 0.193 & 1.560 & 0.198 & 3.181 & 0.472 & 2.181 & 0.246 \\
D9 & 1.170 & -0.080 & 1.218 & -0.094 & 1.452 & -0.648 & 1.480 & -0.698 \\
D10 & 1.128 & 0.156 & 1.177 & 0.122 & 1.339 & -0.539 & 1.413 & -0.595 \\
\bottomrule\end{tabular}
\par\medskip
\par\medskip{\raggedright\scriptsize\noindent Percentage points, averaged across 100 replications. Each replication evaluates annual errors over 2005--2015. ATT is $100\{\exp(\overline\tau_t)-1\}$, the percentage transformation of the mean log effect. Bias is estimated minus true ATT. All methods evaluate the same indicated target group. Monte Carlo standard errors are given in the Online Appendix. Graph/GeoEco: graph estimators; Uniform: equal donor weights; STI: spatio-temporal PRE imputation followed by SCM; STD: direct spatio-temporal prediction; -X: PRE economic covariates.
\par}\end{minipage}\end{table}

%% file: tables/att_mcse_all_existing.tex
\begin{table}[!htbp]\centering\caption{ATT RMSE and Monte Carlo standard errors: ALL.}\label{tab:att-mcse-all}
\begin{threeparttable}\footnotesize\setlength{\tabcolsep}{3pt}
\begin{tabular}{@{}lrrrrrrr@{}}\toprule
DGP & Graph & GeoEco & Uniform & STI & STI-X & STD & STD-X \\\midrule
D1 & \shortstack[r]{0.488\\(0.036)} & \shortstack[r]{0.489\\(0.036)} & \shortstack[r]{3.283\\(0.250)} & \shortstack[r]{0.328\\(0.022)} & \shortstack[r]{0.338\\(0.022)} & \shortstack[r]{0.521\\(0.046)} & \shortstack[r]{0.546\\(0.048)} \\
D2 & \shortstack[r]{0.773\\(0.053)} & \shortstack[r]{0.676\\(0.041)} & \shortstack[r]{3.152\\(0.239)} & \shortstack[r]{0.734\\(0.052)} & \shortstack[r]{0.512\\(0.040)} & \shortstack[r]{2.048\\(0.147)} & \shortstack[r]{1.246\\(0.082)} \\
D3 & \shortstack[r]{2.500\\(0.092)} & \shortstack[r]{1.774\\(0.091)} & \shortstack[r]{2.816\\(0.206)} & \shortstack[r]{2.092\\(0.085)} & \shortstack[r]{0.550\\(0.038)} & \shortstack[r]{2.251\\(0.150)} & \shortstack[r]{1.244\\(0.086)} \\
D4 & \shortstack[r]{0.634\\(0.041)} & \shortstack[r]{0.578\\(0.039)} & \shortstack[r]{1.547\\(0.089)} & \shortstack[r]{0.611\\(0.029)} & \shortstack[r]{0.589\\(0.028)} & \shortstack[r]{1.259\\(0.069)} & \shortstack[r]{1.206\\(0.066)} \\
D5 & \shortstack[r]{0.481\\(0.031)} & \shortstack[r]{0.494\\(0.032)} & \shortstack[r]{0.727\\(0.051)} & \shortstack[r]{0.470\\(0.033)} & \shortstack[r]{0.479\\(0.033)} & \shortstack[r]{1.135\\(0.080)} & \shortstack[r]{1.120\\(0.084)} \\
D6 & \shortstack[r]{0.576\\(0.035)} & \shortstack[r]{0.590\\(0.036)} & \shortstack[r]{3.379\\(0.247)} & \shortstack[r]{0.412\\(0.028)} & \shortstack[r]{0.434\\(0.032)} & \shortstack[r]{0.567\\(0.037)} & \shortstack[r]{0.585\\(0.038)} \\
D7 & \shortstack[r]{0.623\\(0.040)} & \shortstack[r]{0.572\\(0.037)} & \shortstack[r]{2.743\\(0.220)} & \shortstack[r]{0.804\\(0.047)} & \shortstack[r]{0.800\\(0.048)} & \shortstack[r]{2.317\\(0.168)} & \shortstack[r]{2.309\\(0.158)} \\
D8 & \shortstack[r]{0.721\\(0.042)} & \shortstack[r]{0.623\\(0.037)} & \shortstack[r]{2.422\\(0.192)} & \shortstack[r]{0.618\\(0.041)} & \shortstack[r]{0.468\\(0.028)} & \shortstack[r]{1.800\\(0.138)} & \shortstack[r]{1.314\\(0.101)} \\
D9 & \shortstack[r]{0.666\\(0.046)} & \shortstack[r]{0.627\\(0.045)} & \shortstack[r]{1.533\\(0.110)} & \shortstack[r]{0.491\\(0.036)} & \shortstack[r]{0.501\\(0.037)} & \shortstack[r]{1.117\\(0.083)} & \shortstack[r]{1.109\\(0.083)} \\
D10 & \shortstack[r]{0.599\\(0.036)} & \shortstack[r]{0.567\\(0.036)} & \shortstack[r]{1.518\\(0.114)} & \shortstack[r]{0.491\\(0.033)} & \shortstack[r]{0.482\\(0.033)} & \shortstack[r]{1.096\\(0.092)} & \shortstack[r]{1.125\\(0.089)} \\
\bottomrule\end{tabular}\begin{tablenotes}[flushleft]\scriptsize\item Percentage points. Parentheses contain Monte Carlo standard errors over 100 replication-specific annual ATT RMSEs. These are not standard errors for an empirical treatment effect.\end{tablenotes}\end{threeparttable}\end{table}

%% file: tables/att_mcse_post_existing.tex
\begin{table}[!htbp]\centering\caption{ATT RMSE and Monte Carlo standard errors: post only.}\label{tab:att-mcse-post}
\begin{threeparttable}\footnotesize\setlength{\tabcolsep}{3pt}
\begin{tabular}{@{}lrrrrrrr@{}}\toprule
DGP & Graph & GeoEco & Uniform & STI & STI-X & STD & STD-X \\\midrule
D1 & \shortstack[r]{0.926\\(0.069)} & \shortstack[r]{0.971\\(0.073)} & \shortstack[r]{4.002\\(0.293)} & \shortstack[r]{0.730\\(0.049)} & \shortstack[r]{0.805\\(0.054)} & \shortstack[r]{0.888\\(0.065)} & \shortstack[r]{0.943\\(0.070)} \\
D2 & \shortstack[r]{2.769\\(0.208)} & \shortstack[r]{2.092\\(0.172)} & \shortstack[r]{3.877\\(0.312)} & \shortstack[r]{2.763\\(0.225)} & \shortstack[r]{1.519\\(0.104)} & \shortstack[r]{3.220\\(0.249)} & \shortstack[r]{1.771\\(0.130)} \\
D3 & \shortstack[r]{11.873\\(0.360)} & \shortstack[r]{8.199\\(0.396)} & \shortstack[r]{9.910\\(0.486)} & \shortstack[r]{10.186\\(0.360)} & \shortstack[r]{2.145\\(0.160)} & \shortstack[r]{9.136\\(0.427)} & \shortstack[r]{2.292\\(0.170)} \\
D4 & \shortstack[r]{1.508\\(0.096)} & \shortstack[r]{1.337\\(0.092)} & \shortstack[r]{1.845\\(0.115)} & \shortstack[r]{1.513\\(0.085)} & \shortstack[r]{1.415\\(0.081)} & \shortstack[r]{1.771\\(0.112)} & \shortstack[r]{1.772\\(0.106)} \\
D5 & \shortstack[r]{2.047\\(0.148)} & \shortstack[r]{2.145\\(0.157)} & \shortstack[r]{1.629\\(0.125)} & \shortstack[r]{2.186\\(0.168)} & \shortstack[r]{2.244\\(0.170)} & \shortstack[r]{2.364\\(0.181)} & \shortstack[r]{2.465\\(0.185)} \\
D6 & \shortstack[r]{1.941\\(0.124)} & \shortstack[r]{2.025\\(0.129)} & \shortstack[r]{7.669\\(0.442)} & \shortstack[r]{1.274\\(0.097)} & \shortstack[r]{1.381\\(0.097)} & \shortstack[r]{1.404\\(0.113)} & \shortstack[r]{1.477\\(0.112)} \\
D7 & \shortstack[r]{1.572\\(0.112)} & \shortstack[r]{1.441\\(0.114)} & \shortstack[r]{2.926\\(0.245)} & \shortstack[r]{1.738\\(0.121)} & \shortstack[r]{1.627\\(0.107)} & \shortstack[r]{2.841\\(0.224)} & \shortstack[r]{2.698\\(0.214)} \\
D8 & \shortstack[r]{2.527\\(0.190)} & \shortstack[r]{1.958\\(0.158)} & \shortstack[r]{3.951\\(0.334)} & \shortstack[r]{2.584\\(0.208)} & \shortstack[r]{1.560\\(0.118)} & \shortstack[r]{3.181\\(0.259)} & \shortstack[r]{2.181\\(0.159)} \\
D9 & \shortstack[r]{1.202\\(0.079)} & \shortstack[r]{1.116\\(0.075)} & \shortstack[r]{1.759\\(0.137)} & \shortstack[r]{1.170\\(0.074)} & \shortstack[r]{1.218\\(0.079)} & \shortstack[r]{1.452\\(0.111)} & \shortstack[r]{1.480\\(0.114)} \\
D10 & \shortstack[r]{1.273\\(0.090)} & \shortstack[r]{1.243\\(0.090)} & \shortstack[r]{1.558\\(0.114)} & \shortstack[r]{1.128\\(0.080)} & \shortstack[r]{1.177\\(0.079)} & \shortstack[r]{1.339\\(0.103)} & \shortstack[r]{1.413\\(0.109)} \\
\bottomrule\end{tabular}\begin{tablenotes}[flushleft]\scriptsize\item Percentage points. Parentheses contain Monte Carlo standard errors over 100 replication-specific annual ATT RMSEs. These are not standard errors for an empirical treatment effect.\end{tablenotes}\end{threeparttable}\end{table}